\documentclass[aps,pre,final,10pt]{revtex4-2}
\usepackage{amsfonts}
\usepackage{amssymb}
\usepackage{amsmath}
\usepackage{graphicx}
\usepackage{url}
\usepackage{times}

\usepackage[normalem]{ulem}
\usepackage{xcolor}

\begin{document}

\title{\textcolor{black}{Reaction-diffusion} spatial modeling of COVID-19: Greece and Andalusia as case examples}

\author{P. G. Kevrekidis}

\affiliation{Department of Mathematics and Statistics, University of Massachusetts Amherst,
Amherst, MA 01003-4515, USA}

\affiliation{Mathematical Institute, University of Oxford, Oxford, UK}

\author{J. Cuevas-Maraver}

\affiliation{Grupo de F\'{\i}sica No Lineal, Departamento de F\'{\i}sica Aplicada I,
Universidad de Sevilla. Escuela Polit\'{e}cnica Superior, C/ Virgen de Africa, 7, 41011-Sevilla, Spain}
\affiliation{Instituto de Matem\'{a}ticas de la Universidad de Sevilla (IMUS). Edificio
Celestino Mutis. Avda. Reina Mercedes s/n, 41012-Sevilla, Spain, Avda Reina Mercedes s/n, E-41012 Sevilla, Spain}

\author{Y. Drossinos}

\affiliation{European Commission, Joint Research Centre, I-21027 Ispra
  (VA), Italy}

\author{Z. Rapti}

\affiliation{Department of Mathematics and Carl R. Woese Institute for Genomic Biology, University of Illinois
at Urbana-Champaign}

\author{G.A. Kevrekidis}

\affiliation{Department of Mathematics and Statistics, University of Massachusetts Amherst,
Amherst, MA 01003-4515, USA}

\bigskip

\date{\today}

\begin{abstract}
We examine the spatial modeling of the outbreak of
COVID-19 in two regions:\textcolor{black}{the autonomous
community of Andalusia in Spain and the mainland of Greece}.
We start with a {zero-dimensional (ODE-level)
compartmental epidemiological model consisting
of Susceptible, Exposed, Asymptomatic,
(symptomatically) Infected,
Hospitalized, Recovered, and deceased
populations (SEAIHR model).  We emphasize the
importance of the viral latent period (reflected in the exposed population)
and the key role of an asymptomatic population.
We optimize model parameters for both regions
by comparing predictions to
the cumulative number of infected and total number of deaths,
the reported data we found to be most reliable,
\textcolor{black}{via minimizing the
$\ell^2$ norm of the difference between predictions and observed data}.
\textcolor{black}{We consider the sensitivity of model
predictions on reasonable variations of model parameters and
initial conditions, and we address issues of parameter identifiability.}
We model both the pre-quarantine and post-quarantine evolution
of the epidemic by a time-dependent
change of the viral transmission rates that arises
in  response to containment measures.
Subsequently, a spatially distributed
version of the 0D model in the form of
reaction-diffusion equations is developed.
We consider that, after an initial localized
seeding of the infection, its spread is governed by
 \textcolor{black}{the diffusion
(and 0D model ``reactions'')
of the asymptomatic and symptomatically infected populations,which
decrease with the imposed restrictive measures.}
\textcolor{black}{We inserted the maps of the two regions, and we imported
population-density data into the finite-element software package
COMSOL Multiphysics$^\circledR$, which was subsequently used
to solve numerically the model PDEs.
Upon discussing how to adapt the 0D model to this spatial setting,
we show
that these models bear significant potential towards capturing both
the well-mixed, zero-dimensional description} and the spatial
expansion of the pandemic in the two regions.
\textcolor{black}{Veins of potential refinement of the model assumptions towards}
future work are also explored.}
\end{abstract}

\maketitle

\section{The Lay of the Land}

\textcolor{black}{Since December 2019, most countries around the globe have been grappling with
how to best contain the COVID-19 pandemic. This emerging infectious disease is caused by the virus SARS-CoV-2, which belongs to the same family
({\it Coronaviridae}) as the viruses responsible for the Severe Acute Respiratory Syndrome (SARS) identified in 2002 in China
\cite{chowell2003}
 and the Middle East Respiratory Syndrome (MERS) that originated in Saudi Arabia a decade later \cite{breban2013}.}
\textcolor{black}{As of this writing (March 15, 2021)}, the number of confirmed infections throughout the
world has already eclipsed \textcolor{black}{120} million individuals
with well over \textcolor{black}{2.5 million deaths~\cite{WHO_Count}, a number
\textcolor{black}{whose rate of increase has subsided and subsequently increased} as the first wave of infections
(pre summer 2020) gave way to the second wave (fall 2020). A third wave in late
winter/spring is feared.}
\textcolor{black}{The race to identify a suitable vaccine started
  almost contemporaneously with the appearance of the virus, and
  progressed very rapidly, despite subsequent delays
in manufacturing and distributing it~\cite{lev4}.
Numerous vaccines have been developed, some approved
for administration to patients  (possibly segregated
in age groups) by national authorities, others are under testing or
under development.
Naturally, the vaccination of whole populations will take time
and may not be relevant in the short term.
Given the time needed to develop and deploy the distribution of
vaccines,
the so-called ``non-therapeutic interventions''~\cite{lev5,lev3}
have been brought to bear (often strictly so) during the earlier and
even during the ongoing wave of the pandemic. Most notable among them
are
social distancing, self-quarantining (when infected),  lockdown and
severely restricted human mobility, limits on the number of persons in gatherings},
and the use of personal protective equipment  (various forms of face masks)
to mitigate the growth of the number of infections. In an unprecedented
for at least a two-generation setting, more than half of the planet's population
has been under the effect of different levels of such measures.

The urgency of this ongoing and rapidly developing global pandemic
has redirected a significant volume of the research community's
efforts in this particular direction. For biologists/clinicians,
as well as for computational physicists/chemists, a race against time is
underway
to understand the binding properties of the virus and its hacking
of the RNA, the action of its spike protein and how to inactivate {it}
via suitable antibody mechanisms~\cite{oxford}. Aerosol scientists and virologists are
trying to understand the role of expelled, virus-loaded
respiratory droplets in the transmission of the virus, and the importance
of the aerosol transmission mode, as opposed to direct or
indirect contact transmission~\cite{Editorialyd2020},
\textcolor{black}{as well as the importance of forced ventilation and air
purification in indoor environments}. At the same time,
a clear sense of urgent need has emerged for mathematicians
and epidemiologists to consider the spreading of the virus over the
population. The focal points of such studies have been
extremely diverse:  from isolated (or nearly isolated)
entities such as restaurants~\cite{ChineseRestaurant} small villages and cruise ships~\cite{cruise}, and
cities~\cite{beijing},
to states/provinces~\cite{louisiana,england,arenas}, and a large number of
countries~\cite{tsironis,SpainSpatial,country11,brazil,sweden,albania,Italy}, including recently
Greece~\cite{NikosGreece}, aside of
course
from the prototypical examples of Wuhan, China~\cite{mbe,TangRisk},
as well as some of the hard-hit Italian regions such as
Lombardy~\cite{siettos}.
Indeed, as of the present writing there are
around \textcolor{black}{3,700} articles in arXiv,  and \textcolor{black}{14,000} in
medRxiv and bioRxiv centered around the theme of COVID-19/SARS-CoV-2 alone, a rather
staggering number given \textcolor{black}{the eleven months since
the disease being declared a pandemic}.
\textcolor{black}{To mention just a few of these recent works, an inspiring
collection of viewpoints regarding recent developments and lingering challenges
in the mathematical modeling of COVID-19 is given in \cite{vespignani2020}.
Data quality and availability, rare but significant
superspreading events, the role of human behavior and host heterogeneity are just
a few of the hindrances faced by modelers.
An evaluation of various different modeling approaches, including the SIR and SEIR
epidemic models, as well as high-dimensional ones, like the one presented in the
present work, is performed in \cite{anirudh2020}. In \cite{meehan2020}, care
has been taken to address the issue of model limitations due to the amount and
quality of data and uncertainty regarding the fraction of asymptomatic infections
and their role in spreading the  disease. Most modeling reported in these works highlights the
fact that all models are imperfect, but some are still useful \cite{holmdahl2020}.
}

  Within this extremely diverse and rapidly evolving landscape, \textcolor{black}{
  including several variants of the virus \cite{burki2021},}
  our team has identified a niche of significant deficiency in the
  current
  level of modeling. The vast majority of the models
  developed essentially ignores the spatial element, considering
  the country in the form of a well-mixed population that can be
  addressed at the level of {\it ordinary differential equations} of
  the extremely widely used form of
  SIR models and multi-component, as well as multi-age group
  generalizations thereof. \textcolor{black}{Readable and informative reviews
  of the mathematics of infectious diseases and epidemics include the works
  in \cite{MathematicsInfectiousDiseases}.}
  While spatial generalizations of such
  SIR models
  do exist~\cite{theo,reluga2004}, they are often used, \textcolor{black}{but
  not exclusively,} at the level of
  interesting
  models of pattern formation, rather than that of realistic
  population level settings.
 \textcolor{black}{In fact, spatial extensions of SIR models have been extensively
 discussed, see for example \cite{BooksSIRDiffusion}.
 They are typically used to model vector-borne diseases whose vectors
 diffuse, as for example
 mosquitoes that transmit malaria~\cite{Malaria}. In the case
 of SARS-CoV-2 the agents that transport the virus are the respiratory
 droplets~\cite{DropletDynamics} that are closely connected, both in time and space, to
 the infectious individuals.
 The behavior of these droplets on the large spatial scales
 considered here (scales associated with regions) is subsumed to the motion of individuals
that we will consider to be diffusive.
This modeling of individual mobility leads to a set
of reaction-diffusion PDEs that may be used to model the spatio-temporal evolution of the pandemic,
an approach that we follow in this work.
A similar approach , which considers
two interacting and isotropically diffusing populations (susceptible and
infected), was adopted by~\cite{Noble}
to model the spreading
 of the mid fourteen century Black Death Plague in Europe,  with particular interest in
 the propagation velocity of one-dimensional traveling waves. Admittedly, the approximation
 of human mobility as diffusive (and isotropic) neglects
 that human mobility is partially predictable and directed, as
 suggested by numerous recent studies, for example Ref.~\cite{YanScalingLaws}.
 Directed motion can be incorporated in our model via a convection term,
 a term that in this initial work we neglect. The work
 presented herein provides the \textit{necessary framework} to include it,
 as  well as other possible extensions like random but long-range effects (emulating travel),
 or anisotropic diffusion.
 We are not
 aware of any similar PDE simulations of the spreading of an epidemic at the large scales (country-wide)
 considered here \textcolor{black}{(but see \cite{france, italyViguerie}
 for two PDE-based studies, the former applied to France and the latter to
 Lombardy)}. Hence, the limitations and potentialities of this approach
 have not been properly assessed. As we argue, such simulations entail severe computational
 challenges, e.g.,  a numerical simulation
 on the spatial grid of an entire country with the population density
 appropriately gridded, and they avail of novel developments in
 Geographic Information Systems to import the region's geometry and to
 properly populate it by the population density. \textcolor{black}{ Moreover,
 current data in Greece and in Andalusia, the regions we explore,
 do not appear in any source that we are aware of in a spatially distributed form
 at the level and scales presented herein.}
 The model we present argues, for the first time, that it is relevant,
 interesting and computationally accessible
 to simulate
models at the level of a country,
and thus to obtain data to compare at an adequately spatially
resolved level. It is thus a prompt for researchers on the epidemiological
side (or cross-disciplinary collaborations of these with mathematical/physical scientists)
to seek to produce such data.
 More importantly, however,
 it sets the
 stage for a more refined, and more
 realistic, description of the spatio-temporal evolution of
 a disease in terms of convection-reaction-diffusion PDE models.}

 Our aim is to enable a broad scale
  of spatial modeling, at the same time leveraging the
  unprecedented availability of data about this pandemic
  and the spatial connectivities/mobility data
 of the human population. The approach that one can select along
  this niche of spatial resolution of the pandemic is, indeed,
  multifold.
  On the one hand, one can aim to formulate a PDE model incorporating
  the ingredients of a generalized SIR formulation. At the same
  time, a complementary viewpoint that is far less computationally
  expensive but possibly quite informative in its own right is a
  metapopulation network approach in the spirit of the work
  of~\cite{vespignani}.
 \textcolor{black}{While networks and metapopulation studies, see for
 example \cite{Keeling2002,Vespignani2011,Italy}, are useful
 in attempting to examine small numbers of groups of individuals, it is clear
 that these cannot properly capture the scale of an entire country. The number
 of nodes would simply be too huge and it would not be possible to capture the
  tremendous variations of population-density scales. Such models could be well suited to
  study propagation of waves of an epidemic between metropolitan centers,
  but they surely are not well suited to capture different and much smaller
  scales (at least not without significant adaptations). For instance,
  they cannot capture dynamics that happens within a node
  (unless they have further structure) and they
  surely cannot capture dynamics that (spatially)
  happens in-between nodes. The interest in
  such models is to consider transport along links if one is
  interested in long-range transport of a virus by, e.g.,
  airline travel or along highways. Here, we seed viral hotspots and explore
  how the virus will spread locally thereafter.  Our infection initialization
  by hotspots aims to emulate the
  initial long-range transport of infectious individuals, and hence
  it becomes an indirect way to incorporate mobility on a network in the absence of convection.
  In that sense, our technique too requires
  significant adaptations but for a different reason: this is in order
  to properly capture mobility of the population that induces
  the spreading of a pandemic.}

  Moreover, one can envision techniques
  (such as the equation-free modeling framework~\cite{eqnfree})
  that may enable the cross-linking of the above two
  approaches, e.g., the use of metapopulation network
  systems to perform PDE-level tasks. Lastly,
  there exist isolated examples of models that take into account
  the structure of different types of networks.
  A particularly nice example in this direction
  is the work of~\cite{epigraph}, which leverages the availability
  of Enron, Facebook and social graphs in the form of adjacency
  matrix patterns that can be used to represent the connectivity
  within a country's network.

 \textcolor{black}{With these considerations in mind,
 we develop an expanded variant of the classical SIR model (ODE model)
 and then focus on its  PDE spatio-temporal generalization.}
 We incorporate particularities of this virus, such as its latent period,
  i.e., that individuals exposed to the virus may be infected but
  not infectious during the latent period, and the significant fraction of
  infected and infectious individuals that do not develop
  symptoms. In Section II we present the spatial model, analyze first
  its ODE variant that will be used to perform appropriate optimization of its parameters
  in the cases of the country of Greece
  (but {\it without} the
  islands, i.e., the mainland of Greece),  and the Spanish autonomous community of
  Andalusia.
  \textcolor{black}{While our study has focused on both regions,
    for practical purposes, we opted to relegate the presentation and
 discussion of our results for Greece to an Appendix. This renders the presentation of
 our model, our approach, and methodology easier to follow, and more focused.}
 The selection of two \textcolor{black}{seemingly unrelated regions} may appear a bit disparate, yet we argue
  these to be particularly interesting examples. Aside from their
  intrinsic interest to the authors,
  these roughly equally-sized regions, with similar population densities,
  exhibit significantly different, i.e., by an order
  of magnitude, number of deaths,
illustrating the potential impact of different policies.
Upon optimizing the ODE results, we use their output
  to formulate the input of the corresponding PDE framework
  and explain how to set it up within the software package COMSOL.
Following the formulation of both the ODE and PDE approaches in section
II, the results for Andalusia and the respective
interpretations and comparisons with reported date are offered in section III.
  We provide numerical results for key categories such as
  cumulative infections and deaths, comparing the PDE results
  both with the available data for these regions and the associated
  ODE results. Finally, in section IV we summarize our findings
  and present our conclusions offering a number of possibilities
  towards future work. \textcolor{black}{The first Appendix summarizes the calculation of the basic reproduction number by the next-generation matrix approach, whereas the second presents
  model results for the mainland of Greece for well-mixed
  and spatially-distributed populations, mimicking our analysis of the spread of the
  virus in Andalusia.}

  \section{Setup of ODE and PDE Models}
  \label{sec:SetUp}

  We first explain the ODE model which is obtained from the full
  PDE model by removing the  convection and diffusion ``spatial aspects" in the
  convection-reaction-diffusion equations of interest.

  At the level of an ODE formulation, a relevant extension of the standard SIR model
  can incorporate some of the key features of this virus, such as, e.g., that a
  fraction of the exposed population remains  {\it asymptomatic} \cite{arons2020}.
  We thus start with a population of {\it susceptibles} ($S$), which may become
  {\it exposed} ($E$) upon the emergence of the virus within the population.
  This represents the well-documented~\cite{elife} feature that the virus is latent
  within the host for a period of time, before he/she becomes infectious (able to
  transmit the  virus to susceptible hosts).
  After this latent period, exposed individuals, in turn, may become
  {\it asymptomatically infectious} ($A$) at rate $\sigma_A$, or (symptomatically)
  {\it infected} ($I$) at rate $\sigma_I$.
  We assume here that both $A$ and $I$ can interact with the susceptibles $S$ with
    respective rates $\beta_{AS}$ and $\beta_{IS}$ to draw
    new members of the population into the group $E$ of individuals exposed
    to the virus. We note that the transmission rates $\beta$
    incorporate the total population size (ODE model) or
    the total population density (PDE Model).
A fraction of hosts in the $I$ class may need hospitalization, thus giving rise to
a population of {\it hospitalized} ($H$) at rate $\gamma M$. Among these, a
fraction responds to the  treatments,
thus leading ultimately to a population of {\it recovered} ($R$) at a rate $(1-\omega) \chi$.
At the same time, the seriously ill who are hospitalized yield also a number of {\it deceased} ($D$) at a rate
$\omega \psi$. Asymptomatically infected  hosts recover at a rate $M_{AR}$
(i.e., {\it asymptomatic recovered})
and move into class $AR$ and the seriously ill recover at a rate
$(1-\gamma)M$.
While $AR$ could, in principle, be merged with $R$, in our view,
  it is meaningful to maintain these two populations separate since
  $R$ is a measurable quantity within available COVID-19 data (and,
  hence, comparable to model predictions),
  while $AR$ is not. Notice that the above constants reflect both the population
fraction partition (e.g., $\omega$ vs $1-\omega$) and the (inverse) time scales
(e.g., $\chi$ vs. $\psi$) for transition between subgroups.

A weak effect of net change of the population due to other birth or
other mortality factors ($-\mu S$) can be incorporated in the susceptibles and
 can be adequately assessed from census data, yet we do not
 incorporate
 it in the $D$ population aiming to evaluate purely the deaths stemming
 from COVID-19.
 {Here, we briefly note two points. The pool of
 susceptible
 individuals is not significantly affected (over the time scale of our
 study)
 from this term which can be safely neglected for our purposes.
 Secondly, one can include such a term in the rest of the populations
 involved
 in our study. However, the (underlying) health conditions often involved in such
 mortality often lead to complications in the concurrent presence of
 COVID-19 and when this leads to mortality, the latter is attributed
 to COVID-19. Hence, we do not include such a separate term in the
 rest of the equations.}

\begin{figure}[!htb]
\centering
\includegraphics[width=0.8\textwidth]{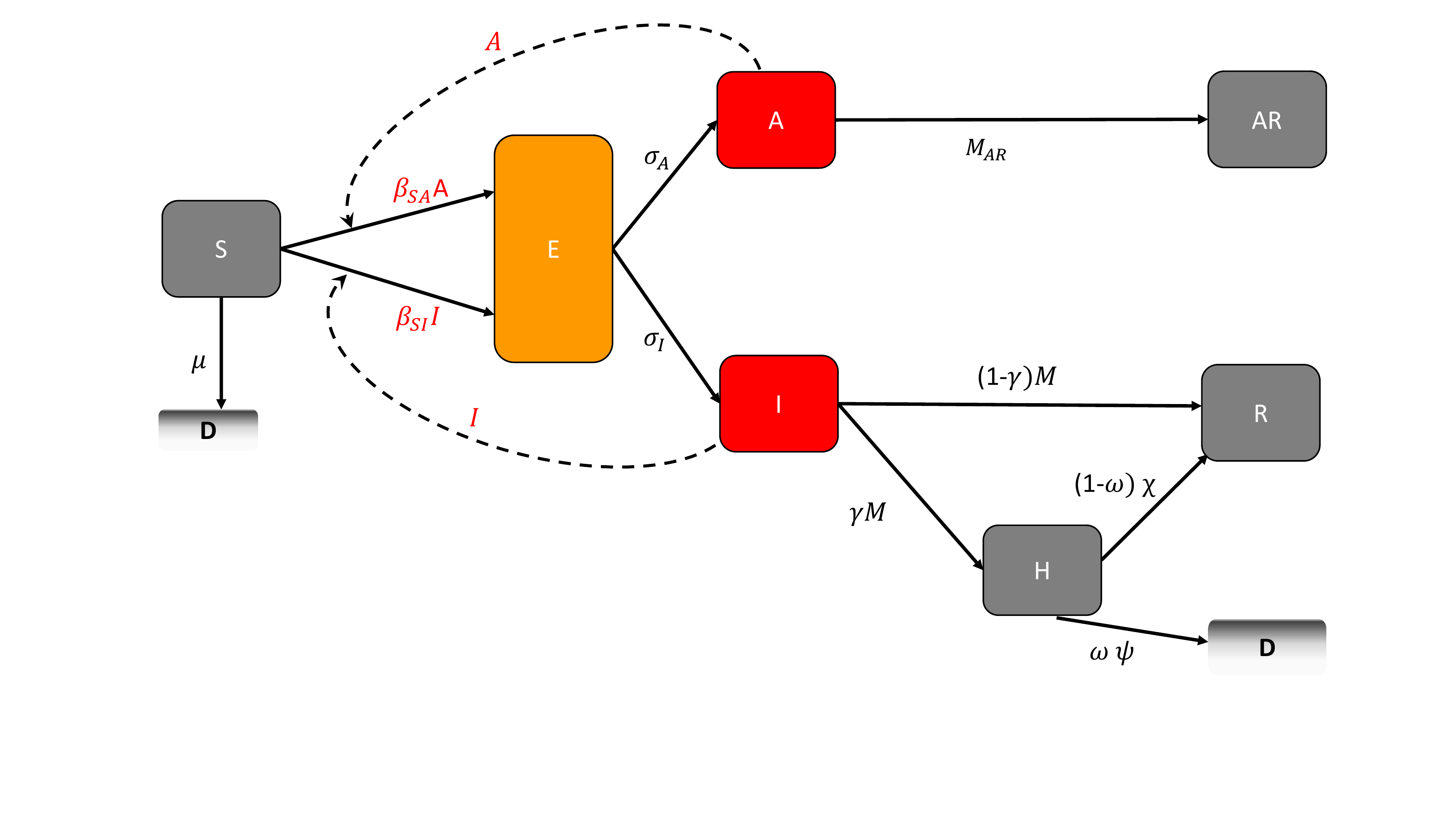}
\caption{Schematic diagram of the SEAIHR model. \textcolor{black}{The dashed lines denote
the interaction of the infectious populations, Asymptomatic and symptomatically Infected,
with the Susceptible population that leads to infection.}}
\label{fig:SEAIHR}
\end{figure}

 The above {specify}  the ``ODE parameters'' within the group; these
reflect
processes that happen either in an averaged way  at the ``well mixed'' level (when no
spatial dependence is assigned) of the ODEs or processes that happen locally at
every
point in space for the PDEs. We will return to this when we
discuss parameter conversions in the next section.
 The relevant populations and rates of conversion
 can be seen in a self-contained form in Fig.~\ref{fig:SEAIHR}.
\textcolor{black}{The ODE version of the proposed model, described in
the figure, is similar to the SEAIHR model used to model the
 transmission of  the MERS coronavirus in the Republic of Korea~\cite{SEAIHRversionMERS},
 but slightly distinct in its treatment of the asymptomatically recovered, the recovered,
 and the hospitalized who, in the current model, do not transmit the virus, as
 they are expected to be in isolation.}

  The relevant population model at the PDE level reads:
 \begin{eqnarray}
 S_t &=& \nabla \left(\mathfrak{D}_S \nabla S\right) -(\vec{v} \cdot \nabla) S-
 \beta_{SA} S A -\beta_{SI} S I {- \mu S} ,
    \label{eqn1}
    \\
    E_t&=& \nabla \left( \mathfrak{D}_E \nabla E\right) -(\vec{v} \cdot \nabla) E +
    \beta_{SA} S A + \beta_{SI} S I - (\sigma_A + \sigma_I) E ,
    \label{eqn2}
    \\
    A_t & =& \nabla \left( \mathfrak{D}_A \nabla A\right) -(\vec{v} \cdot \nabla) A+\sigma_A E - M_{AR} A
 ,    \label{eqn3}\\
AR_t & =& \nabla \left( \mathfrak{D}_{AR} \nabla AR \right) -(\vec{v} \cdot \nabla) AR +M_{AR} A ,
\label{eqn4}\\
    I_t&=&  \sigma_I E - M I ,
    \label{eqn5}
    \\
    H_t & =& \gamma M I -(1-\omega) \chi H - \omega \psi H ,
\label{eqn6}
    \\
    R_t&=& \nabla \left( \mathfrak{D}_R \nabla R\right) + (1 -\gamma) M I + (1-\omega)\chi H ,
    \label{eqn7}
    \\
    D_t&=& \omega \psi H .
    \label{eqn8}
  \end{eqnarray}

We now turn to the PDE properties of the model involving spatial spreading of the pandemic.
Initially, we note that we do not anticipate that
  infected (which should be self-quarantined), hospitalized (or at
  stages
  thereafter) will have a diffusivity, i.e., $\mathfrak{D}_I= \mathfrak{D}_H=0$
  in the initial installment of the model.
  As regards the $R$ and $AR$, in principle they can have a diffusivity
  (although there is a period of recovery), yet since it is fair to
  assume
  that these populations have immunity in the immediate interval
  after their infection, we can assign $\mathfrak{D}_R= \mathfrak{D}_{AR}=0$.
  However, an interesting possibility within
 the
 model is the inclusion of population time-dependent
 diffusion, \textcolor{black}{possibly anisotropic},
 and also directed (along the direction of the velocity $\vec{v}$)
 motion, \textcolor{black}{as considered, for example,
 in~\cite{FastDiffusion} where a laboratory case
 of epidemic propagation along lines of
 fast diffusion is presented to model the spreading of a virus along a highway.}
 As regards the remaining populations, it may be tempting
 to examine nonlinear variants where the diffusivity
 is larger, e.g., where the population is larger, reflecting the
 existence of a well-established transportation/mobility network. Nevertheless,
 in the present work, we will initiate relevant considerations by
 assuming constant diffusivity of the susceptibles, the exposed,
 and the asymptomatics. The latter are the {\it key}, given their
 mobility
 and spatial spreading for the corresponding spreading of the pandemic
 in the context of Eqs.~(\ref{eqn1})-(\ref{eqn8}).

 An additional important decision that can be incorporated at the
 level
 of the PDE model concerns
the functional form of the {\it directional velocity}
   $\vec{v}$. In principle,  this can be used to capture ``daily practices''
   (e.g., going to work, spending time there, commuting back
   and resting practices), but also longer temporal or
   spatial scales (e.g., trips from
   city to city, or  country to country).
   Motivated partially by the colloquial understanding of some
   of the case examples considered such as the spreading
   of the pandemic in Greece~\cite{greecewiki}, at
   the present level, we opt not to incorporate these effects
but simply allow diffusion to perform the relevant spreading.
The idea  within a given region
then is  that arriving infected individuals, e.g., from international travel, form  local
   hotspots within the $S$ population and we examine the diffusional
   spreading effect of the virus in the presence of the above
  local viral dynamics. \textcolor{black}{Hence, the initial infectious
  seeding within the susceptible population is a rough approximation
  intended to emulate
  long-range transport in the absence of convection.} We will see that this approach is
  not unreasonable given
   the  results that we obtain for the spreading of the PDE
   results with both the ODE ones and the data available online for
   the cumulative infections and the deaths within the regions of interest.
Naturally, it is hoped that this will be a seed study towards a
further refinement of such considerations on the basis of more
accurate
spatial data for the spreading of the disease.

   In the results given in the next section, we have selected
   as our illustrative example the autonomous community of Andalusia within Spain.
   \textcolor{black}{The example of the mainland of Greece is presented
   in Appendix~\ref{app:Greece}.}
   While these examples
   may seem somewhat disparate, they bear some significant
   advantages as regards their nature and their comparison.
   First off, they are regions of similar populations of about 8-10
   million inhabitants.
   Greece has been praised  in international media~\cite{time}
   regarding its handling of the \textcolor{black}{first-wave} COVID-19 crisis and the
   effectiveness and promptness of the associated social-distancing
   measures. Additional relevant features of this region include
 (a) day 0 of the infection \textcolor{black}{first wave}
 and (b) the origin of the localized
 events thereof could be successfully identified, as well as
 (c) strict lockdown effects went into place early on.
 Another
 example at the opposite end with very significant numbers
 of infections and deaths is Spain.
 However, here there is a significant set of complications.
 Not only is Spain far larger in spatial and population size,
 but importantly for the number of reported cases and especially
 the number of deaths, there is no universally accepted way of
 reaching the relevant conclusive numbers across the 17 different
 autonomous communities. For all of the above reasons, and also for reasons of
 clearer comparison of comparable sizes (and also for ones of
 intrinsic interest to the authors, admittedly), we selected
 the autonomous community of Andalusia.

 Having selected our target regions, the next complication is to
 formulate the solution of Eqs.~(\ref{eqn1})-(\ref{eqn8}) at the level
 of the autonomous community/country as a ``two-dimensional spatial grid''.
That is one significant complication
 toward spatial modeling which we have addressed by utilizing the
 finite
 element package COMSOL Multiphysics$^\circledR$ \cite{Comsol}. We have inserted the regions' map
 as a geometry within COMSOL and proceeded subsequently to form
 a triangulated mesh of the computational domain.

 The next and also rather complex step is to formulate a population
 as an initial condition of susceptibles within the relevant grid.
 Here, we have
 leveraged tools from the large scale geographic project World Pop~\cite{worldpop}.
 This methodology encompasses census data and enables via random forest
 models~\cite{randomf} the generation of a gridded prediction of the population
 density
 at a resolution of about 90 m.  We have imported this type of
 data within our spatial country grids
 and via interpolation we are in a position to simulate models
 of the type of Eqs.~(\ref{eqn1})-(\ref{eqn8}) with arbitrary choices
 of parameters, and, in principle, also initial conditions. This is,
 in our view, a significant combined asset (the spatial grid of
 a region
 combined with an interpolated over this grid realistic representation
 of population census data) towards modeling spreads of epidemics.

 The crucial next step, within this line of modeling the
 spreading of the epidemic, is to identify
 suitable parameters, similarly to what has been done in numerous
 earlier studies~\cite{mit,another} at the ODE level.
 To do so, \textcolor{black}{we utilized a nonlinear optimization algorithm
 such as the constrained minimization, {\tt fmincon} function within
 Matlab.}
 \textcolor{black}{We determined the optimal model parameters by
 minimizing the Euclidean distance $\mathcal N$ ($\ell^2$ norm)  between
 the time series generated by the model, identified by the
 subscript ``num", and the
 corresponding ``observed" (data) time series, identified
 by the subscript ``obs",}
 \textcolor{black}{
 \begin{equation}\label{eq:norm1age}
 {\mathcal N}=\sum_i^{t_{\textrm{fit}}^{\textrm{end}}} \Big{(} |\log(C_\mathrm{num}(t_i))-\log(C_\mathrm{obs}(t_i))|^2+
 |\log(D_\mathrm{num}(t_i))-\log(D_\mathrm{obs}(t_i))|^2\Big{)}
\end{equation}
where the index ``$i$" identifies a point in the time series.}
The parameters were optimized to reproduce
 the time series of the reported total number of infected cases ($C(t) = I(t) + H(t)
 +R(t) +D(t)$, the total number of "cases") and total number of deceased ($D(t)$).
 We found these two time series to provide the most reliable data.
 Specifically, for the case of Greece we note some nontrivial lapses in the apparent
curation of the data. Particularly noteworthy is the case of the
recovered individuals in~\cite{greecewiki}. The data must
evidently be significantly inaccurate, as the number of recovered
individuals appears to stay fixed at 53 between March 29 and April 5,
only then to jump entirely abruptly to 269 recovered, only to stay
there between April 6 and April 29, then to jump on to 1374.
Admittedly, the unprecedented circumstances were straining the
data collection process, yet it is particularly important to provide
accurate
data to modelers to calibrate adequately the models
towards
the future spreading of the pandemic.
It is these two data columns (total cases and deaths) from~\cite{greecewiki} that we thus compare
to our 0D model for Greece {and the columns that were used in the parameter
optimization.}

 As expected for this large parameter space, initial parameter choices and parameter
 constraints affect the parameters resulting from the optimization algorithms.
 \textcolor{black}{In the next section, where we present the ODE parameters, we
 discuss a number of sensitivity studies as well as related
 issues of parameter identifiability.} In addition, we argue that the
 suggested median values are biologically and socially reasonable, and
 that they are in line with a number of features known about the SARS-CoV-2
 transmission.

 At the level of parameters within a certain individual
 and how the virus acts on it ``on average'', i.e., as concerns
 parameters such as $(\sigma_A,\sigma_I,M,\gamma, M_{AR}, \omega,
 \chi, \psi)$, we preserve the same values at the PDE level as at the
 ODE one. The  transmission rates $\beta$ are more complicated.
 Keeping in mind that at the PDE level the quantities, $S$, $E$, etc.
 are no longer populations, but rather population densities~\cite{MargueriteSpatialDynamics} which
 integrate over the region's spatial surface (through the respective
 surface integrals) to the true population of each category, we can
 immediately infer that the units of such densities are proportional
 to $l^{-2}$ where $l$ is a characteristic length-scale of the
 analysis.
 In that vein, the $\beta$'s need to be multiplied by $l^2$ to
 adapt dimensionally
 between the ODE and the corresponding PDE model.
 Indeed, we found these to be the most complicated parameters
 to select at the level of the PDE model, as we will explain in the
 discussion of the results below. It is important to bear in mind that while the
 results below are given for these two regions our aim is to develop a set
 of tools that could be in principle used, alongside with data for the
 pandemic from different countries~\cite{WHO_Count}, to perform similar
 analyses of other regions.

\section{Computational Results}
\subsection{ODE model: Well-mixed populations}

\begin{figure}[!ht]
\centering
\begin{tabular}{cc}
\includegraphics[width=.45\textwidth]{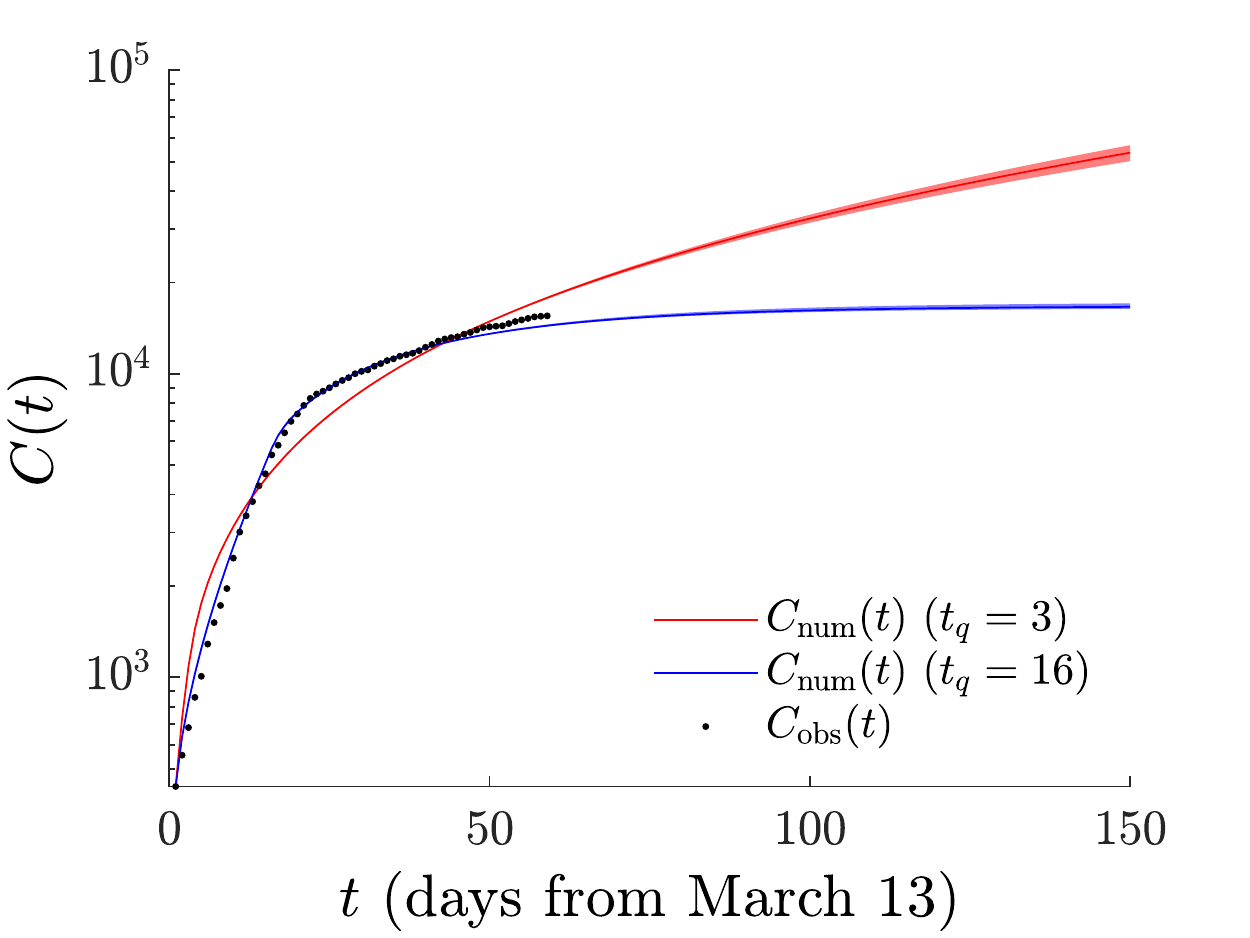} &
\includegraphics[width=.45\textwidth]{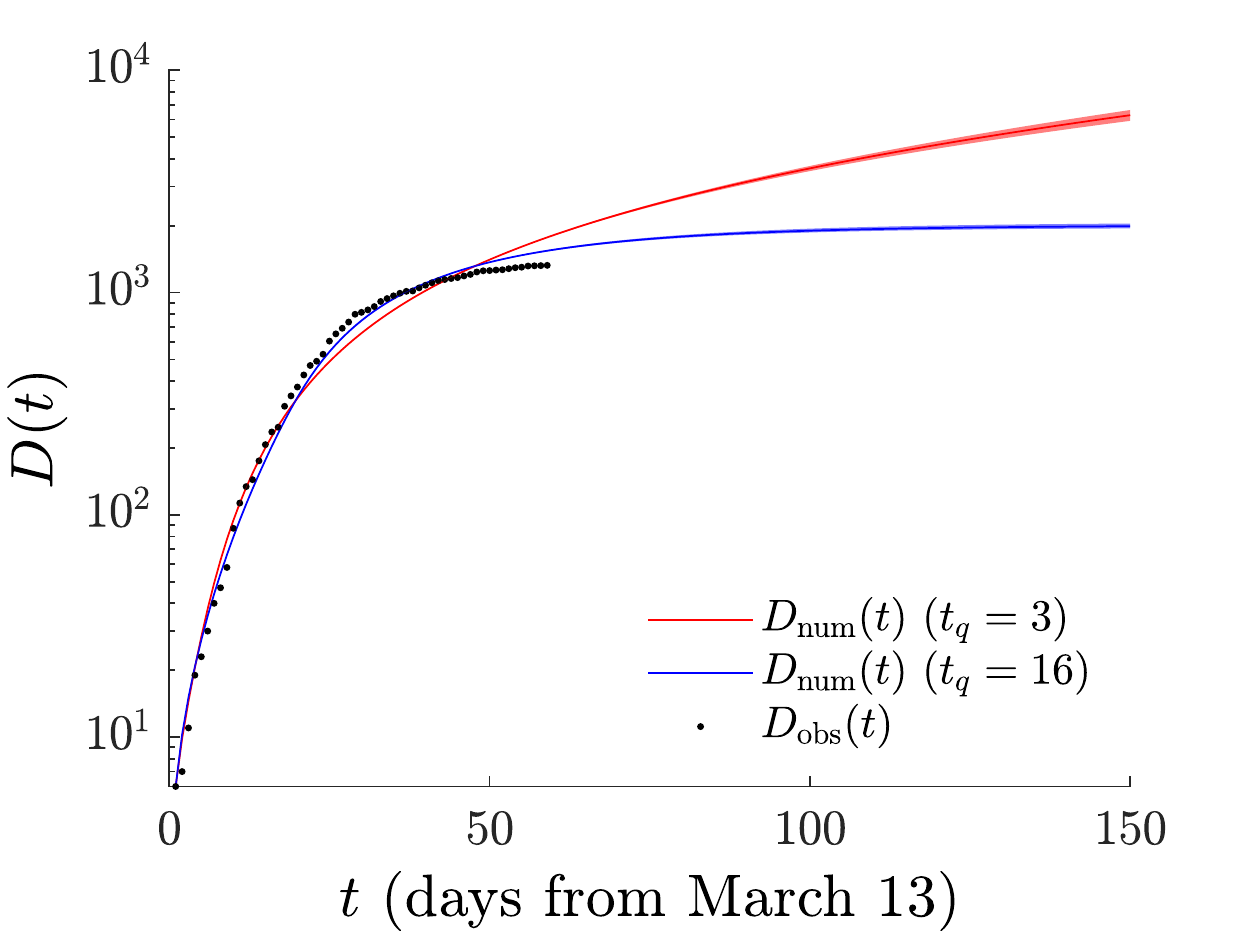} \\
\includegraphics[width=.45\textwidth]{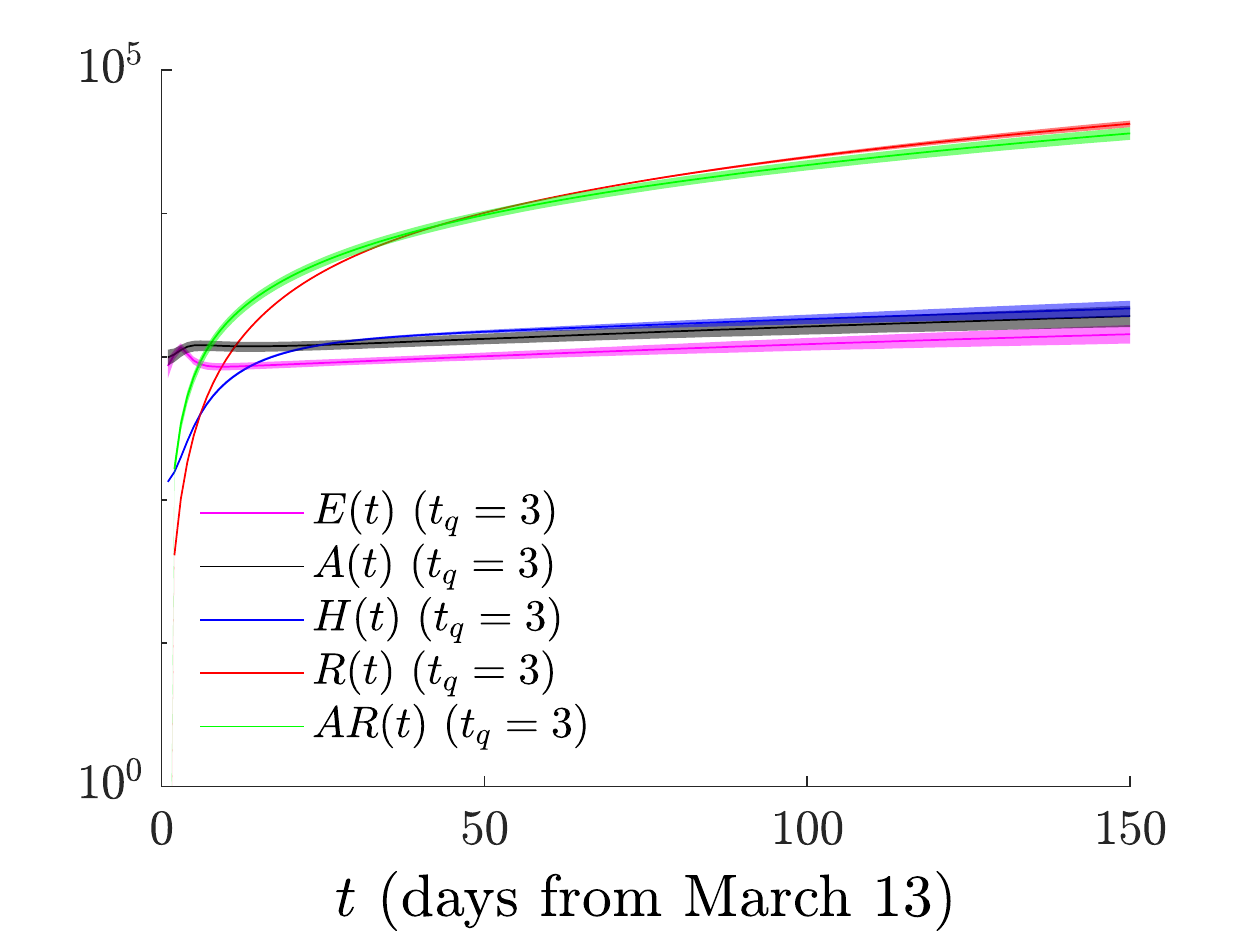} &
\includegraphics[width=.45\textwidth]{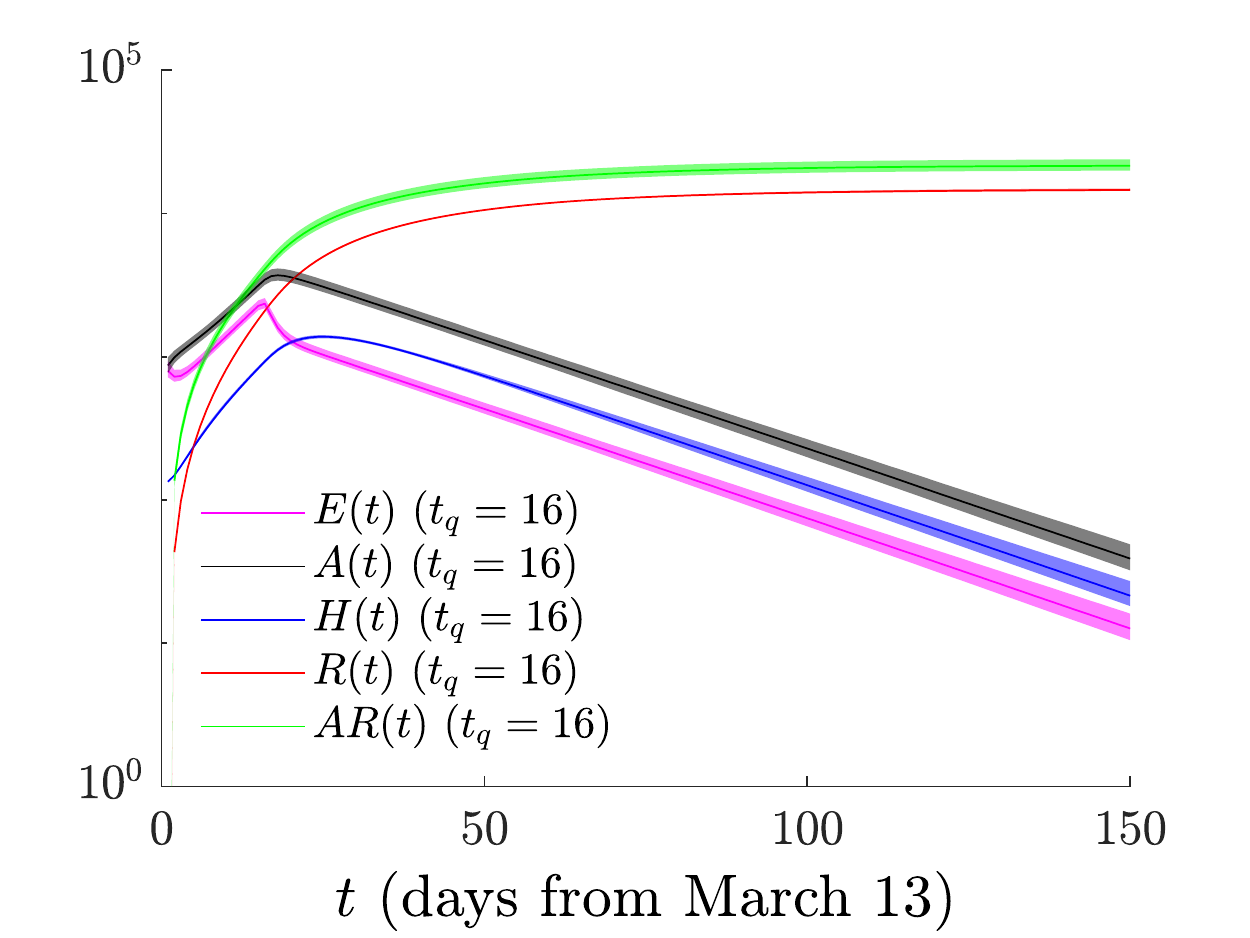} \\
\end{tabular}
\caption{{(Color online.)} 0D model for Andalusia with fitting to
official data from March 14, 2020 ($t=t_{\textrm{init}}=1$) to May 11, 2020
($t_{\textrm{fit}}^{\textrm{end}} = 59$). Official confinement started
on March 16, 2020 ($t_q=3$).
Top panels show the official data (black dots) and simulations: red line {(top solid line at t=150)} for $t_q=3$ (scenario one)
and blue line  {(bottom solid line at t=150)} for $t_q=13$ (scenario two, quarantine starting on March 29, 2020).
Left top panel: Confirmed cases $C(t)=I(t)+R(t)+H(t)+D(t)$; Right top panel: Number of deaths $D(t)$.
Bottom panels show the other populations. Bottom left panel shows these populations for $t_q=3$
{(distinguishable at t=150 from top to bottom: recovered $R(t)$, asymptomatic
recovered $AR(t)$, hospitalized $H(t)$, asymptomatic $A(t)$, and exposed $E(t)$)}.
The bottom right panel shows them for $t_q=16$
{(distinguishable at t=150 from top to bottom: asymptomatic
recovered $AR(t)$, recovered $R(t)$, hospitalized $H(t)$, asymptomatic $A(t)$, and exposed $E(t)$)}.
In all the panels, shaded regions correspond to the interquartile range for each quantity, whereas the full line corresponds to simulations with the median parameter (and initial-condition) values.}
\label{fig:Andalusia0D}
\end{figure}

We start the exposition of our results by discussing what we will
refer to as the ``0D'' model (the version of Eqs.~(\ref{eqn1})-(\ref{eqn8}) without space
dependence) for Andalusia. \textcolor{black}{Data for the evolution of
the pandemic in Andalusia were obtained from the Andalusian-Goverment
COVID-19 site~\cite{datos_Junta}.} The relevant results  are given
in Fig.~\ref{fig:Andalusia0D}. \textcolor{black}{We obtained the optimal (best-fitting)
0D-model parameters for Andalusia (and Greece)  from 2,000 optimizations that
compared model predictions to jointly the number of cumulative
infected and the deceased, as shown in Eq.~(\ref{eq:norm1age}).
For each optimization, the initial guess for each parameter and initial condition was
uniformly sampled within a pre-specified range. The upper and lower limits
were used as boundaries in the
constrained minimization algorithm (implemented in Matlab via the
\texttt{fmincon} function).
The parameter ranges were
determined from epidemiological information.
We note that at the initial time of model fitting, the number
of exposed and asymptomatic individuals is not known.
We thus optimized (and varied) their ratio to the initially infected $I(0)$, a number  that
was obtained by subtracting the officially reported number of deaths, recovered,
and hospitalized from the (reported) number of cases.
The sensitivity of the predicted
model parameters when the ratio of $\beta_{AS}$ to $\beta_{IS}$ is allowed to vary within
a specified interval ([$0.2, 2$]), in steps of 0.02, and parameter-identifiability
issues are discussed at the end of this Section.} 

\begin{table}[!htb]
\centering
\caption{ODE parameters for Andalusia: optimal (best-fitting), median and interquartile
range, and variation range used in the optimization algorithm. Initial parameters and
initial-condition guesses were uniformly sampled within these ranges.}
\label{tab:And_parameters}
\begin{ruledtabular}
\begin{tabular}{ccccc} 
& & Median (interquartile range) & Median (interquartile range) & Initial value\\ 
& & $(t_q=3)$ & $(t_q=16)$ & \\ 
\hline 
Population & $N$ & \multicolumn{2}{c}{8,414,240} & \\ 
Initial populations & ($I_0, H_0, D_0$) & \multicolumn{2}{c}{(297, 134, 6)} & \\ 
Non COVID-19 death rate [per day] & $\mu$ & \multicolumn{2}{c}{$5.79 \times 10^{-7}$} & \\ 
Transmission rate, $S \rightarrow I$ [per day] & $\beta_{IS}$\footnote{The transmission rates
$\beta$ have to be divided by $N$ when used in the ODE model.} & 0.35 (0.28--0.41) & 0.32 (0.29--0.35) & $c \in U[0,1]$ \\ 
Transmission rate, $S \rightarrow A$ [per day] & $\beta_{AS}$\footnotemark[1]  & 0.55 (0.50--0.60) & 0.28 (0.26--0.30) & $c \in U[0,1]$ \\ 
Lockdown effect, $S \rightarrow I$ & $\eta_{IS}$ & 0.28 (0.24--0.35) & 0.39 (0.36--0.42) & $c \in U[0,1]$ \\ 
Lockdown effect, $S \rightarrow A$ & $\eta_{AS}$ & 0.48 (0.43--0.56) & 0.38 (0.35--0.41) & $c \in U[0,1]$ \\ 
Latent period, $E \rightarrow A$ [days] & $1/\sigma_A$ & 4.52 (4.36--4.69) & 2.97 (2.87--3.05) & $1/k, \, k  \in U[2,7]$ \\ 
Incubation period, $E \rightarrow I$ [days] & $1/\sigma_I$ & 3.15 (2.98--3.33) & 3.77 (3.54--4.14) & $1/k, \, k  \in U[2,7]$ \\ 
Infectivity period [days] & $1/M$ & 6.06 (5.98--6.12) & 5.97 (5.89--6.04) & $1/k, \, k  \in U[5,12]$ \\ 
Recovery period (asymptomatics), $A \rightarrow AR$ [days] & $1/M_{AR}$ & 6.15 (6.08--6.21) & 6.85 (6.75--6.94) & $1/k, \, k  \in U[5,12]$ \\ 
Recovery period (hospitalized), $H \rightarrow R$ [days] & $1/\chi$ & 8.39 (8.23--8.54) & 6.74 (6.60--6.85) & $1/k, \, k  \in U[5,20]$ \\ 
Period from hospitalized to deceased, $H \rightarrow D$ [days] & $1/\psi$ & 9.38 (9.22--9.57) & 8.33 (8.17--8.53) & $1/k, \, k  \in U[5,20]$ \\ 
Conversion fraction ($I \overset{\gamma}{\longrightarrow} H$, $I \overset{1-\gamma}{\longrightarrow} R$) & $\gamma$ & 0.58 (0.57--0.59) & 0.55 (0.54--0.57) & $c \in U[0.25,0.75]$ \\ 
Conversion fraction ($H \overset{\omega}{\longrightarrow} D$, $H \overset{1-\omega}{\longrightarrow} R$) & $\omega$ & 0.25 (0.24--0.26) & 0.25 (0.25--0.26) & $c \in U[0.1,0.5]$ \\ 
Initial population fraction, exposed & $E_0/I_0$ & 2.90 (2.39--3.27) & 2.69 (2.42--3.14) & $c \in U[1,5]$ \\ 
Initial population fraction, asymptomatic & $A_0/I_0$ & 3.33 (2.92--3.77) & 2.93 (2.59--3.34) & $c \in U[1,5]$ \\ 
Diffusivity, $S$ [km$^2$/day] & $\mathfrak{D}_S$ & --- & 10 & \\
Diffusivity, $E$ or $A$ [km$^2$/day] & $\mathfrak{D}_E$ or $\mathfrak{D}_A$ & --- & 100 & \\
\end{tabular} 

\end{ruledtabular}
\end{table}

Upon performing the optimizations
we find that the fitting yields the results summarized
in Table~\ref{tab:And_parameters}.
\textcolor{black}{We show the median parameters, as well as the interquartile range,
and the range of variation used to sample the parameters  (and initial conditions).
Model predictions (with median parameter values, solid blue or red  lines) are compared
graphically to data (black dots) in Fig.~\ref{fig:Andalusia0D}. Model output sensitivity
to parameter (and initial-condition) variations
is represented by the shaded regions.}

\begin{table}[!htb]
\centering
\caption{Time sequence of events and simulation times.}
\label{tab:TimeSequence}
\begin{ruledtabular}
\begin{tabular}{ccccc}
Country & Initial simulation time ($t_{\textrm{init}} = 1$) &
Imposed lockdown ($t_q$) &
Effective lockdown ($t_q$)
& Last fitting day ($t_{\textrm{fit}}^{\textrm{end}}$) \\
& & (Scenario one) & (Scenario two) & \\ \hline
Andalusia & March 14, 2020 & March 16, 2020 ($t_q = 3$) &
March 29, 2020 ($t_q= 16$) &
May 11, 2020  ($t_{\textrm{fit}}^{\textrm{end}} = 59$) \\
Greece & March 12, 2020 & March 24, 2020 ($t_q = 13$) &
April 3, 2020 ($t_q = 23$) &
May 11, 2020  ($t_{\textrm{fit}}^{\textrm{end}} = 61$) \\
\end{tabular}

\end{ruledtabular}
\end{table}

The optimal parameters were obtained for two
scenarios (see Table~\ref{tab:TimeSequence}). The first scenario considers that
restrictive measures (quarantine, lockdown) in Andalusia
were strictly enforced on March 16, 2020. To account for the
change in parameters induced by the lockdown, we imposed a time dependence
on the transmission rates $\beta$
\begin{subequations}
\begin{equation}
  \beta_{IS}(t)=\beta_{IS} \left [ \eta_{IS}+(1-\eta_{IS})\frac{1-\tanh[2(t-t_q)]}{2} \right ]
\end{equation}
\begin{equation}
  \beta_{AS}(t)=\beta_{AS} \left [ \eta_{AS}+(1-\eta_{AS})\frac{1-\tanh[2(t-t_q)]}{2} \right ]
\end{equation}
\label{eq:TimeDependentBeta}
\end{subequations}
so that the transmission rates $\beta_{IS}$ and $\beta_{AS}$ decrease
by a factor $\eta_{IS}$ and $\eta_{AS}$ (respectively)
relatively abruptly at the time $t_q$ the lockdown was imposed.
The transmission rates effectively \textcolor{black}{incorporate}  the rate of contact of susceptible
individuals with infectious individuals (infected or asymptomatic in our model)
that leads to exposure
to the virus. In fact, the transmission rates $\beta$ may be
expressed as the product of the average daily contacts (contact rate)
a susceptible has with any individual times the probability of infection given a contact:
{the probability of infection is proportional to the viral load (viral concentration in
the respiratory-tract fluid) of expelled respiratory droplets~\cite{Editorialyd2020}.
\textcolor{black}{Hence, government-imposed restrictive measures, e.g.,
mobility restrictions, social-distance requirement, face-mask wearing,
and limitations on the number of persons in a gathering, are expected to
decrease the transmission rates (for both infected and asymptomatics),
an effect that is reflected in the $\eta$'s.}

\textcolor{black}{The two top panels of Fig.~\ref{fig:Andalusia0D} illustrate
that how well we capture the data for Andalusia
depends on the scenario chosen, i.e., when the quarantine is imposed
(the events time series is summarized in Table~\ref{tab:TimeSequence}).
The red solid line {(top line at t=150)} and shaded region correspond to
imposing the quarantine at $t_q =3$ (scenario one,
modeling the beginning of the quarantine almost immediately when it
officially occurred), whereas the blue solid line {(bottom line at t=150)} and shaded
region present data (median and range) for
$t_q=16$, scenario two. Both scenarios reproduce
reasonably well the number of the fatalities
(top right panel), scenario two better, but not so the number of cases. Moreover,
the difference between predictions and data increases with simulation time,
with the predictions of scenario one becoming progressively very high and dissonant with
the trend of the data.}
There is a
characteristic feature in the top left panel (cases) that the first scenario
fails to capture: there is an ``angle'' in the semi-logarithmic
plot associated with the curbing of the observed cumulative number of infections $C(t)$
due to containment measures.
It is apparent that the attempt to
capture the data, due to the relevant mismatch in the associated angle,
leads to far more significant deviations. While model
prediction seems to minimize the distance to the data by
over-predicting $C(t)$ initially, and under-predicting it later, it clearly starts
over-predicting the trend of the quantity towards the end of the
available (fitted) data time-series. This results in predicted cumulative infections of the
order of several (more than 4) tens of thousands. A similar
over-prediction
seems to develop in $D(t)$ leading to nearly five thousand
deaths, while the data seem to clearly tend to values below that.

If we were to shift arbitrarily the time of the application of
the quarantine data by 13 days later ($t_q = 16$), we note a nontrivial difference.
While we are not
missing on $D(t)$ (in fact, the fit is more accurate),
we capture accurately the angle in the $C(t)$ data.
This suitable
shift of the quarantine time clearly does a far better job in capturing the
actual trends of both $C(t)$
and $D(t)$ with the $C(t)$ lying
between
$10^4$ and $2 \times 10^4$ and, correspondingly, $D(t)$  staying
below $2 \times 10^3$ \textcolor{black}{at the end of the five-month simulation period.}
Table~\ref{tab:And_parameters} clearly
illustrates the source of the discrepancy at a parametric level:
compare the medians reported in the first and second columns.
\textcolor{black}{The most noteworthy difference is that while for
scenario one (no  reproduction of the angle) $\beta_{AS} > \beta_{IS}$ this inequality
reverses in scenario two. A concomitant change occurs in the ratio of
exposed who turn asymptomatics ($\sigma_A/(\sigma_I+\sigma_A)$) from 0.41 (scenario one)
to 0.56 (scenario two). The relative importance of these changes
is discussed at the end of this section where parameter identifiability is addressed.
Lastly, note the slight change in the recovery period $\chi$.
A potential interpretation of this admittedly
somewhat arbitrary shift of the quarantine-parameter imposition
may be that at the model level such measures have an immediate,
essentially instantaneous effect, while in the realistic country data,
there is a time lag before this switch in the number of contacts
(due to lockdown) has a perceptible effect, depending on how fast individuals adapt to
the imposed restrictions.}

\textcolor{black}{Since the second scenario reproduces more satisfactorily
the observed data we discuss (with the proviso mentioned at the end of this section)
the biological and societal
significance of its median parameters  in the
second column of Table~\ref{tab:And_parameters}.}
\textcolor{black}{For instance, the median latent period is
approximately 3 days (2.976), whereas the median incubation
period is approximately 4 days (3.77), in reasonable agreement with
the values reported in~\cite{elife}, 3 and 5 days, respectively}.
The value of $M_{AR}$ suggests
a time scale of nearly 7 days (\textcolor{black}{6.85}) for the recovery of
asymptomatics.
On the other hand, $M$ suggests a time scale of about \textcolor{black}{6 (5.97)} days
for those with symptoms to potentially need hospitalization.
\textcolor{black}{The median timescales associated with leaving the hospitalized
compartment imply almost 7 days ($1/\chi = 6.74$) to recovery
and almost 8.5 days ($1/\psi = 8.33$) to fatality.
Hence, the approximate recovery period for mild cases is
about 6 days, and for severe cases approximately 13 days,
while Ref.~\cite{elife} estimates recovery periods of approximately 2 weeks for mild
cases and approximately 6 weeks for the quite severe cases.}
The value of $\gamma$ roughly suggests a half-half split between those
recovering directly versus those needing some form of hospitalization.
The value of $\omega$ suggests that among
those needing hospitalization nearly 75\% recover, while only
25\% die. \textcolor{black}{As discussed above, there is no way to evaluate
the initial population of exposed $E_0$ and asymptomatics $A_0$.
We thus opted to introduce their
fraction to the initial population of infected $I_0$ as two additional
parameters (referred to as initial conditions), $E_0/I_0$ and $A_0/I_0$
in the optimization.}
Our optimization yields initially (at the beginning of the
simulations) \textcolor{black}{approximately equal ratios
of exposed and asymptomatics to the infected.}

\textcolor{black}{The relative transmission rates of asymptomatics and
infected, as well as the ratio of their populations, merit a comment.
The reported median values satisfy $\beta_{IS} > \beta_{AS}$, i.e.,
infected (with symptoms) are predicted to be more
infectious than asymptomatics.
We surmise that the contact rate would be significantly smaller
for the (expected to be) self-isolating infected individuals, than
for the asymptomatics who continue their life, not knowing that they
are carrying SARS-CoV-2 (and most importantly that they are infectious).
Hence, their higher transmission rate would imply
a higher emitted viral load.}
\textcolor{black}{Related to the transmission rates is
the ratio of asymptomatics to symptomatically infected.}
Assuming that the latent time scale
of the virus is similar for asymptomatics and infected
(as is reasonable to assume), the fraction of turning asymptomatic
versus turning infected ($\sigma_A/\sigma_I$) is \textcolor{black}{$1.27$.}
\textcolor{black}{Equivalently, the ratio of becoming asymptomatic to
the total number of exposed ($\sigma_A/(\sigma_A + \sigma_I)$) is about $0.56$
and that to becoming symptomatically infected $0.44$.}
This ratio reflects the importance of asymptomatics~\cite{arons2020} in
the transmission of SARS-CoV-2,
a particularly important feature that differentiates
it from the transmission of other respiratory viruses like influenza and SARS-CoV-1.
Lastly, after lockdown measures are imposed, \textcolor{black}{we find
that both populations are equally affected, $\eta_{IS} \approx \eta_{AS}$,
possibly because mobility restrictions, and the
associated decrease in the average number of daily contacts,
apply equally to both populations.}

One final observation at the level of data rather
than at that of the model is the significance of the early imposition
of restrictive measures. In Spain these measures were taken when already
the number of cumulative infections and deaths was significantly
higher than the corresponding numbers in Greece when the decision
was taken. This ultimately
appears to have led the smaller of the
two regions (Andalusia having 8.4M inhabitants)
to have an order of magnitude larger losses of life  and infections
than the larger of the two regions (Greece having 10.7M inhabitants).

We also calculated the basic reproduction number $R_0$ reflecting the
number of cases expected to be produced by one infectious case in a
fully susceptible population. This is used to estimate how the
epidemic developed initially. We used the next-generation matrix approach
(see Appendix~\ref{app:R0}).
The pre-quarantine basic reproduction number \textcolor{black}{(for
the second scenario, $t_q = 16$),
is $R_0 = 1.91 (1.86 - 1.95)$, where we report both the
median and the interquartile range.}
The calculated basic reproduction number
is close to the epidemiologically determined range of
$2-4$~\cite{elife}, a range that encompasses the variation of the basic
reproduction number in space and time.
Although our focus is not on the calculation of $R_0$, which is of
principal
interest to a wide range of studies regarding SARS-CoV-2, it is worth
noting that a post-quarantine
effective reproduction number may be calculated
for the lockdown-decreased transmission rates.
We find $R_{\textrm{eff}} = 0.763 (0.72 - 0.74)$
reflecting the decline of the epidemic spreading under the lockdown measures.
\textcolor{black}{A similar calculation with the scenario-one parameters yields
a pre-quarantine $R_0 = 2.65$ $(2.38 - 2.90)$, and, interestingly,
a post-quarantine effective reproduction number
$R_{\textrm{eff}} = 1.04$ $(1.03 - 1.04)$  a value that suggests the epidemic
has not been effectively controlled (as implied by the predicted
evolution of the epidemic in Fig.~\ref{fig:Andalusia0D}, top left).}

\begin{figure}[!htb]
\centering
\begin{tabular}{cc}
\multicolumn{2}{c}{\includegraphics[width=.45\textwidth]{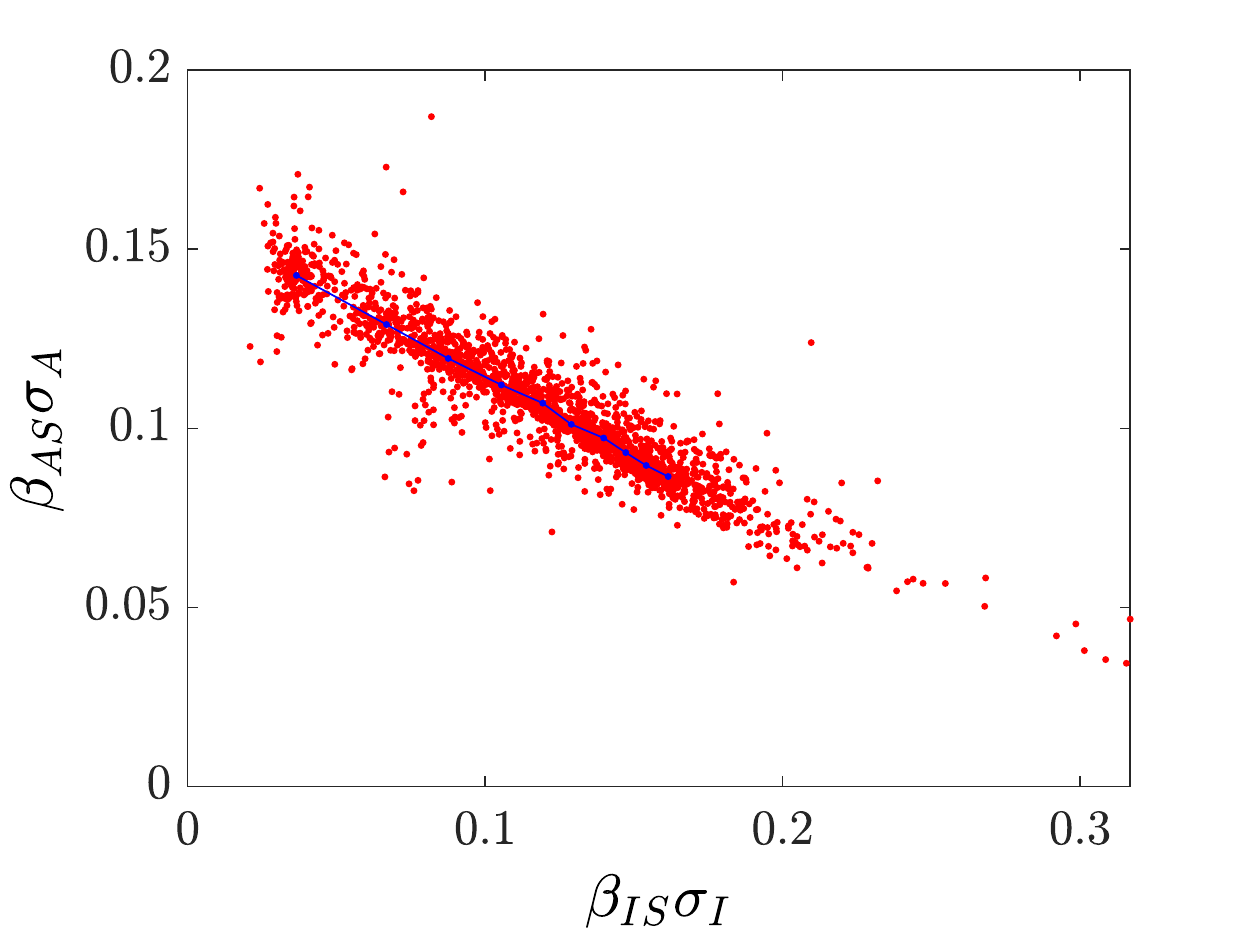}} \\ 
\includegraphics[width=.45\textwidth]{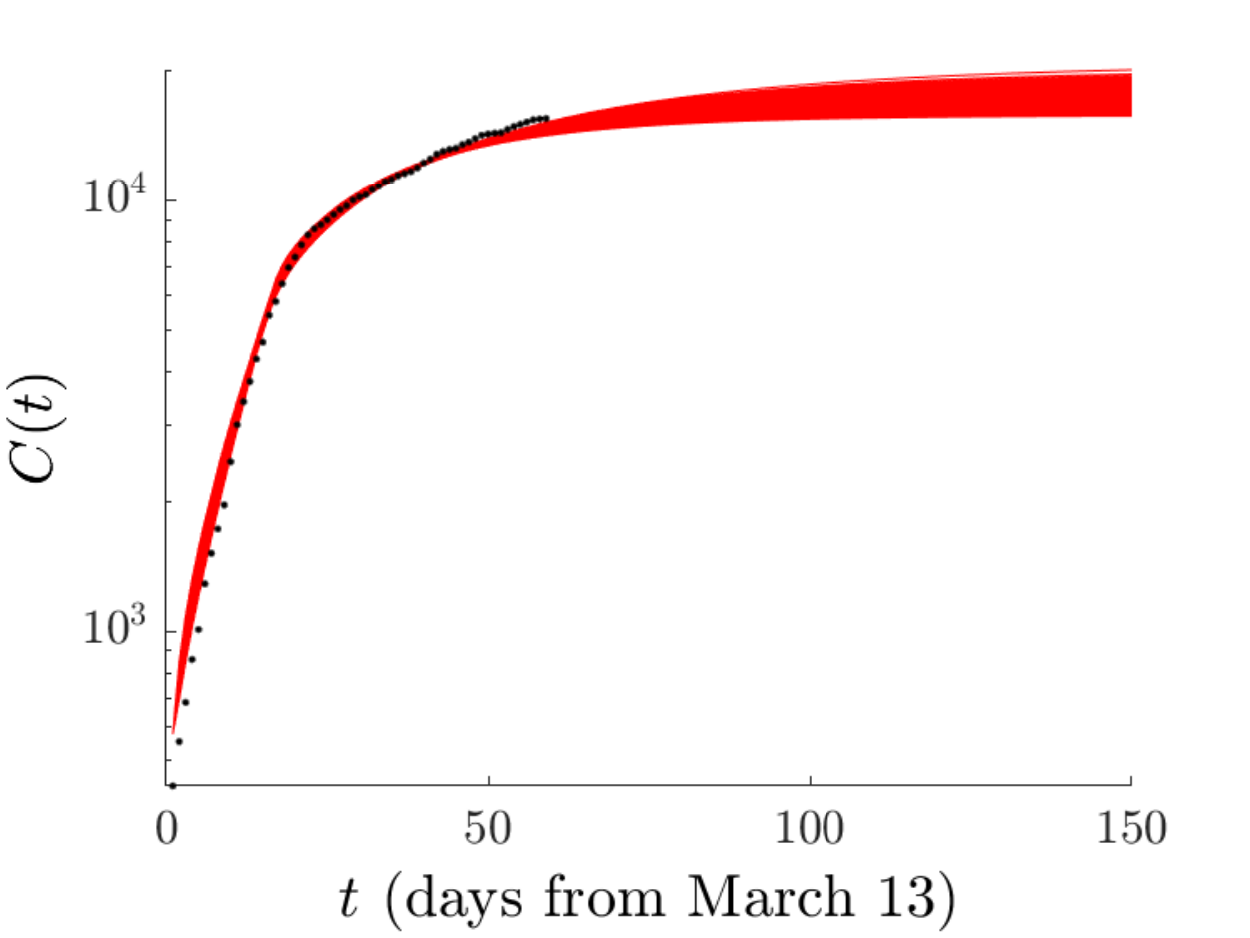} &
\includegraphics[width=.45\textwidth]{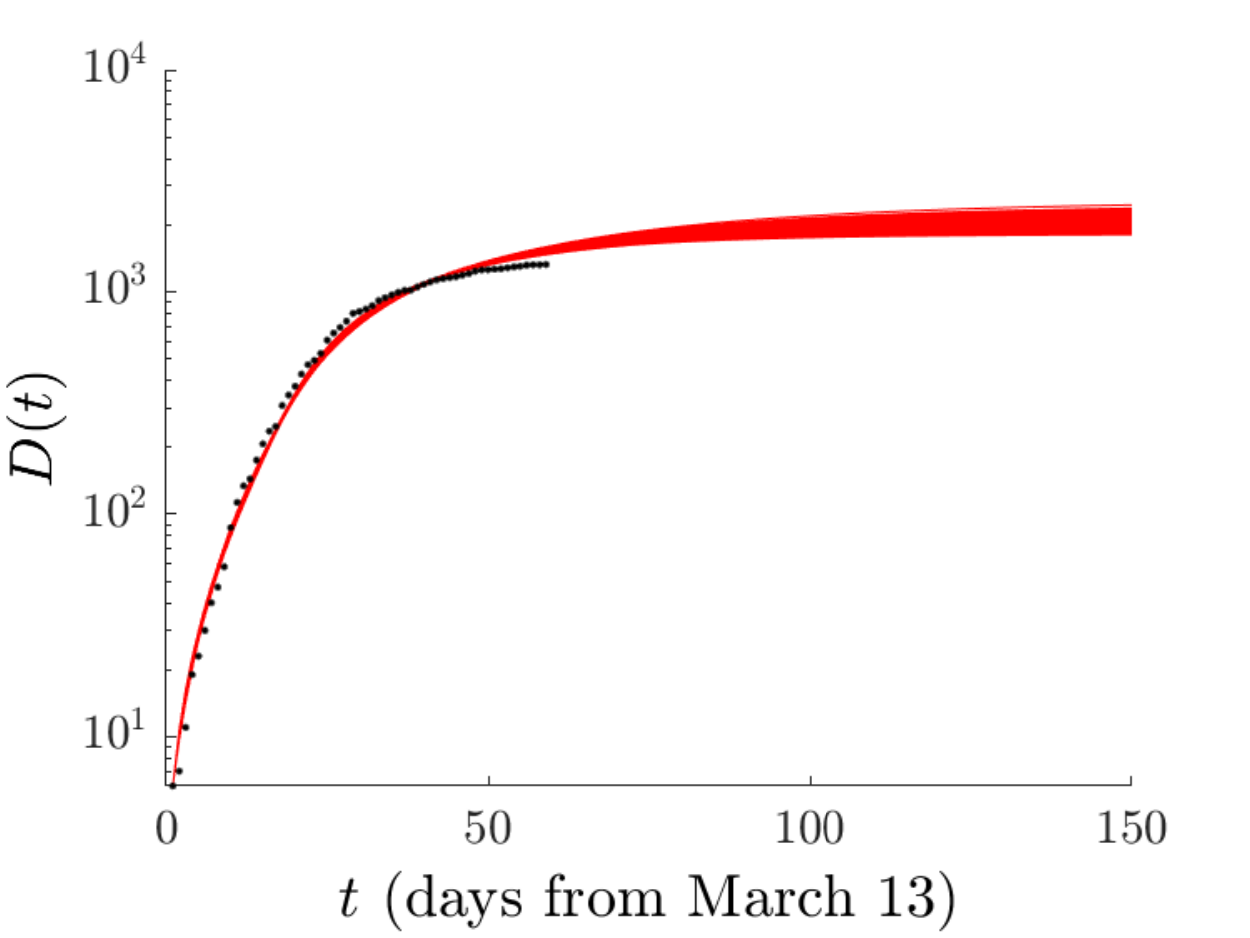}
\end{tabular}
\caption{\textcolor{black}{{(Color online.)} Sensitivity analysis for Andalusia ($t_q=16$). The top panel shows the inverse
relationship between $\beta_{AS} \sigma_A$  and $\beta_{IS} \sigma_I$ for a series
of optimizations (200) performed by uniformly sampling model parameters  and
ten fixed equidistant ratios of $\beta_{AS}/\beta_{IS}$ within the range [0.2, 2].
Blue dots are simulation results with the medians of the optimized
values (the line is a guide to the eye), while the
red dots are the results of all the optimizations.
The two bottom panels compare the predicted future number of cases (left panel, red bundle)
and fatalities (right, red bundle) for all the optimizations shown in the top panel; black
dots are the reported numbers (and used for model fitting). Despite the wide variation of
transmission rates (times the associated time scales), we observe a
rather small uncertainty in the forward, model predicted evolution of the pandemic,
an indication that even though some model parameters are
not independently identifiable, they enable an adequate predictor for
the
quantities of epidemiological interest.}}
\label{fig:And_sens}
\end{figure}

\textcolor{black}{We conclude the analyses of the ODE model by commenting on
the identifiability of model parameters.
We partially
addressed it through the previously presented parameter sensitivity analysis
with 2000 optimizations. This procedure
led to the determination of the median values and their interquartiles. In addition,
and as argued in~\cite{MathBio2021}, one approach to specifying confidence
intervals is through the Hessian of the variation of the
Euclidean norm Eq.~(\ref{eq:norm1age}), the objective
function of our optimizations, with respect to model
parameters~\cite{Farmer2008}. Specifically, if we denote model parameters by $\boldsymbol{\theta}$
the Hessian is $H_{ij} = \partial^2 {\cal N} / \partial \theta_i \partial \theta_j$,
suggesting that if it remains invariant to parameter changes,
these parameters would not be identifiable (since their changes  would
not modify the optimized norm).
Alternatively, as discussed in~\cite{Farmer2008}, the inversion
of the Hessian leads to the confidence intervals associated
with each parameter. When we carried out this programme for
a model similar to the 0D model presented here~\cite{MathBio2021},
we found that
the Hessian was singular: in fact, it had two zero eigenvalues.
Our above line of argumentation (expanded upon in~\cite{MathBio2021})
suggests that these two ``zero-cost" eigendirections are closely
connected to the identifiability of the model, and specifically that
a number of parameters associated with these eigendirections are not
independently identifiable. One eigendirection
is easily specified through inspection of the model.
The three parameters $\omega, \psi$ and $\chi$ may be
easily combined to two $\kappa_1 = (1-\omega) \chi$ and
$\kappa_2 = \omega \psi$.}

\textcolor{black}{The second combination of parameters that defies
  identifiability is less immediately transparent.
  However, we get a hint of the other zero-eigenvalue
eigendirection, and
the associated not-independently-identifiable parameters
by comparing columns one versus two in Table~\ref{tab:And_parameters}
(and also from the runs of Fig.~\ref{fig:And_sens}; see especially the
top panel thereof).
When the time the lockdown was imposed is modified (going from
scenario one to scenario two), the transmission rates
and the time scales
shift from $\beta_{IS} < \beta_{AS}$ and
$1/\sigma_A < 1/\sigma_I$ (whose ratio, as discussed earlier,
determines the fraction of turning  asymptomatic to turning
symptomatically infected to be $0.70$)
to the $\beta_{IS} > \beta_{AS}$ and $1/\sigma_A > 1/\sigma_I$
(with the corresponding fraction becoming $1.27$). Alternatively,
for $\beta_{AS}>\beta_{IS}$ the fraction of exposed turning
asymptomatic is smaller than when  $\beta_{AS} < \beta_{IS}$,
i.e., the larger the asymptomatic transmission rate the smaller their fraction.
This 	inverse relation becomes quantitative in the top panel
of Fig.~\ref{fig:And_sens} where we note that as
$\beta_{AS} \sigma_A$ decreases $\beta_{IS} \sigma_I$
increases. We chose these two parameter combinations 
as they appear naturally in the two summands of the basic
reproduction number, Eq.~(\ref{eq:R0}).
The optimizations, whose results are reported in
the figure, were performed as previously discussed (i.e.,
parameters and initial conditions were uniformly sampled within
their range of variation) with an additional
constraint on the ratio of the two transmission
rates $\beta_{AS}/\beta_{IS}$. We chose their ratio
to vary between $0.2$ to $2$, sampled in ten equidistant values.
For every ratio of the transmission rates we performed 200 optimizations.
The {solid} blue line denotes the relationship for the median parameters
for each choice of $\beta_{AS}/\beta_{IS}$, the red dots
correspond to the optimal parameters for each optimization.
Monitoring the results obtained in Fig.~\ref{fig:And_sens},
it is natural to conclude $\beta_{AS}$ and $\beta_{IS}$
are not independently identifiable. Instead, there is effectively
a monoparametric freedom (associated with the singular eigendirection)
connecting these two parametric combinations.}

\textcolor{black}{For this wide range of parameters we also
compare the predicted number of cases (bottom left panel
Fig.~\ref{fig:And_sens}) and fatalities (bottom right panel) to the reported
numbers. Importantly, we note that
even though the transmission rates times the associated inverse time scales
may vary significantly due to their non-identifiability,
the predicted number of cases and fatalities does not
(compare also to Fig.~\ref{fig:Andalusia0D}).
This is, indeed, a manifestation of the singular eigendirection:
a wide range of model parameters provides an
equally good predictor of the total number of cases and fatalities.
Hence, the uncertainty in the identification of model parameters,
and their non-identifiability, has a relatively small effect on the predictions
of the model.  We believe this provides
convincing evidence of the predictive ability of the model
and its accuracy.
A far more relevant question is not the specification of model parameters
and their confidence intervals, but how does the
flexibility to specify them, as allowed by the
singular eigendirection, modify model predictions.
We find that the optimally determined
model parameters provide a reasonable, within a given range, estimate of the modeled
quantity, even though due to the
non-identifiability of the model the model parameters may vary.
It is also an interesting direction to explore what additional pieces
of data (such as, e.g., on the asymptomatic infections) may render
the model identifiable, enabling a more precise identification of the
relevant parameters.}

\subsection{PDE model: Spatially distributed populations}

We now turn to the PDE simulations.
Relevant results for the autonomous region of Andalusia
may be found in Fig.~\ref{AndalusiaPDE} for the same
diagnostics as for the 0D model. However, now,
we complement them with the space-time evolution
simulations of
Figs.~\ref{fig:AndalusiaPDE_I}--\ref{fig:AndalusiaPDE_C}
{that will be compared also in what follows
with the data of the map of Fig.~\ref{fig:AndalusiaSpatialDist}}.

\begin{figure}[!htb]
\centering
\begin{tabular}{cc}
\includegraphics[width=.45\textwidth]{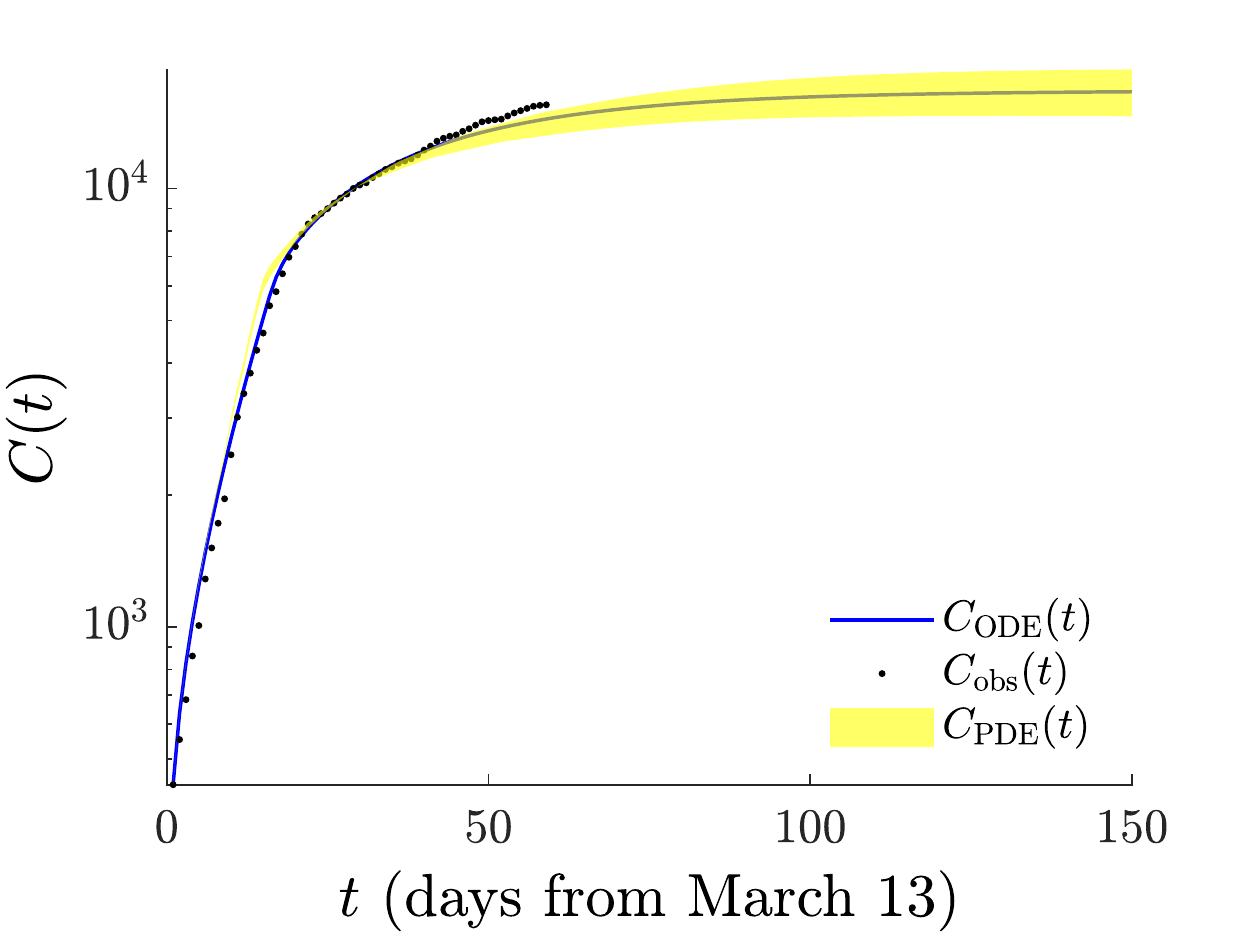}  &
\includegraphics[width=.45\textwidth]{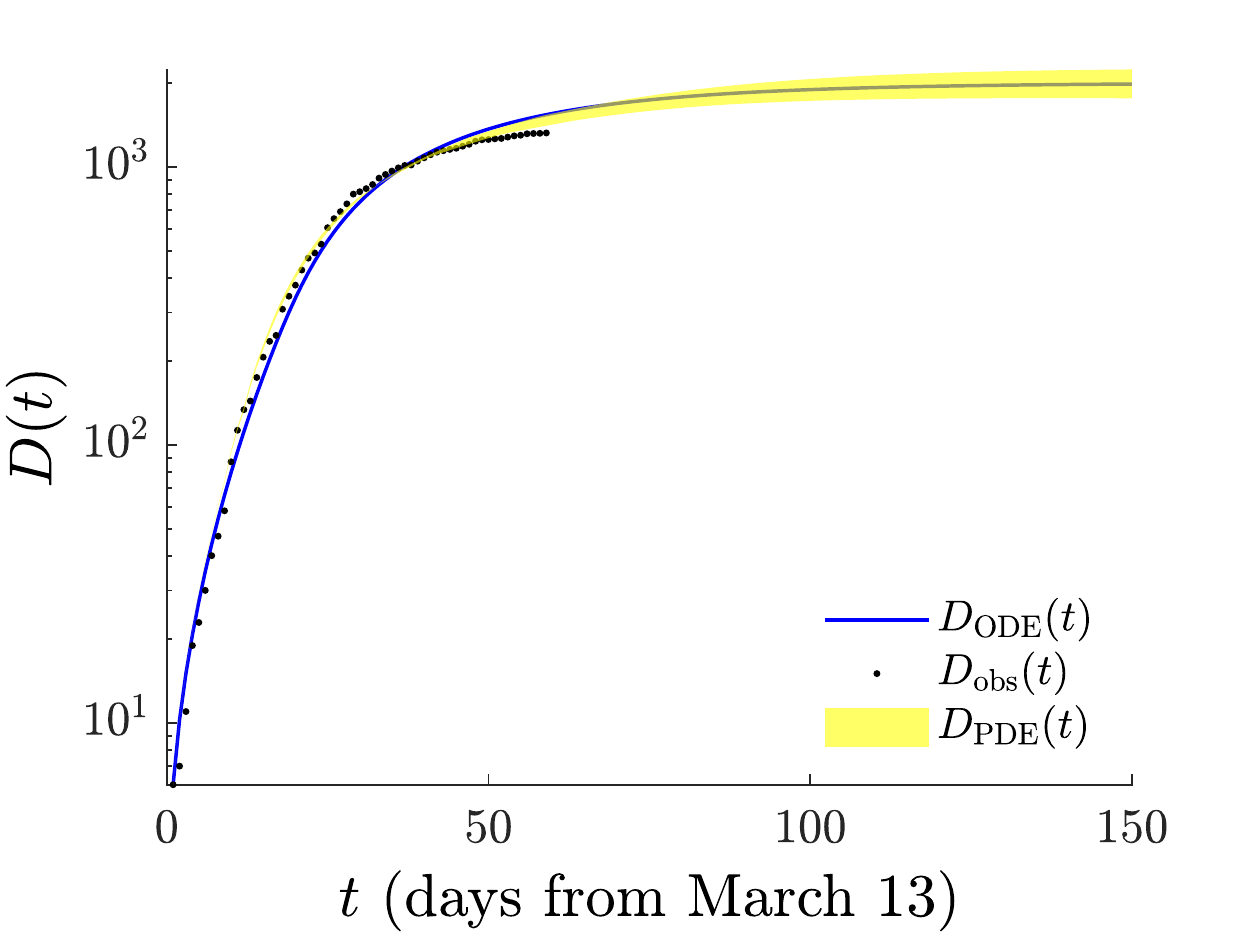} \\
\multicolumn{2}{c}{\includegraphics[width=.45\textwidth]{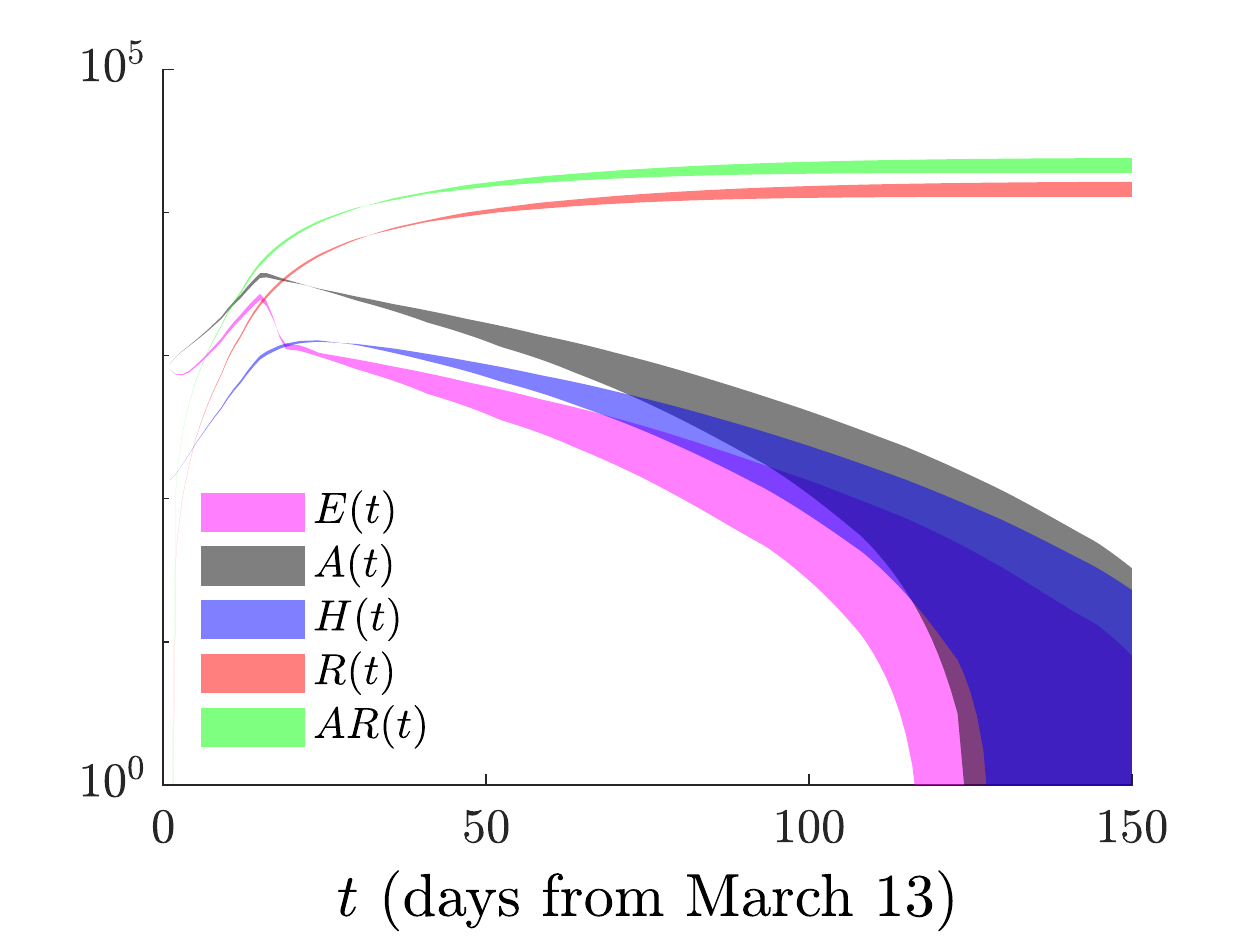}}
\end{tabular}
\caption{{(Color online.)} PDE model for Andalusia with fitting to official data from March 14, 2020 ($t=t_{\textrm{init}}=1$)
to May 11, 2020. Confinement time starts on March 29, 2020 ($t_q=16$, only scenario two
is plotted).
The solid blue line (top two panels) reproduces the 0D simulation, cf. Fig.~\ref{fig:Andalusia0D}.
The median parameters
shown in Table~\ref{tab:And_parameters} (second column) are used, except for the
transmission rates that have been scaled by $\xi \epsilon [0.00480, 0.00489]$.
The optimal scaling factor $\xi$
increases as the diffusion-coefficient reduction factor
$\eta_D$ increases in the interval $[0.25, 0.50]$ in steps of $0.05$.
Shaded regions are delimited by the optimal plots for $\eta_D=0.5$ and $\eta_D=0.25${; their order at $t=150$ is the same
as that of the bottom right panel of Fig.~\ref{fig:Andalusia0D}.}}
\label{AndalusiaPDE}
\end{figure}

We first explain how we selected the model parameters and how
we initialized the PDE model of Eqs.~(\ref{eqn1})-(\ref{eqn8})
and then we discuss the numerical results, emphasizing their advantages and deficiencies.
At the regional (spatial)  level, we
must adapt the 0D model parameters.
In the PDE model, we retained
the same \textcolor{black}{median} parameters as the optimized 0D (ODE) model parameters
starting with the $\sigma$'s and beyond in Table~\ref{tab:And_parameters}.
This is because they involve processes
occurring at the level of a single individual, i.e., ``locally'',
and hence we do not expect them to change at the country
level in the transition from the ODE to the PDE model.
\textcolor{black}{In addition, we kept the same reduction
factor of the transmission rates $\eta_{IS}$ and $\eta_{AS}$ to
model the effect of restrictive measures on them.}

\begin{figure}[!ht]
\centering
\begin{tabular}{cc}
\includegraphics[width=.45\textwidth]{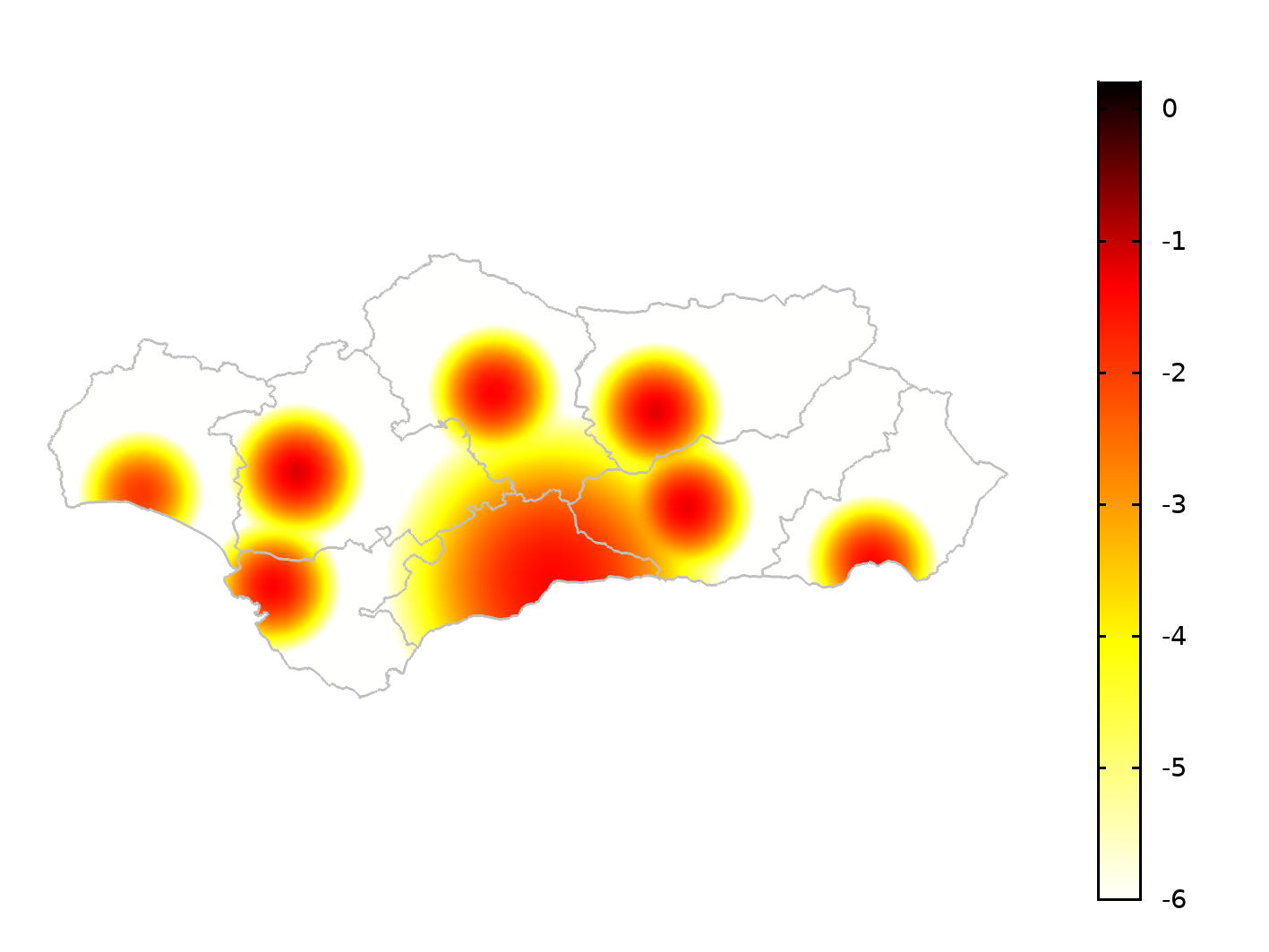} &
\includegraphics[width=.45\textwidth]{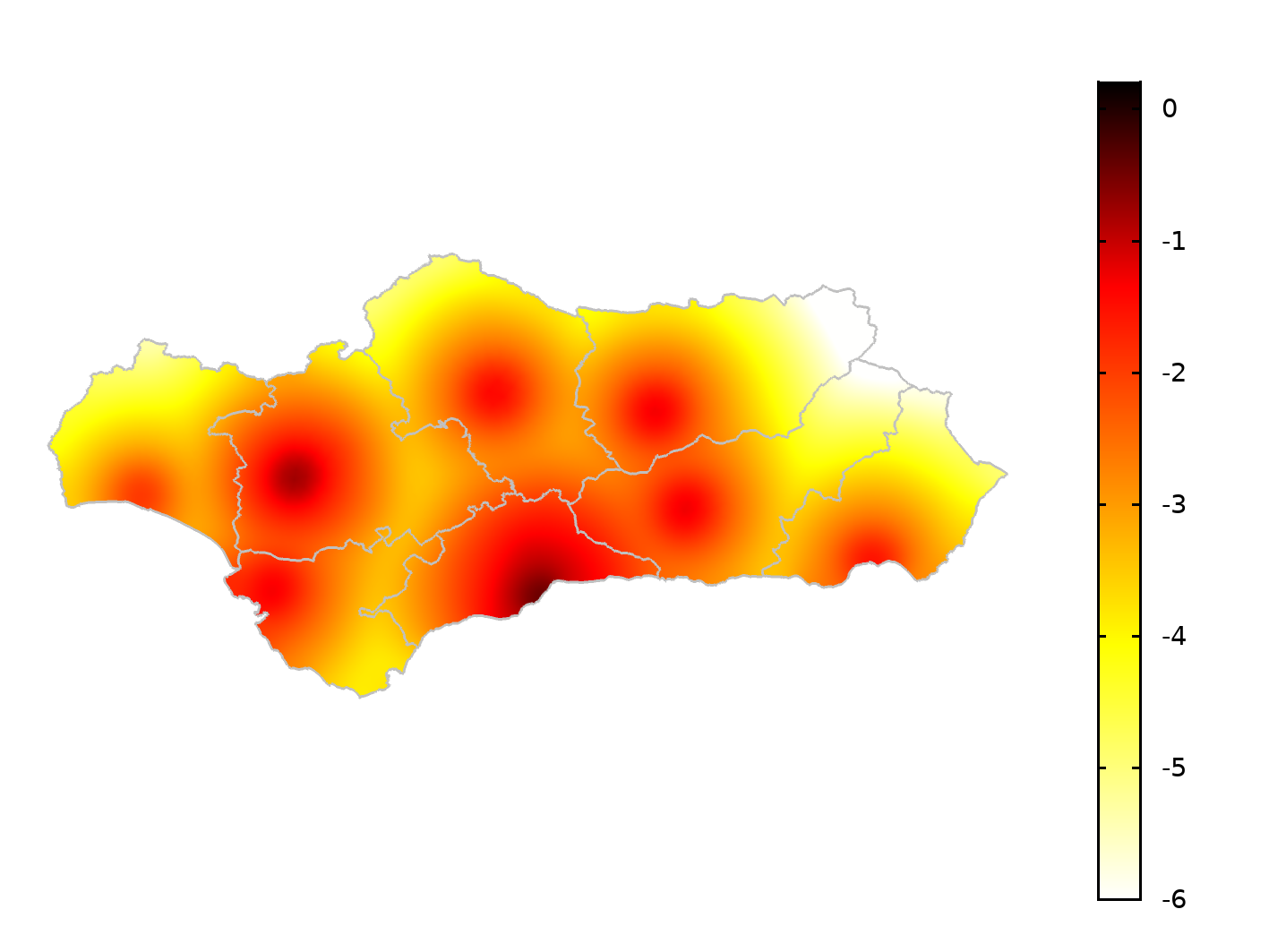} \\
\includegraphics[width=.45\textwidth]{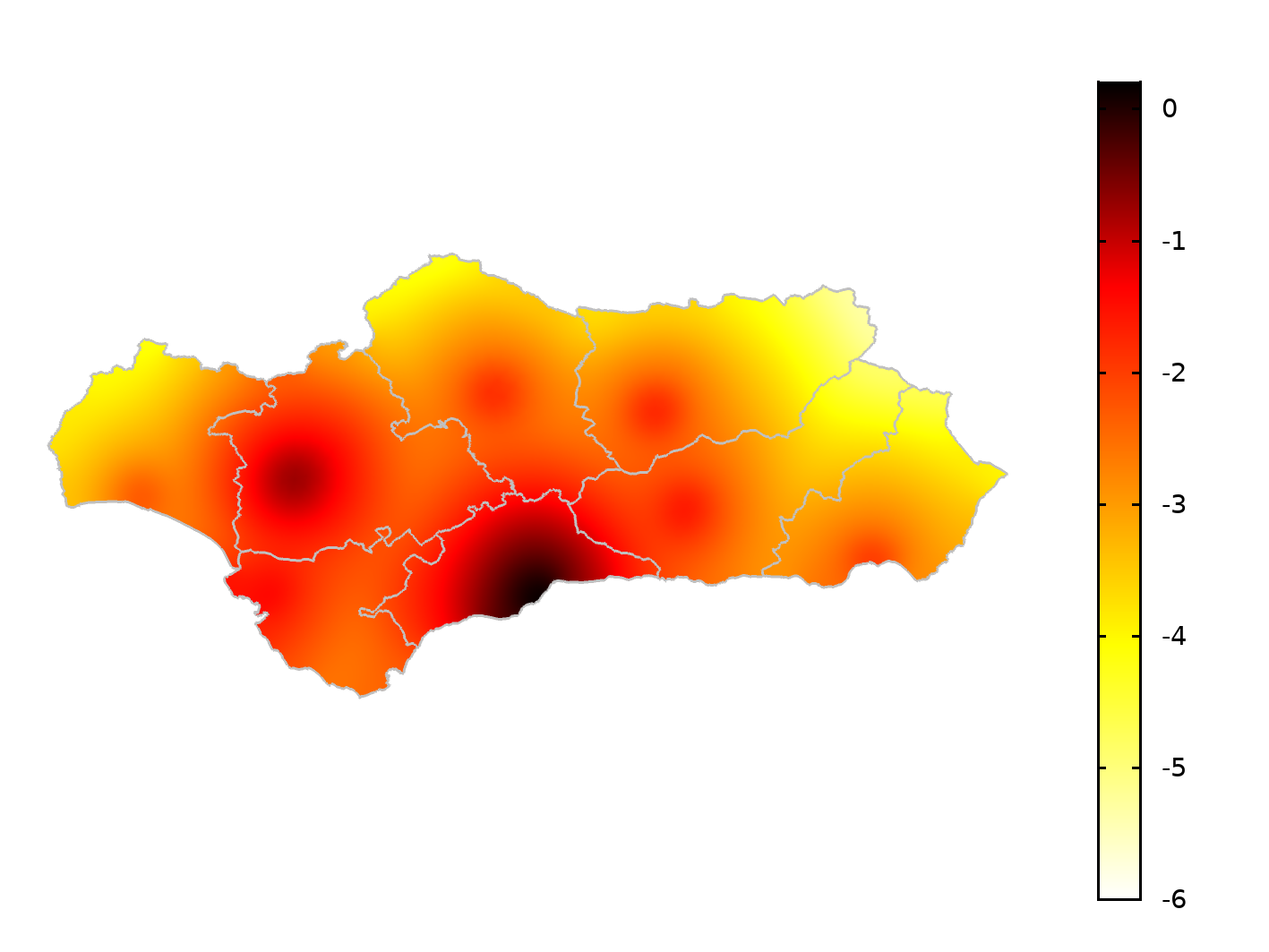} &
\includegraphics[width=.45\textwidth]{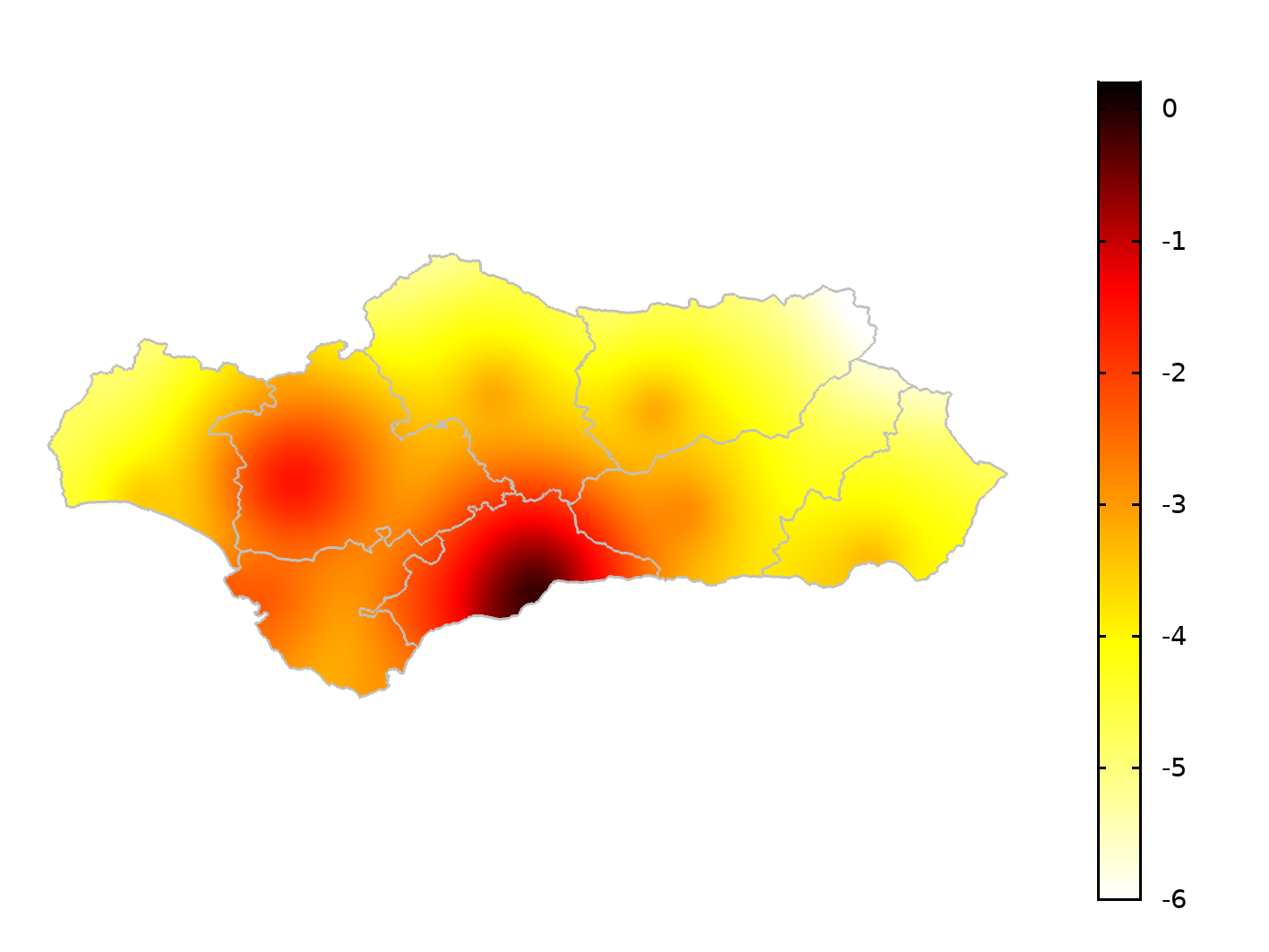} \\
\end{tabular}
\caption{{(Color online.)} Evolution of the Andalusian infected population density $\log_{10}I(x,y,t)$
for $t=1$ day (top left, March 14, 2020), $t=6$ days (top right, March 19, 2020),
$t=16$ days (bottom left, March 29, 2020),
and $t=47$ days (bottom right, April 29, 2020).
\textcolor{black}{Scenario two ($t_q=16$),  reduction of
diffusion coefficients by $\eta_D=0.3$ and scaling factor $\xi = 0.00480$.}
{A logarithmic (base 10) colorbar scale is used.}}
\label{fig:AndalusiaPDE_I}
\end{figure}

\begin{figure}[!ht]
\centering
\begin{tabular}{cc}
\includegraphics[width=.45\textwidth]{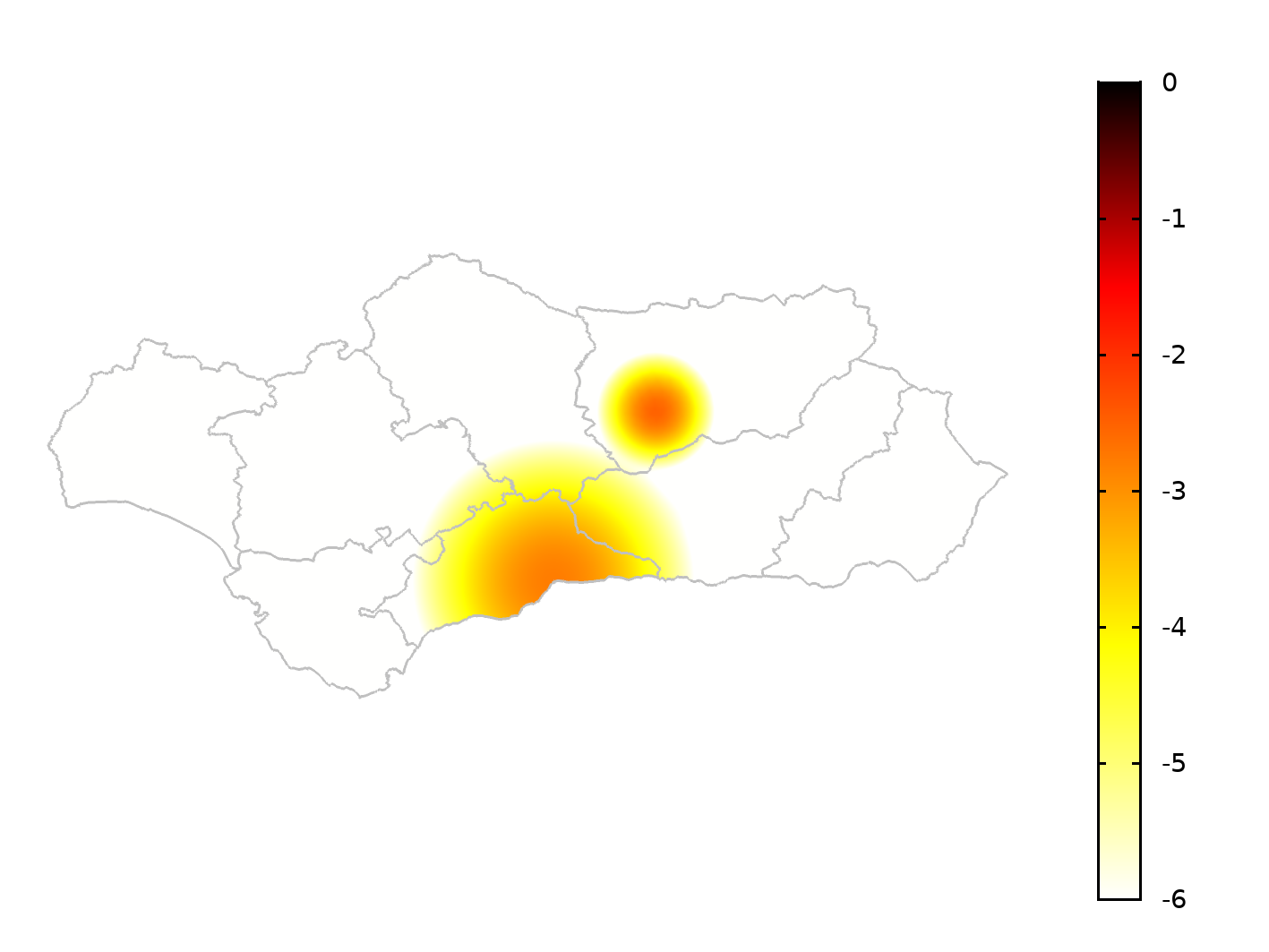} &
\includegraphics[width=.45\textwidth]{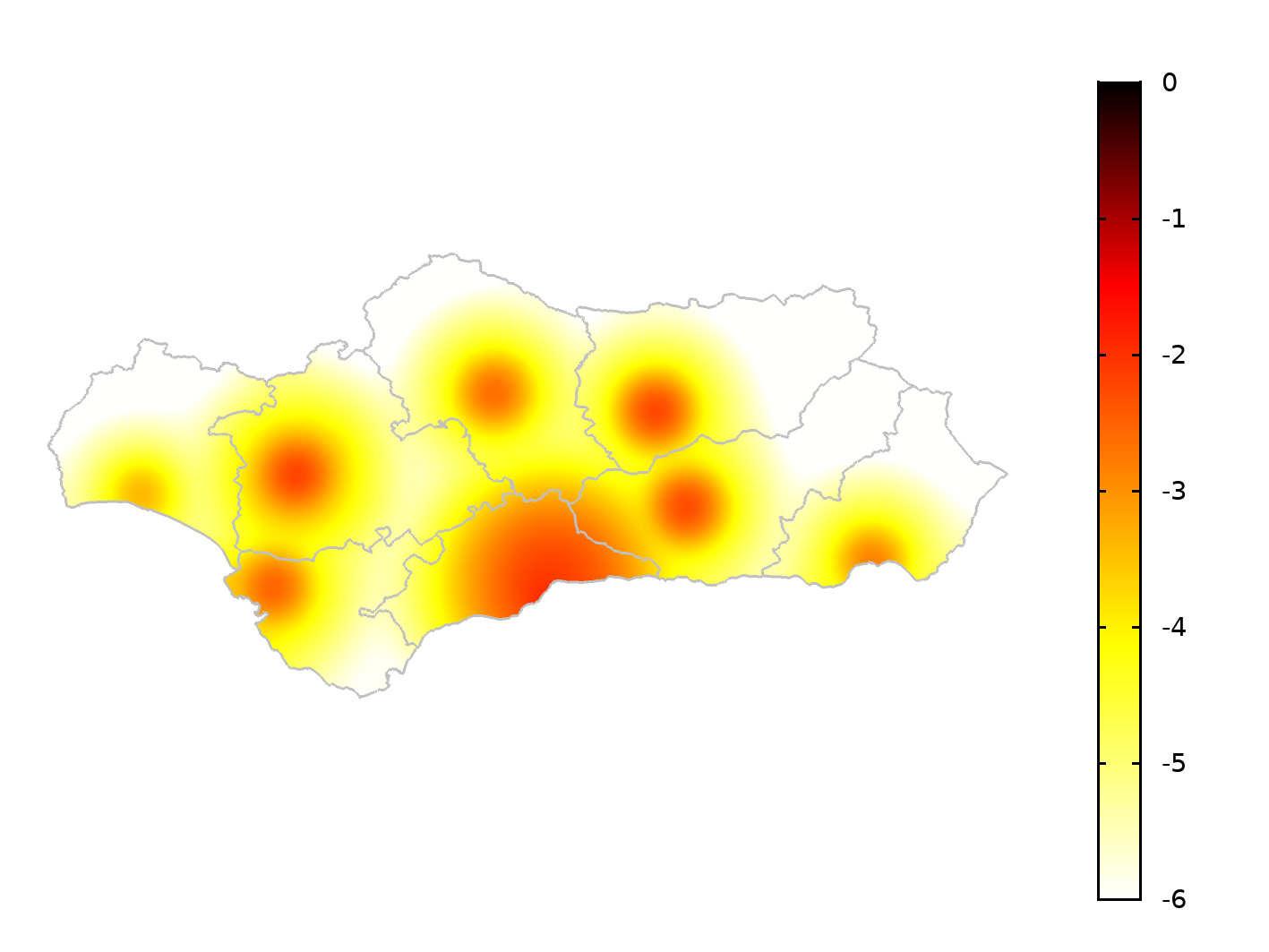} \\
\includegraphics[width=.45\textwidth]{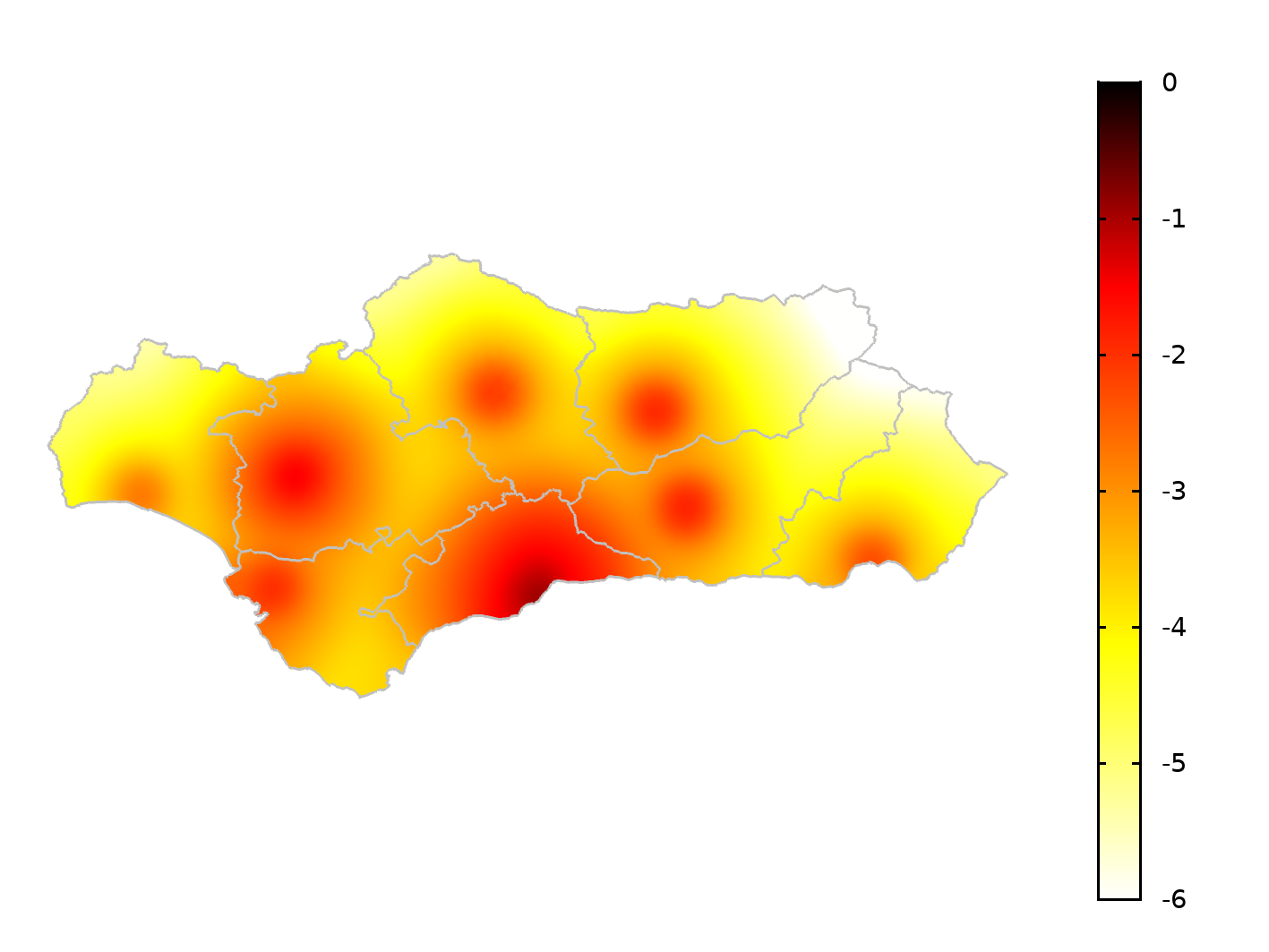} &
\includegraphics[width=.45\textwidth]{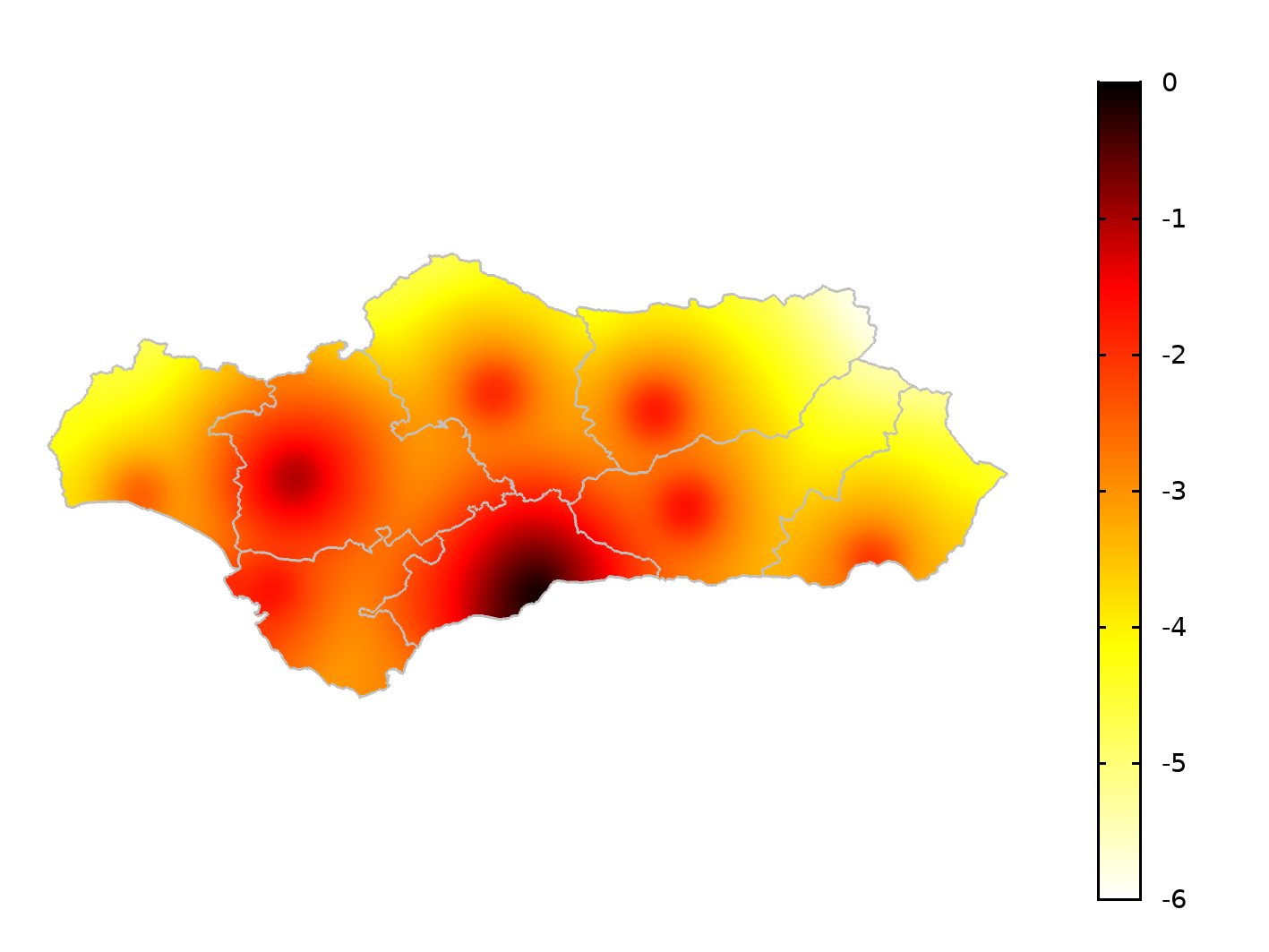} \\
\end{tabular}
\caption{{(Color online.)} Evolution of the Andalusian fatalities population density $\log_{10}D(x,y,t)$
for $t=1$ day (top left, March 14, 2020), $t=6$ days (top right, March 19, 2020),
$t=16$ days (bottom left, March 29, 2020),
and $t=31$ days (bottom right, April 29, 2020).
\textcolor{black}{Scenario two ($t_q=16$), decrease of
diffusion coefficients by $\eta_D=0.3$ and scaling factor $\xi = 0.00480$.}
{A logarithmic (base 10) colorbar scale is used.}}
\label{fig:AndalusiaPDE_D}
\end{figure}

\begin{figure}[!ht]
\centering
\begin{tabular}{cc}
\includegraphics[width=.45\textwidth]{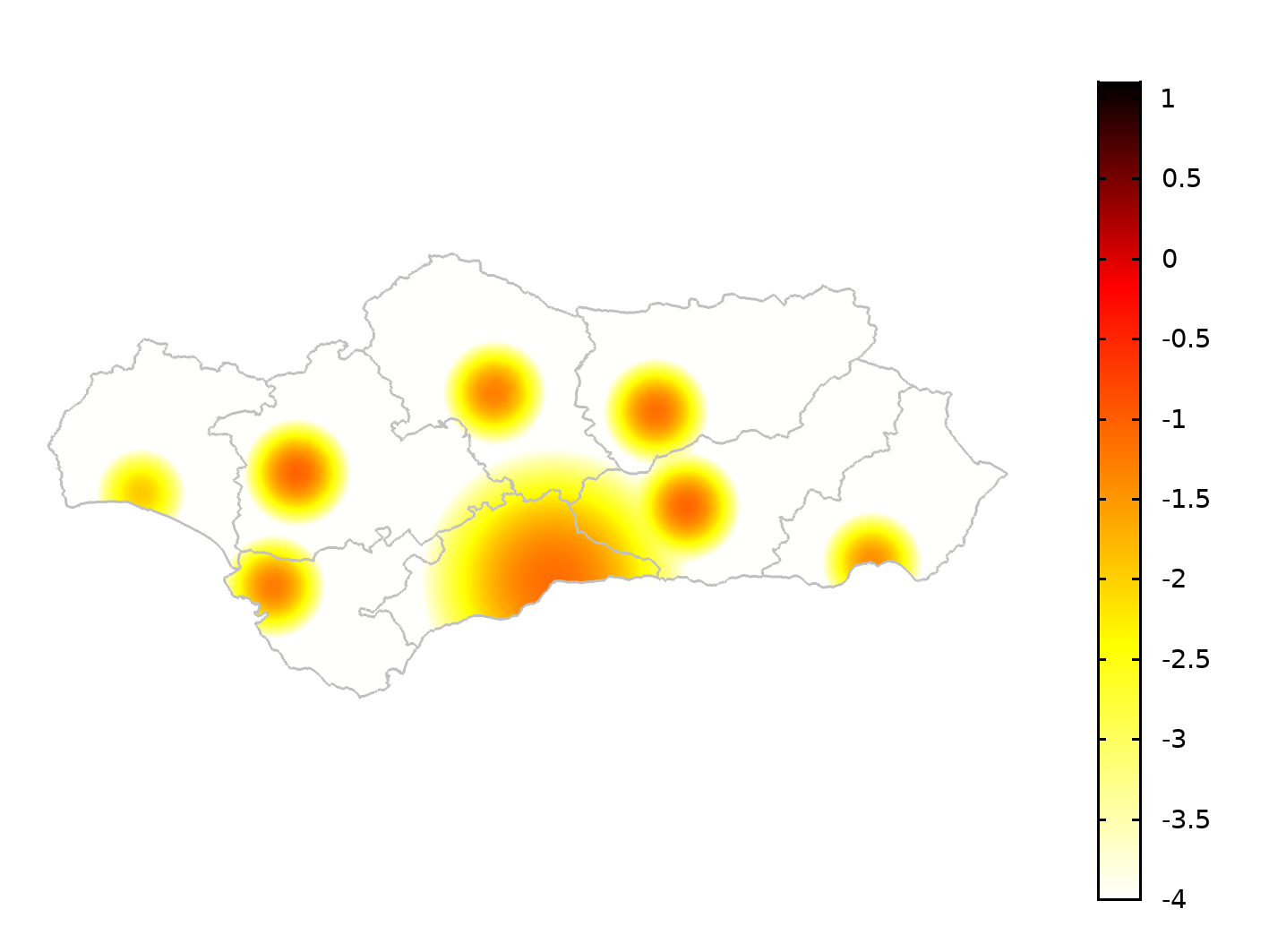} &
\includegraphics[width=.45\textwidth]{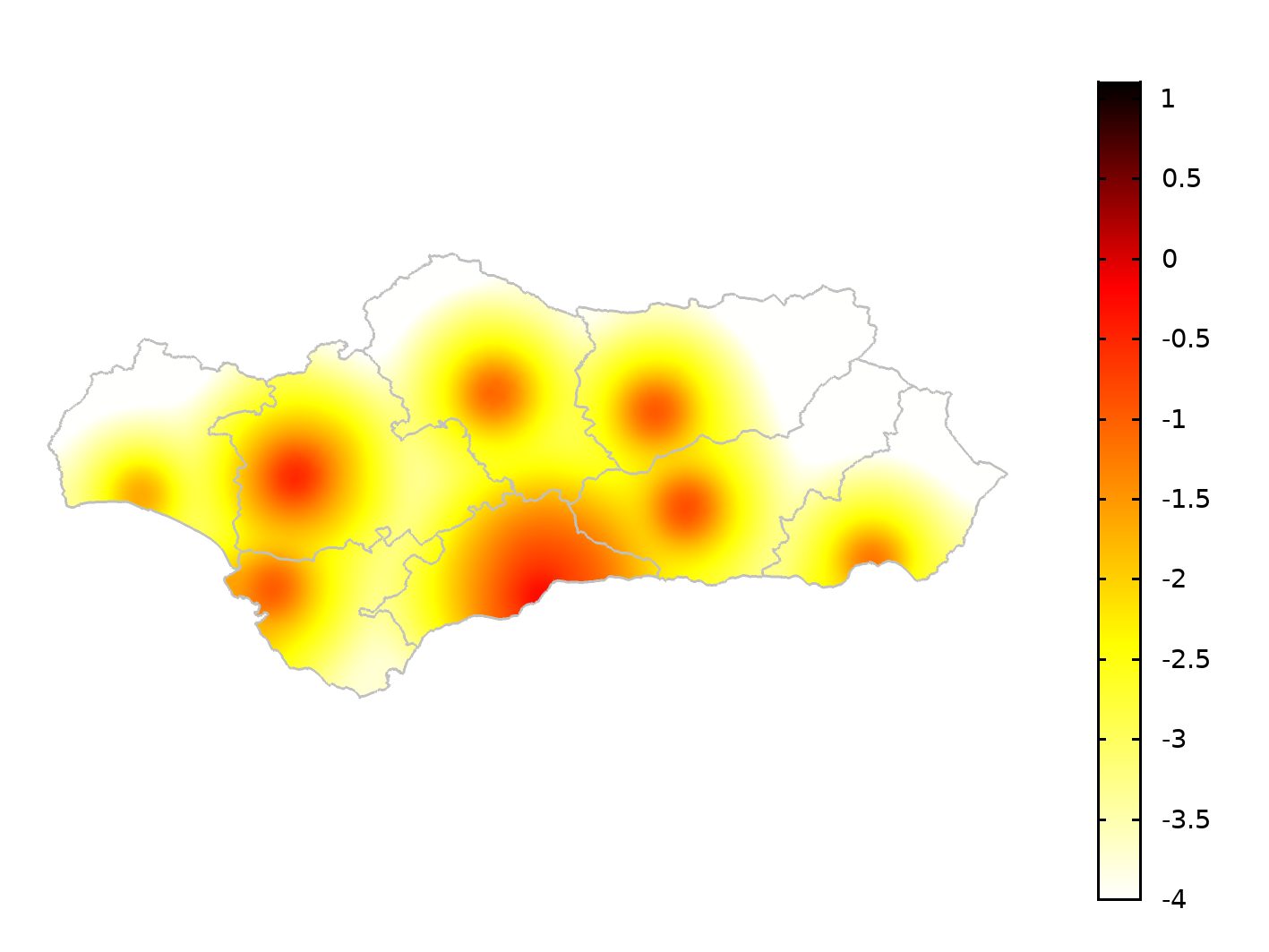} \\
\includegraphics[width=.45\textwidth]{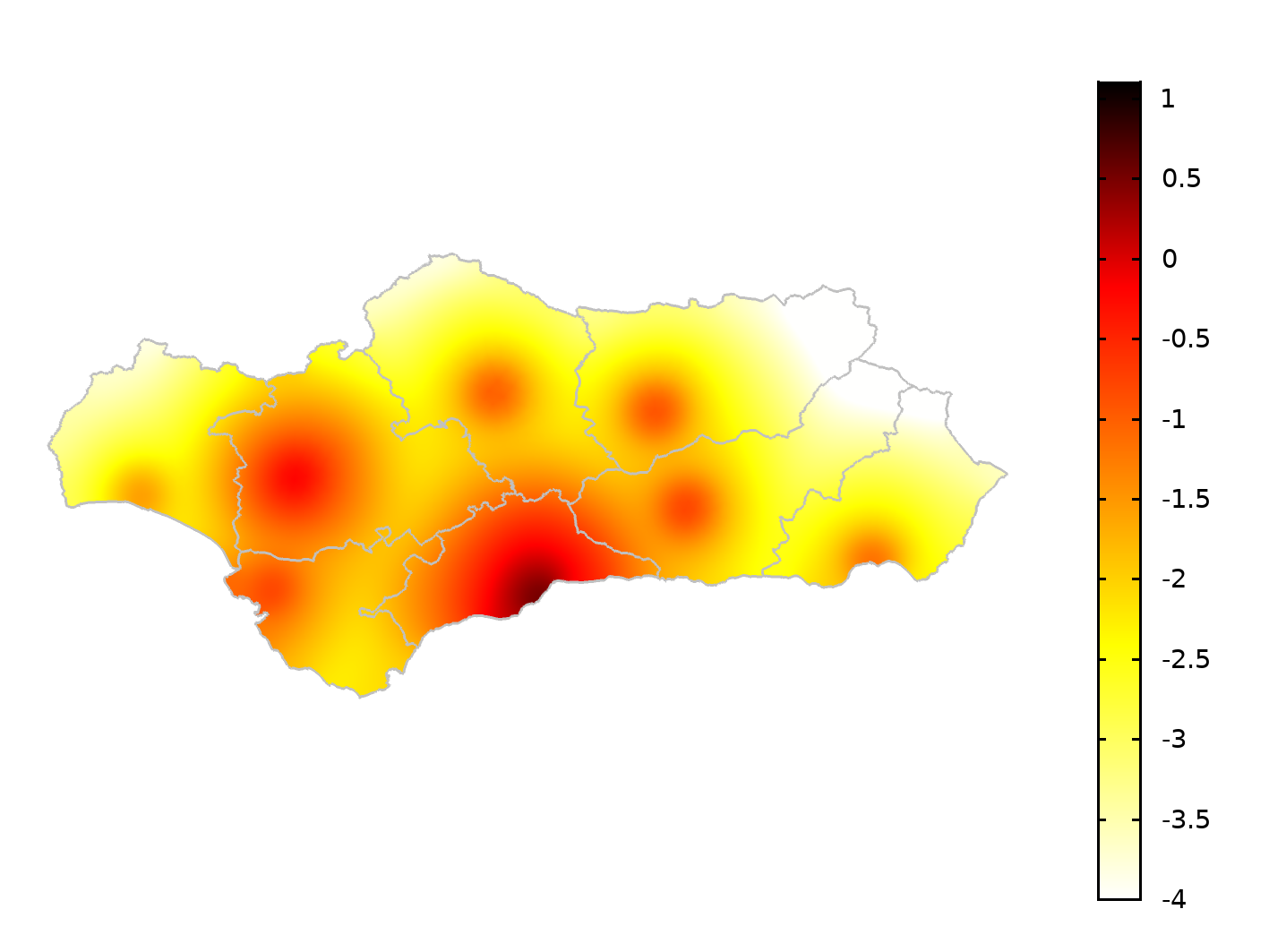} &
\includegraphics[width=.45\textwidth]{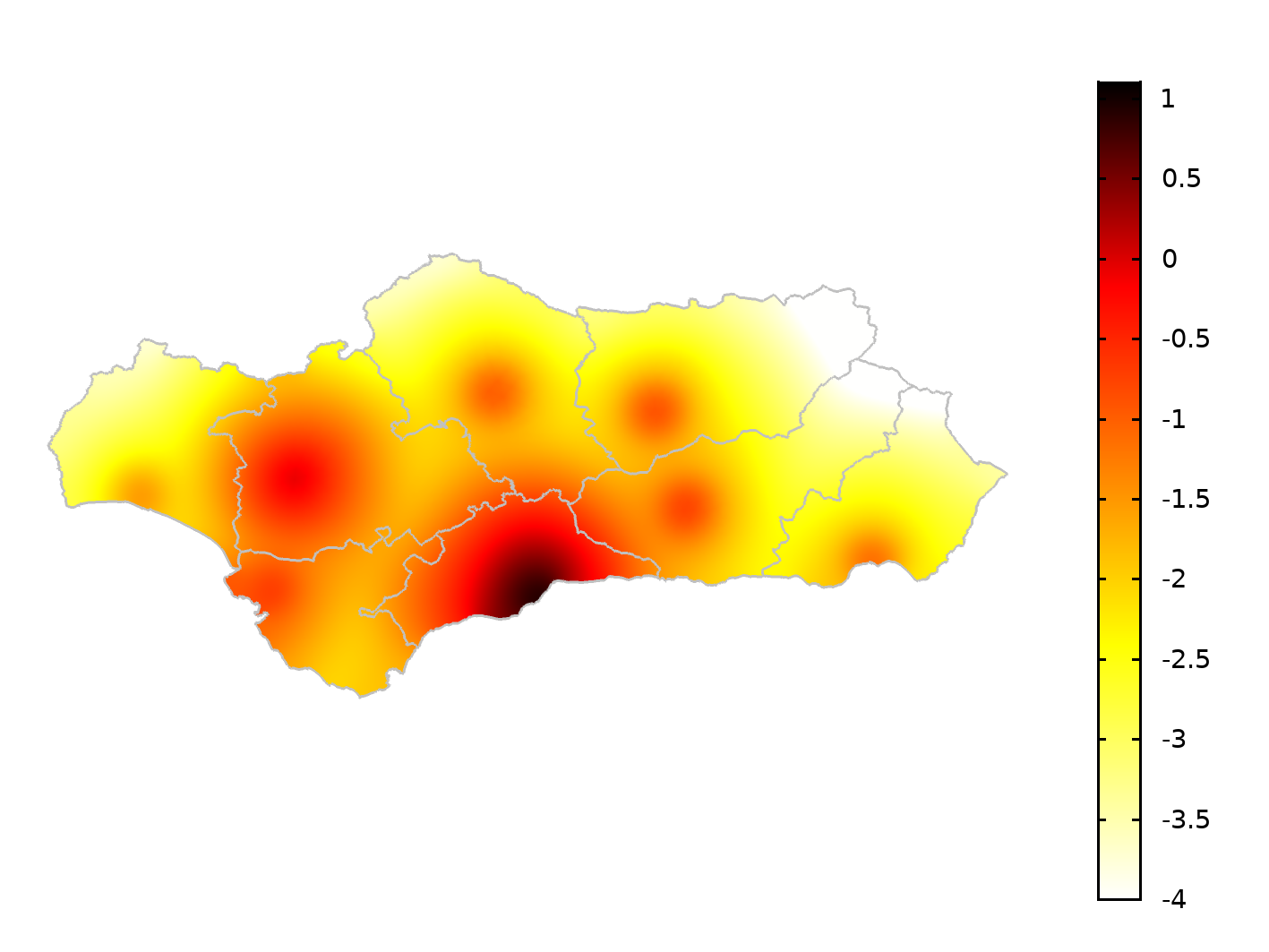} \\
\end{tabular}
\caption{{(Color online.)} Evolution of the Andalusian confirmed case density $\log_{10}C(x,y,t)$
for $t=1$ day (top left, March 14, 2020), $t=6$ days (top right, March 19, 2020),
$t=16$ days (bottom left, March 29, 2020),
and $t=47$ days (bottom right, April 29, 2020).
\textcolor{black}{Scenario two ($t_q=16$), reduction of the
diffusion coefficients by $\eta_D=0.3$ and scaling factor $\xi = 0.00480$.}
{A logarithmic (base 10) colorbar scale is used.}}
\label{fig:AndalusiaPDE_C}
\end{figure}

\begin{figure}[!htb]
\centering
\begin{tabular}{cc}
\includegraphics[width=.45\textwidth]{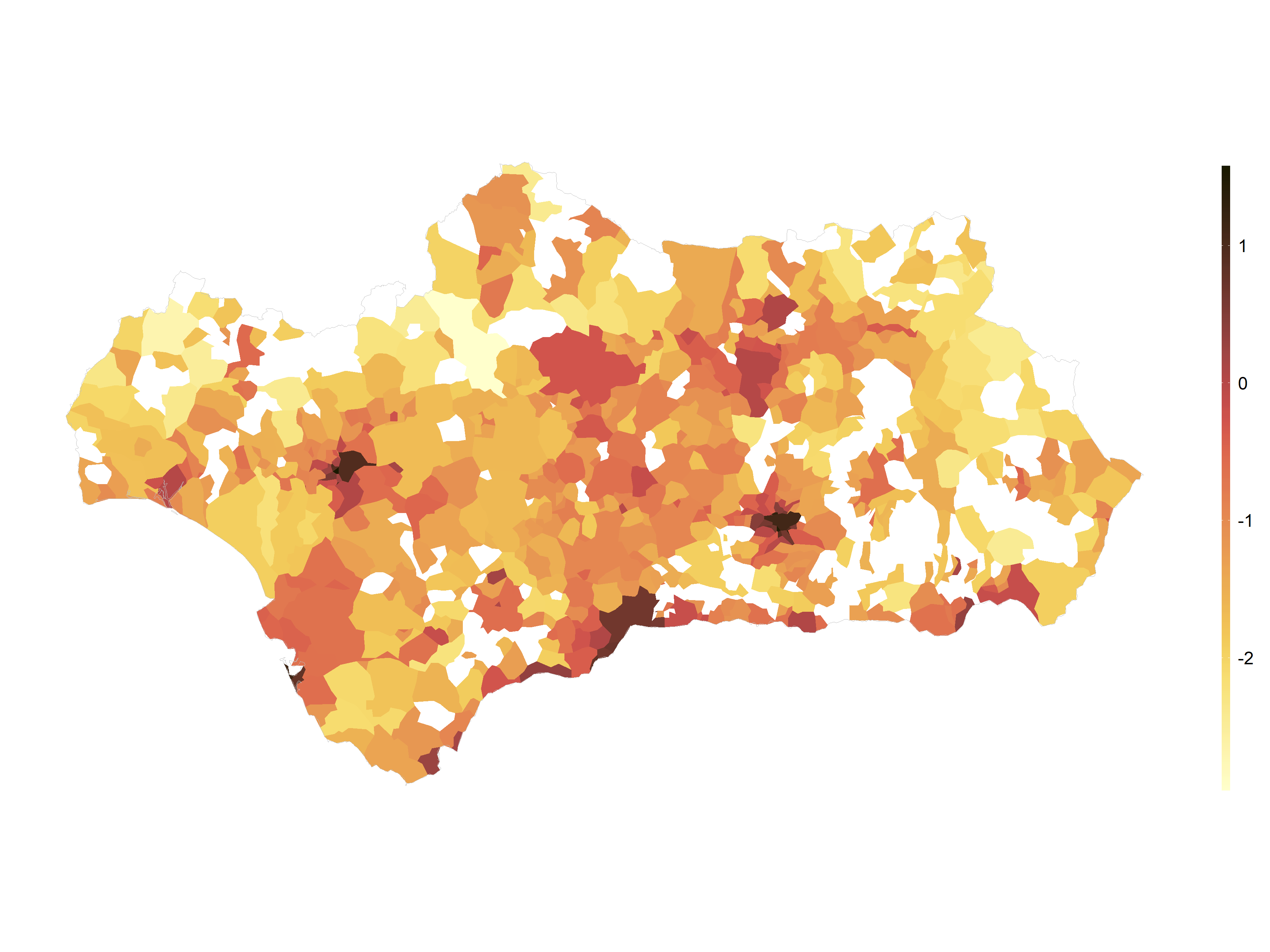}
\end{tabular}
\caption{\textcolor{black}{{(Color online.)} Map of the COVID-19 outbreak in Andalusia as of April 29, 2020 ($t=47$) reproduced
with data and their manipulation via an R code from Ref.~\cite{AndalusiaSpatial}.
{Logarithm (base ten) of the number of confirmed cases per squared kilometer at each municipality
denoted by color (logarithmic colorbar scale)}.
Compare and contrast with the bottom right
panel at time $t = 47$ in Fig.~\ref{fig:AndalusiaPDE_C}.}}
\label{fig:AndalusiaSpatialDist}
\end{figure}

On the contrary, we do not expect this to be the case for
the transmission rates $\beta$.
They depend
on the interaction between individuals since they may be expressed as the
product of the daily average number of contacts times the
infectious disease transmission probability~\cite{Editorialyd2020}.
At the ODE level, the presence of $S$ and $A$ or $I$
immediately leads to the conversion of susceptibles to
exposed. At spatial (region or county) level, this effect does not occur
{\it homogeneously} as it does at the ODE level,
but rather in a {\it distributed} way.
As the population is (spatially) distributed over the country
in a highly heterogeneous way,
the
ODE $\beta$'s have to be modified to obtain their ``spatially averaged'' variant.

We obtained these spatially averaged transmission rates
by first keeping the {\it ratio} of the $\beta$'s the same
as that of the ODE, but scaling each one by a scaling factor $\xi$.
The transition from the ODE to the PDE transmission
rates involves the introduction of two length scales.
The first reflects the transition from the number of individuals (e.g.,
$S$, $I$, $E$, etc.) in
the ODE description to spatial densities of individuals in the PDE description;
the other length scale reflects the transition from a
spatially homogeneous to a spatially distributed model.
We obtained their product by noting that the
$\beta$'s have to be multiplied~\cite{MargueriteSpatialDynamics} by an effective
inverse density $l^2/N$, $N$ being the country population,
\textcolor{black}{i.e., by multiplying the ODE transmission rates by
the scaling factor $\xi \equiv l^2/N$.}
The product length scale $l$ defines
an effective spatial scale over which the ODE
transmission rates need to be rescaled to obtain the corresponding
PDE transmission rates.
\textcolor{black}{The scaling factor was determined by minimizing the $\ell^2$ norm
specified in Eq.~(\ref{eq:norm1age}). For these optimizations we kept all
model parameters at their median values, while diffusivities were
varied as subsequently
discussed in Eq.~(\ref{eq:TimeDependentD}).}

In addition to the decision regarding the scaling of the $\beta$'s,
an important decision is that of the selection of the diffusivities.
Recall that in the present first work  we decided to avoid
attempting to model convection effects, but rather mostly focus
on the role of diffusion. We assume that
most of the populations relevant to the infection which have not
developed
any symptoms, namely the asymptomatics and the exposed, diffuse
with a diffusivity of $\mathfrak{D}_c=100$ km$^2$/day (i.e., associated
with a characteristic spatial scale of about $10$ km). Our
motivation for this choice is that in this small population (for the regions and
data considered) associated
with the infection, it is relevant to include a wider spatial spread
of their motion to enable (through their contacts) the infection to
spatially spread. On the other hand, for the far larger population of
susceptibles,
we assign a smaller diffusivity ($0.1 \mathfrak{D}_c$)
\textcolor{black}{since we consider that the mobility of susceptibles does not
significantly change
their distribution~\cite{FastDiffusion}. In essence the
much larger susceptible population provides a background  for the relative motion of
asymptomatic and exposed carriers of the virus.}
The rest of the populations (most notably,
$I$, $H$, and $D$, since the immunity of $R$ and $AR$ renders their
diffusion inconsequential) are assumed to be highly localized/self-isolating and
hence bear, for our purposes, a vanishing diffusivity.
For all the non-vanishing diffusivities, we assume that the
quarantine reduces them to a fraction $\eta_D$ of their original value (see Table~\ref{tab:And_parameters}) in a similar ramped
form as before for the transmission rates:
\begin{equation}
  \mathfrak{D}(t)=\mathfrak{D} \left \{ \eta_{D}+(1-\eta_{D})\frac{1-\tanh[2(t-t_q)]}{2} \right \}
  \label{eq:TimeDependentD}
\end{equation}

\begin{figure}[!htb]
\centering
\begin{tabular}{cc}
\includegraphics[width=.45\textwidth]{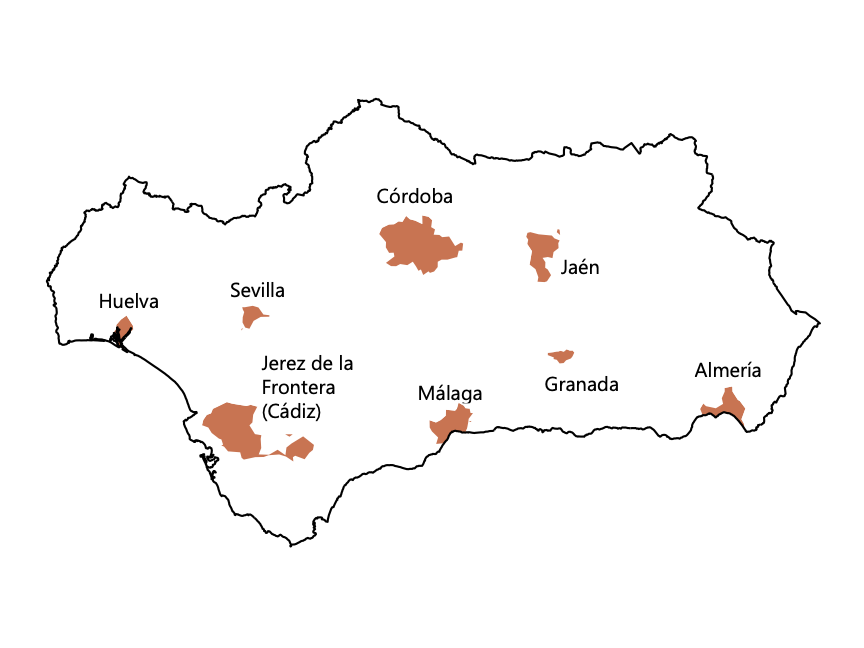}
\end{tabular}
\caption{{(Color online.) Map of Andalusia showing the most populated cities of each province. Notice that the name of
these cities coincides with the name of the province, except for the province of C\'adiz, where the city
of Jerez de la Frontera is more populated than the city of C\'adiz.}}
\label{fig:AndalusiaCities}
\end{figure}

We should also describe the initialization of the model.
We selected to populate initially eight key ``hotspots'' of
the infection as they arose in Andalusia. The
selected areas (Almer\'{\i}a, C\'{o}rdoba, Huelva, Granada, Ja\'en, Jerez de la Frontera, M\'alaga and Sevilla)
correspond to the most populated cities of each province of
the autonomous community {(see Fig.~\ref{fig:AndalusiaCities} for a map indicating the location of these cities)}.
Initial values for the
infections and deaths were provided by the Andalusian Government (``La Junta de Andaluc\'{\i}a'') \cite{datos_Junta}.
We defined an infection radius of $10$ km around the center
of each hotspot, within which we placed the source of infection
to initialize the epidemic,
what we refer to as  ``blobs'' of infection.
These epicenters of infection were modeled via Gaussian
profiles whose spatial (variance) scale was selected to be the infection
radius; their amplitude was chosen such that the total number of infections,
deaths, recoveries and hospitalizations, as calculated via the surface
integrals of the associated densities through the region, be the same
as the one reported in the original data. The population of
asymptomatics
and exposed was, similarly to the ODE optimization, selected to be
proportional to the infected one with the proportionality ratios $A_0/I_0$ and
$E_0/I_0$ maintained as those of the ODE.

With all these choices, the PDE model was run {\it without} optimizing at the PDE level \textcolor{black}{the
median parameters that are not expected to depend
on spatial scales. The quantity that we varied was the
diffusion-coefficient
reduction factor
$\eta_D$ in steps of $0.05$ in the interval $[0.25, 0.50]$ (six simulations in total).
For each simulation the scaling factor $\xi$ was determined by minimizing the
$\ell^2$ norm, as previously discussed.
We used the second scenario parameters ($t_q = 16$) since this
choice reproduced better the data and the  flattening of the epidemic curves
with the imposition of the lockdown.}

The comparison of the spatially-integrated PDE results to the data for Andalusia and the
ODE prediction, shown in Fig.~\ref{AndalusiaPDE}
is quite promising.
We show the observed data as black dots, the 0D-model
predictions as the solid blue line, and the spatially integrated
results of the PDE model as the shaded region. \textcolor{black}{The bottom
and top of the shaded region are enclosed by the curves
corresponding to $\eta_D = 0.50$ and $\eta_D = 0.25$ (when
the optimal scaling factor $\xi$ is used). We found that
both $C(t)$ and $D(t)$ asymptote to lower values for the optimal
scaling factor as $\eta_D$ increases. As in
the case of the 0D model, the change in the transmission
rates and diffusion coefficients as a result of the
lockdown leads to a flattening of the
epidemic curves. An additional remark on the reported and predicted number of cases,
and on the counting of asymptomatics, is in order
(Fig.~\ref{AndalusiaPDE}, top left panel). As reported in~\cite{AndalusiaAsymptomatics}
starting on April 13, 2020 ($t=32$) the reported number of cases includes asymptomatic individuals,
i.e., susceptibles that have tested positively to the presence of the virus. This might
explain the underprediction of cases after $t=32$.}

Clearly, the spatial model can do an adequate job in
capturing both the cumulative infections and the number of deaths (with the caveats to be
given in the discussion below). Notice that in the bottom row of the figure,
we illustrate the surface integrals of each of the density of $E$, $A$, $H$, $R$ and $AR$ as a function of time,
representing the evolution of the pandemic at the ``integrated'' level of the entire country
in an illustration similar to the one that we typically obtain from the ODE models.

In addition,
we complemented the spatially-averaged results
by the space-time evolution simulations of Figs.~\ref{fig:AndalusiaPDE_I}--\ref{fig:AndalusiaPDE_C}.
For the reported cases in the figures, the scaling
factor multiplying the  $\beta$'s is $\xi = 0.00480$, corresponding to a characteristic
scale $l \approx 0.200$km.
\textcolor{black}{A comment on the scaling factor $\xi$ is in order.
For the different choices of the diffusion-coefficient reduction factor $\eta_D$ we determined
the optimal scaling factor to be in  the range $\xi \epsilon [0.00480, 0.00489]$, for
the lower value of which we obtain the reported  characteristic length scale ($l =\sqrt{\xi N} \approx 0.200$ km).}
\textcolor{black}{It is tempting to associate the scaling factor,
and in particular $\xi = 0.00480$, with the inverse of the ``lived" population density, the population density
perceived by a randomly chosen individual~\cite{LivedDensityCovid19}.
According to~\cite{LivedDensityAndalusia} the inverse lived
density for
Andalusia is approximately 0.0046,
remarkably close to the smaller value of the scaling-factor range. A similar observation holds for the
scaling factor for Greece which, in turn, suggests that this is an
important insight (and not a serendipitous occurrence) as concerns
the ``translation'' of the 0D model coefficients into the PDE ones.
}

Of course, the PDE model has considerable additional
information through its spatial resolution. In
Figs.~\ref{fig:AndalusiaPDE_I}--\ref{fig:AndalusiaPDE_C} we can see the
spatio-temporal evolution of the infections in Andalusia (i.e., the spatial distribution at a few
snapshots over time), the deaths and the cumulative infections $C(x,y,t)$,
respectively. \textcolor{black}{We also produced movies of the corresponding
evolution that can be found in~\cite{movies}.}
We can observe how the biggest
fractions of the infections remain in the most populated cities
of Sevilla and M\'{a}laga, and that the provinces of Huelva
and Almer\'{\i}a are those with the smallest number of infections,
in accordance with the {actual status} of the pandemic~\cite{datos_Junta}.
\textcolor{black}{The predicted spatial distribution of the total confirmed cases $C(t)$
may be compared to the data shown in Fig.~\ref{fig:AndalusiaSpatialDist}, where
officially reported data are presented per municipality~\cite{AndalusiaSpatial}.
The comparison is favorable as both figures show that M\'alaga is the hardest hit municipality,
followed by Sevilla. In addition, the predicted number of cases in Granada, C\'ordoba, Ja\'en
and Jerez de la Frontera (in decreasing number of cases), which are lower than in the previous two cities,}
{shows the same decreasing trend as that manifested in the reported number
  of cases shown in Fig.~\ref{fig:AndalusiaSpatialDist}.}
\textcolor{black}{The predictions for the number of deaths in the different
provinces shows a slightly different behavior from the observed data.
If the fatalities spatial density is integrated over
each province, the predicted final number
of deaths is higher in M\'alaga than in any other province.
From the data, the final number of deaths is almost the same in M\'alaga, Sevilla and Granada.
This suggests that modeling human mobility in these provinces solely
by diffusion, and specifically by the chosen diffusion coefficients,
cannot fully account for this dispersion of the number of fatalities.
This may arise, within our reaction-diffusion model, possibly
due to the number and intensity of the chosen hotspots of infection
or to the requirement that
a diffusion coefficient of different value should have been chosen.}

\section{Conclusions, Discussion and Future Work}
\label{sec:Conclusions}

In this work we presented a platform for establishing a
compartmental epidemiological
model both at the level of ODEs (0D, no spatial dependencies)  in
line with numerous earlier
works, as well as at the {spatially distributed}
level of PDEs
{to study the spatio-temporal spreading of COVID-19.} The regions of interest
were the mainland of Greece and the Spanish autonomous region of Andalusia for which there
has been a small number of studies.
As regards Greece, there are some probabilistic~\cite{tsironis,newer},
 some network-based approaches for time-series
analysis~\cite{newer0}, \textcolor{black}{and some based on the SIR variant
SEAIR~\cite{NikosGreece}.}
Studies that focus on Spain also examine Andalusia
as a case example using either probabilistic~\cite{arenas} or
POD-based decomposition techniques~\cite{andal}.
Our effort has been to explore a model of the
{SEIR variety} 
that incorporates some of the particular biological features
of the SARS-CoV-2 virus~\cite{elife}, such as its latent period,
and the potential to generate a significant fraction of asymptomatic
hosts, which, in turn, play a crucial role in spreading the
infection. {The resulting SEAIHR model
involves a number of populations: Susceptible,
Exposed, Asymptomatics, symptomatically Infected, Hospizalized,
Recovered, and deceased.}
We found that for the regions of interest the
model reproduces the epidemiological data that we determined to be
most reliable, namely the data on the cumulative infections
and especially the number of deaths.
Naturally, more accurate data including also spatially resolved ones
(on the spatial scale of our PDE model)
would be helpful towards the improved calibration of the results
offered herein.

{We modeled both the early, pre-quarantine,
stage of the epidemic, as well as its development at a later stage
when containment measures had been enforced.
The effect of quarantine on the spreading of the disease was imposed
via a time-dependent   (on the time scale of a day) change
of the transmission rates \textcolor{black}{and of the diffusivities
  (the latter in the PDE model).}
While initiating a quarantine roughly when it was imposed
yields more acceptable results in Greece, in Andalusia this is less so.
\textcolor{black}{In fact, for both regions, simulations
reproduced more accurately the observed data, and in particular
they captured the ``angle'' indicating the curbing of
the infection due to government-imposed intervention measures, 
if a time-lag is imposed on the application of the (instantaneous in
the ODE model) quarantine set of parameters. We, thus, considered
two scenarios, corresponding to different delays in
imposing the quarantine: the second scenarios reproduced
the reported data better.}
{We note that \textcolor{black}{this seemingly artificial
time shift in imposing the lockdown reflects the fact that
model parameters change over  a short time scale upon
the introduction of lockdown measures,
whereas the effect of social intervention measures (self-quarantine, social
distancing, face masks, etc) appears to arise later in
the epidemiological data.}

We determined model parameters via
optimizing model predictions with respect to reported
total number of (infected) cases and number of deaths.
\textcolor{black}{The optimization algorithm minimized the Euclidean
distance between model predictions and observed data. For the 0D model, we performed
2000 optimizations for model parameters and (the unknown ones
among the) initial conditions
uniformly sampled within specified ranges. Median and interquartile
ranges for model parameters were determined. Additional
sensitivity analyses were performed to conclude that
combinations of parameters, specifically the product of
the asymptomatic transmission rate times the inverse latent period,
is non-identifiable.
This parametric combination is intimately connected to the product of the infected
transmission rate times the inverse incubation period: in fact, our
results suggest that the relation between these two products
can be well approximated via a straight line of negative slope.}
We interpreted both the median time scales of, e.g.,
the conversion of exposed to asymptomatics and (symptomatically)
infected, and the fraction of, e.g., hospitalized that
lead to recoveries or deaths. \textcolor{black}{We found them, for both countries
and the second scenario, to be in reasonable
agreement with current epidemiological estimates.
Our median results reinforce the feature prevalent in
numerous studies about the importance of asymptomatics
in the transmission of the SARS-CoV-2 virus, cf.~\cite{arons2020,he2020},
a particularity of this coronavirus. The asymptomatic
transmission rate $\beta_{AS}$ was found to be smaller than the symptomatically
infected rate $\beta_{IS}$, coupled to
the fraction of exposed evolving to asymptomatics being larger than those evolving to
symptomatically infected.}
In \cite{li2020} it was reported that asymptomatic infectious
hosts may account for up to 86\% of cases, thus further supporting our prediction
of their importance in the spread of the disease. \textcolor{black}{We remark
that of the four cases
studied only one (Andalusia, scenario one, early imposition of lockdown measures)
had $\beta_{AS} > \beta_{IS}$ (and a lower asymptomatic to infected
split), a case that did not reproduce accurately the data: in the more accurate simulations
of scenario two the inequality was inverted. Once again, however, we
caution the reader that issues of identifiability prevent us from
assigning
a particular weight to the findings about the relative size of
$\beta_{AS}$ vs. $\beta_{IS}$, other than their corroborating
the central role of asymptomatics in the transmission of SARS-CoV-2.}

We then utilized the \textcolor{black}{median} parameters in a spatially
distributed, \textcolor{black}{reaction-diffusion}  model. Here, we overcame the major challenges
of formulating a mesh with the boundaries of a region
within the software package COMSOL and also leveraged state-of-the-art geographical
methods such as the World Pop project (for population mapping based on
census data)  to set up distributed simulations of the
pandemic
spreading in the geographical domain.
\textcolor{black}{We consider this computational effort a significant
and necessary non-trivial step for the eventual inclusion of more realistic
long-range human mobility modeling via the inclusion, e.g., of convection
or other modeling of directed-motion of individual populations.}
We pondered on how
to adapt the parameters of the ODE model to the PDE framework
and argued that ``onsite'' (i.e., single-individual) parameters
can be maintained the same. We also explained the challenge of
adapting contact parameters (such as the transmission rates) to
the level of the country: this process involves issues of homogeneity at the ODE
level vs. substantial heterogeneity at the PDE level. We
also made a first series of assumptions at the
level of convection (neglected herein) and isotropic diffusion (selected as the
primary mechanism for disease spreading herein) to
explore the time-resolved dynamics at the country/autonomous community setting.

At the level of our distributed simulations, there exist some
promising results. We were able to seed the infection at some
of its key epicenters and observe it to produce infections,
recoveries,
deaths, etc., over the entire region. \textcolor{black}{The ``hotspot" seeding
at various locations is an indirect attempt to model the
movement of infectious individuals as is, e.g., considered
by metapopulation or network models.} At the cumulative level
of the region, surface integrations enabled comparisons with the
collected data at the regional level  
yielding reasonable
correspondence between model results and the observed
\textcolor{black}{cumulative} epidemiological
reality.
\textcolor{black}{Moreover, the model appears to be promising
  towards capturing some
of the spatial features of the infection progression: for instance,
visual comparison of model predictions with reported spatially distributed
data for the cumulative infections shows that the model reproduces
the persistence of infections in highly populated areas, albeit with
a possible time lag.}
\textcolor{black}{At the spatial level} we find
that the infection persisted the longest in regions of very
high population density.
We believe that this effort paves the way for a distributed
observation
of the relevant spreading, but it also has some weaknesses,
challenges, and improvements that are worth considering
in future steps.
\textcolor{black}{As stated in the Introduction, the
spatio-temporal modeling of the epidemic by reaction-diffusion PDEs,
and specifically with isotropic diffusion being the dominant
mechanism of spatially spreading the virus is an \textit{ important
first} step towards developing a continuous description of disease spreading
where human mobility is modeled at a fine spatial scale. This approach
should be contrasted to discrete network-based metapopulation models
and the length scales considered in these models. A distinct
advantage of the continuous model over the meta-population one is that the former
can model interaction at a finer and more extended inter-nodes scale.}

It would be especially useful in the context of the
present
pandemic of unprecedented information flow~\cite{WHO_Count} to have
easily accessible temporally and spatially resolved data for the
evolution
of the pandemic in different regions. Such "seeding" in a
distributed
way (rather than the colloquial seeding at hotspots performed herein)
would build into the model an accurate spatial distribution of
infected
population, and, hence, would be far closer to the country's pandemic
evolution. Indeed, there is another challenge that is arguably even
more significant.
\textcolor{black}{Diffusion as a mechanism for spreading a disease
is traditionally associated with diseases that have specific transmission
characteristics, as for example vector-borne malaria
that is transmitted by mosquitoes \textcolor{black}{\cite{zhao2011}}.
Herein, we considered
diffusion as a proxy for short-term human mobility
(which via its interplay with nonlinear contact interactions
provides a mechanism for spreading the disease), relegating long-range
transport
to the initial seeding of the infection.}
Yet, admittedly it is not
sufficient
for expanding the infection at the scale of the country as our results show, at
least
not via realistic spatial and temporal scales of individual mobility.
In particular, it has not escaped our attention that this type of spreading does
not account for the directed motion of individuals (possibly infected
ones) from the city to the country, or from one city to another for
pleasure or business. This is especially important for travel
(and hence infection transport) at a longer spatial scale (rather than the
shorter one enabled by diffusion).
\textcolor{black}{We note that a form of anisotropic diffusion was used
to model disease spreading primarily
along highways in~\cite{FastDiffusion}.}

This suggests that some form of a probabilistic element needs to be
inserted in the model. One possibility that we are exploring
is the spatial distribution of the initial condition of the asymptomatic
population.
This may generate infections in a more spatially distributed way,
leading
to the spatial expansion of the pandemic throughout the country
in a more consistent way with the observed data~\cite{greecewiki}.
A perhaps even more significant or possibly complementary perspective
worth considering is, naturally, a probabilistic
one.
In addition to deterministic processes like diffusion or convection
(which
is  worth integrating in a subsequent version of the model), it seems relevant
to include a probabilistic gain and loss term reminiscent of (a
long-range variant of) the {conservative}
Kawasaki
dynamics~\cite{red} at the level of spins. This type of term would generate
infections in a probabilistic way (possibly with a probability weighed
upon the region's population density) by allowing individuals to
effectively
``perform trips'' through the country, i.e., disappearing from one
location and reappearing (within a short time scale of less
than a day for the regions of interest) in another.

As also discussed in the Introduction, there are other ways
by which to bypass the practicalities of the application of PDEs
at the level of a country. One of the canonical ones involves
the application of the theory of networks in the realm of
metapopulation
models in a way similar to the work of~\cite{vespignani}.
Such approaches are already being brought to bear, as in the work
of~\cite{epigraph} or \cite{arenas} and are certainly also worth expanding upon
and refining, as well as comparing with the data available in the
context of the SARS-CoV-2 virus. Building such networks
for the examples of Greece and Andalusia considered
herein (and of course beyond) also constitutes a worthwhile
direction of future research. Clearly, further efforts at the level
of data collection and curation, at the level of model setup and
validation, and then at the level of optimization and utilization
for prediction are needed. Our hope, however, is that the approach
proposed herein is an initial step towards putting
together a number of relevant
tools to enable going beyond the 0D approach of ODE models
and gradually  considering in more detail the expansion of a
pandemic at a combined spatial and temporal level.

\appendix

\section{Next-generation calculation of the basic reproduction number $R_0$}
\label{app:R0}

We will use the next generation matrix approach
of the system
of equations Eqs.~(\ref{eqn1})-(\ref{eqn8}) without the
spatial term to find $R_0$.
In particular, we set up the vectors:
\begin{align*}
\mathcal{F} =\left(
\begin{array}{c}
\beta_{SA} S A + \beta_{SI} S I \\
0 \\ 0\\ 0\\ 0\\ 0\\ 0 \\ 0
\end{array}
\right),~~\mathcal{V} =
\left(
\begin{array}{c}
(\sigma_A + \sigma_I) E  \\
-\sigma_A E + M_{AR} A \\
-\sigma_I E + M I\\
-\gamma M I +(1-\omega) \chi H + \omega \psi H \\
\beta_{SA} S A +\beta_{SI} S I\\
M_{AR} A\\
-(1-\gamma) M I - (1-\omega) \chi H\\
-\omega \psi H
\end{array}
\right).
\end{align*}
The idea is that we rearrange the compartments so that the
infectious/infected compartments $E, A, I, H$,  appear first. We then place
$S, AR, R, D$. If we calculate $\mathcal{F} - \mathcal{V}$, it should yield a
reordered version of the vector field that
describes our disease system.

We then focus on the $4$ infectious/infected compartments and ignore the rest.
We find  the Jacobians of $\mathcal{F}, \mathcal{V}$ with respect to $E, A, I, H$
in  the order in which they appear. This  will yield two $4 \times 4$ matrices:
\begin{align*}
&F= \left(
\begin{array}{cccc}
0 & \beta_{SA} S^*& \beta_{SI} S^*& 0\\
0 & 0& 0 &0 \\
0 & 0& 0 &0 \\
0 & 0& 0 &0 \\
\end{array}
\right),~~V=
\left(
\begin{array}{cccc}
(\sigma_A+\sigma_I) & 0 &  0 & 0\\
-\sigma_A & M_{AR}&  0 &0 \\
-\sigma_I & 0&  M &0 \\
0 & 0& -\gamma M & \omega \psi +d +(1-\omega) \chi  \\
\end{array}
\right)
\end{align*}
The basic reproductive number is the spectral radius of $F V^{-1}$ which in our case is
\begin{align}
R_0 = \frac{\beta_{SA} S^* \sigma_A}{(\sigma_A +\sigma_I)M_{AR}} +
\frac{\beta_{SI} S^* \sigma_I}{(\sigma_A +\sigma_I)M}.
\label{eq:R0}
\end{align}
This result is in accordance with epidemiological intuition:
the first contribution to $R_0$ is proportional to $\beta_{SA}$ and $S^*$,
namely, the transmission rate and total susceptible population $S^*$.
It is also proportional to  $\sigma_A/(\sigma_A+\sigma_I)$, namely the
fraction of  exposed hosts becoming
infectious, yet asymptomatic, $A$.
Finally, it is inversely
proportional to the loss rate $M_{AR}$ of the infectious asymptomatic class $A$.
The second contribution to $R_0$ is analogous to the first one and stems from
the second  mode of transmission, i.e., through contact with $I$.

\section{Spatial modeling of Greece}
\label{app:Greece}

\subsection{ODE model: Well-mixed populations}

We present the ``0D'' model
predictions for Greece
in Fig.~\ref{fig:Greece0D}.
The data we consider~\cite{greecewiki} start on March
12, 2020 when losses of life started to occur and the cumulative number
of infected (total number of cases)
was already a bit over 100 individuals. We follow the evolution of the pandemic till May 11, 2020
using official data up to this point to
optimize model parameters and initial conditions.
As in the case of Andalusia,
we minimized the combined Euclidean distance of model results and observed data
for the total number of cases ($C(t) = I(t) + H(t) + R(t) + D(t)$) and
the total number of deceased ($D(t)$), see Eq.~(\ref{eq:norm1age}).
A few remarks on the quality of the data and our
choice of the most reliable time series (total
number of reported cases and fatalities) are presented
in Section~\ref{sec:SetUp}, following Eq.~(\ref{eq:norm1age}).
We performed 2,000 optimizations to obtain the parameters shown
in Table~\ref{tab:Gre_parameters}. We present the medians of all
model parameters and initial conditions, as well as the interquartile range and the range of variation of the initial values of parameters.

\begin{figure}[!ht]
\centering
\begin{tabular}{cc}
\includegraphics[width=.45\textwidth]{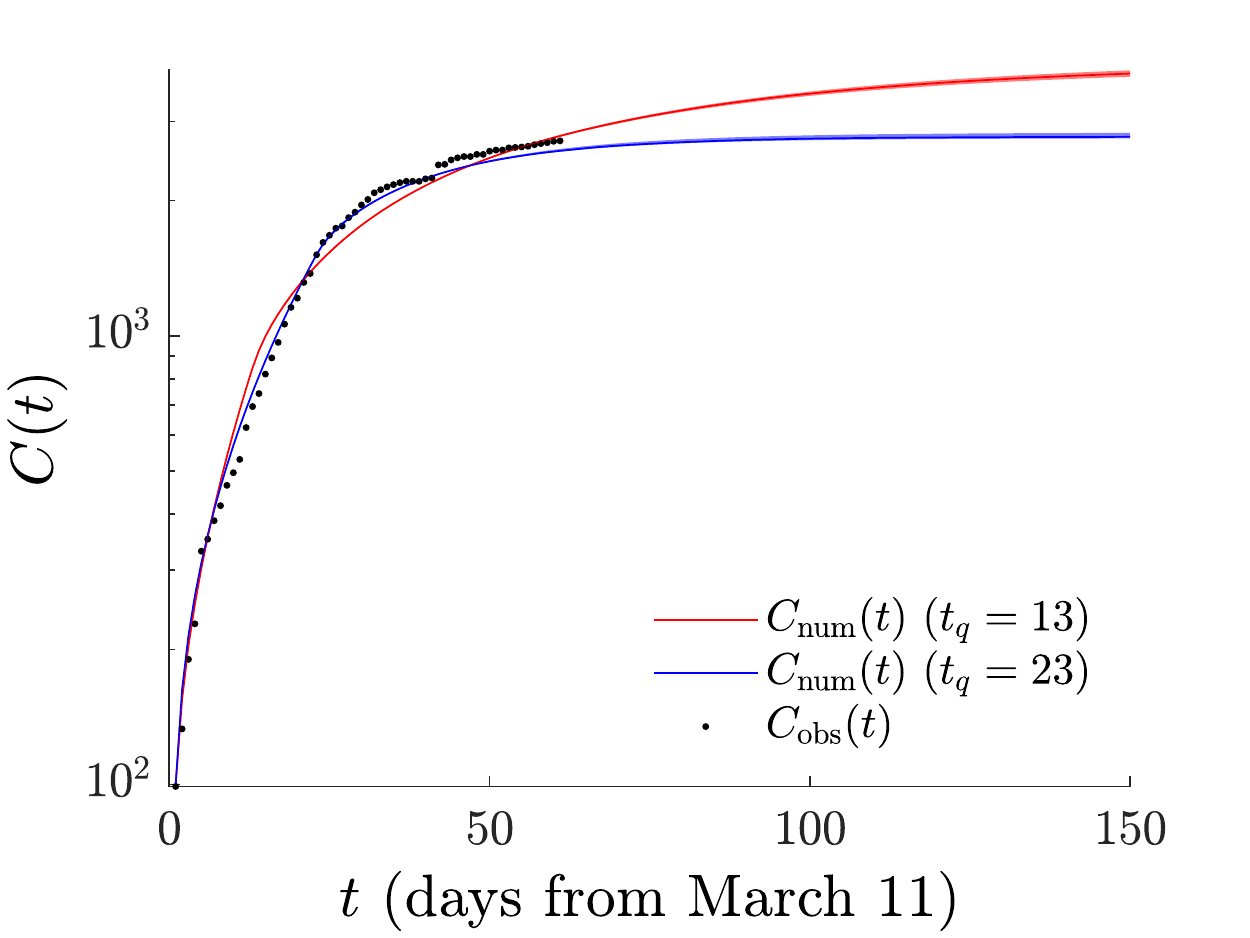} &
\includegraphics[width=.45\textwidth]{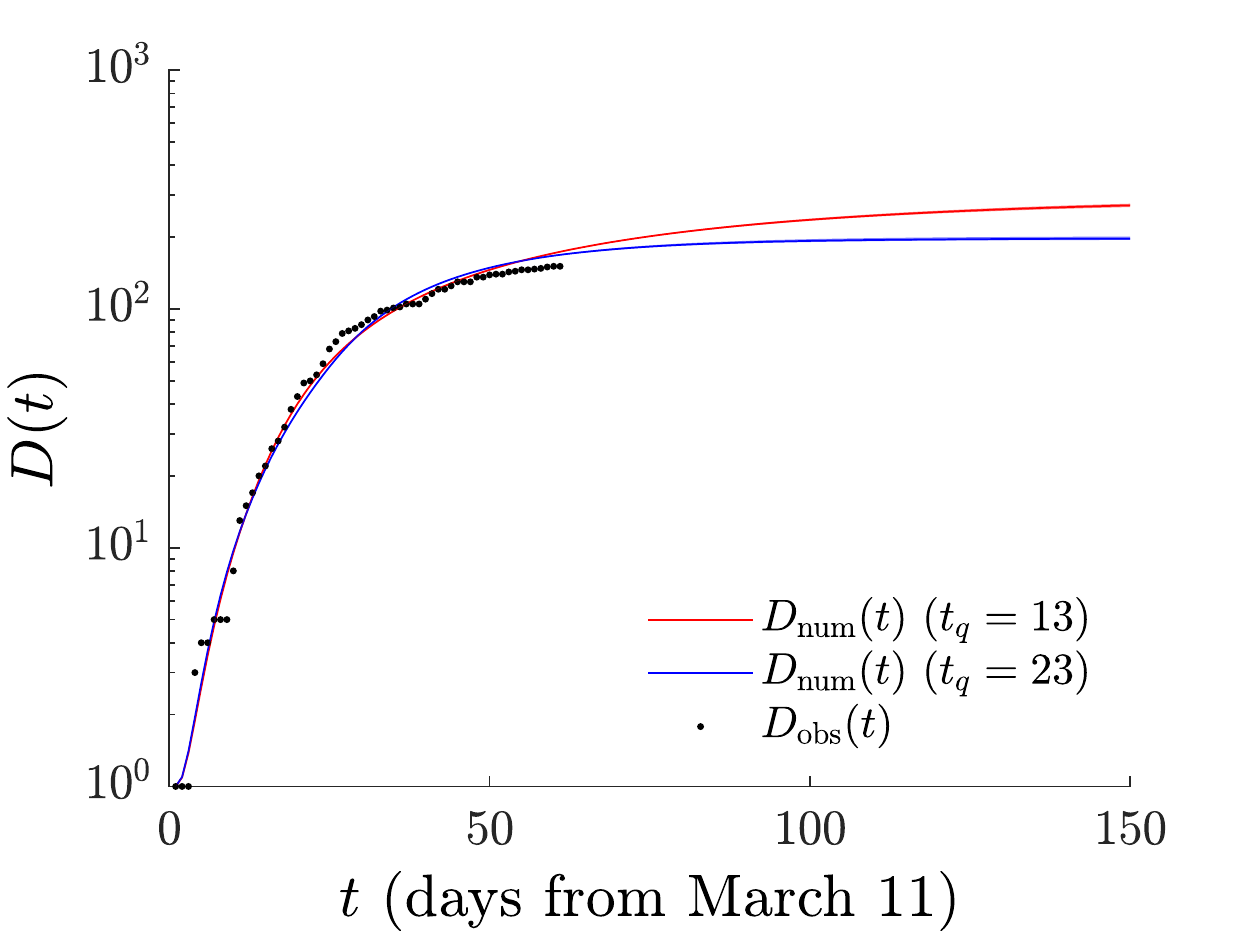} \\
\includegraphics[width=.45\textwidth]{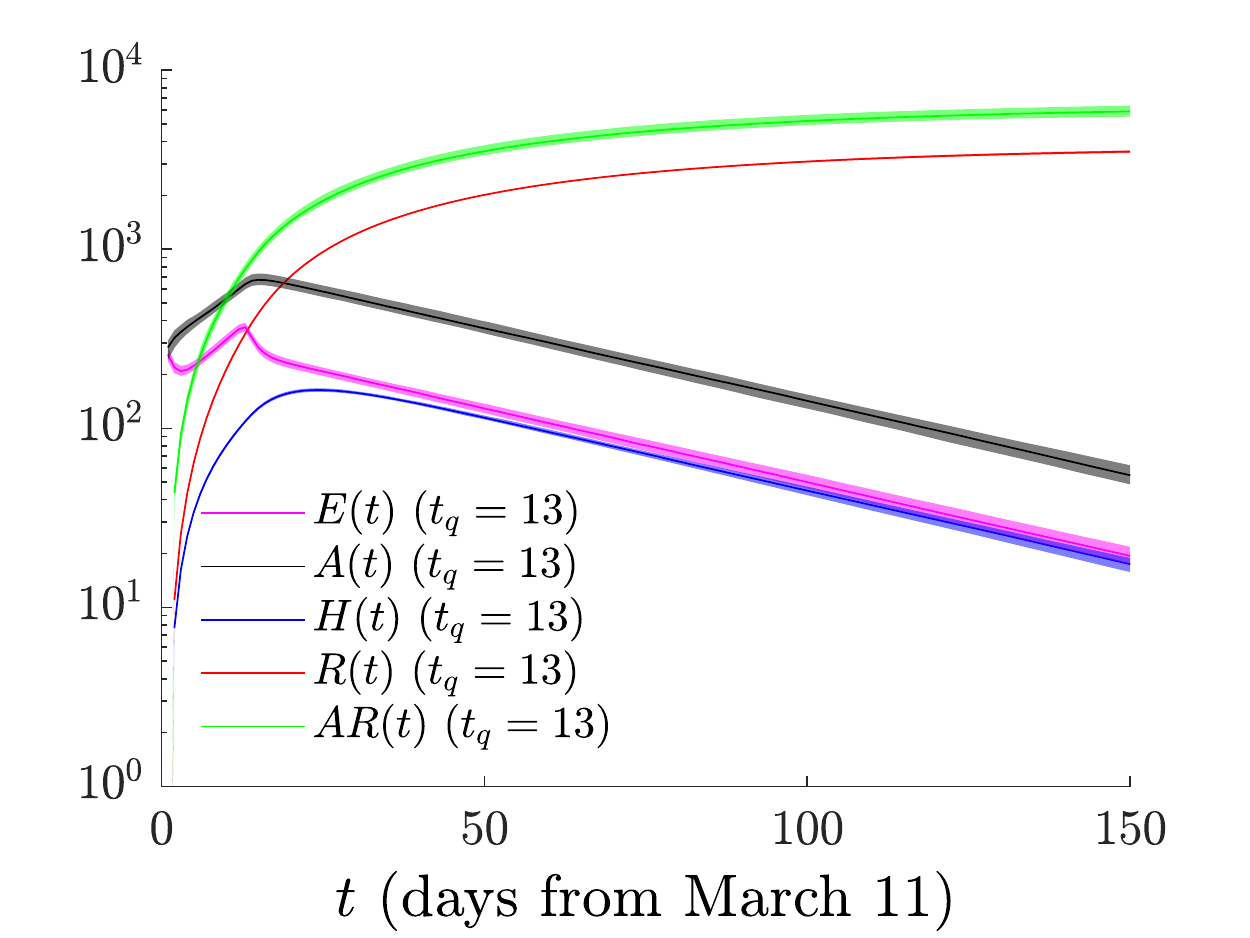} &
\includegraphics[width=.45\textwidth]{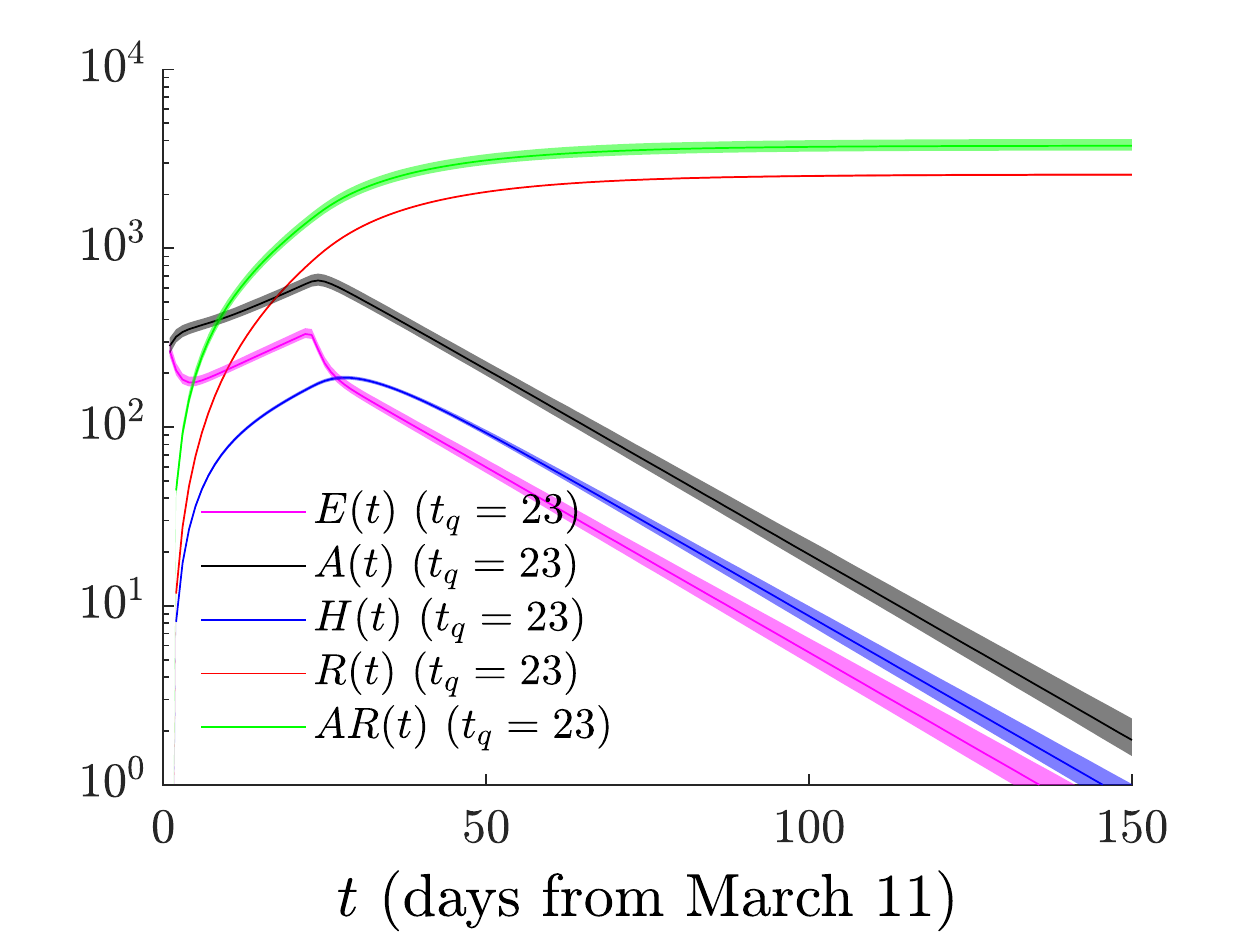} \\
\end{tabular}
\caption{{(Color online.)} 0D model for Greece with fitting to official data from March 12 ($t=t_{\textrm{init}}=1$) to May 11, 2020
($t_{\textrm{fit}}^{\textrm{end}} = 59$). Official confinement started
on March 22, 2020 ($t=11$).
The top panels show the official data (black dots) and simulations: red line
{(top solid line at t=150)} for $t_q=13$ (quarantine starting on March 24, 2020)
and blue line {(bottom solid line at t=150)} for $t_q=23$ (quarantine starting on April 3, 2020). Left top panel: Confirmed cases $C(t)=I(t)+R(t)+H(t)+D(t)$; Right top panel: Number of deaths $D(t)$.
The bottom panels show the other populations,
{(distinguishable at t=150 from top to bottom: asymptomatic
recovered $AR(t)$, recovered $R(t)$, hospitalized $H(t)$, asymptomatic $A(t)$, and exposed $E(t)$)}.
The bottom left panel shows these populations for $t_q=13$  and the
bottom right panel for $t_q=23$. In all panels, shaded regions correspond to the interquartile range for each quantity, whereas the full line corresponds to simulations with the median parameter (and initial-condition) values.}
\label{fig:Greece0D}
\end{figure}

\begin{table}[!ht]
\centering
\caption{ODE parameters for Greece:
optimal (best-fitting), median and interquartile
range, and variation range used in the optimization algorithm. Initial parameters and
initial-condition guesses were uniformly sampled within these ranges.}
\label{tab:Gre_parameters}
\begin{ruledtabular}
\begin{tabular}{ccccc} 
& & Median (interquartile range) & Median (interquartile range) & Initial value \\ 
& & $(t_q=13)$ & $(t_q = 23)$ & \\ 
\hline  
Population & $N$ & \multicolumn{2}{c}{10,768,477} & \\ 
Initial populations & ($I_0, H_0, D_0$) & \multicolumn{2}{c}{(117,0,1)} & \\ 
Non COVID-19 death rate [per day] & $\mu$ & \multicolumn{2}{c}{$8.49 \times 10^{-6}$} & \\ 
Transmission rate, $S \rightarrow I$ [per day] & $\beta_{IS}$\footnote{The transmission rates
$\beta$ have to be divided by $N$ when used in the ODE model.} & 0.31 (0.29--0.33) & 0.24 (0.23--0.25) & $c \in U[0,1]$ \\ 
Transmission rate, $S \rightarrow A$ [per day] & $\beta_{AS}$\footnotemark[1] & 0.21 (0.19--0.22) & 0.18 (0.17--0.19) & $c \in U[0,1]$ \\ 
Lockdown effect, $S \rightarrow I$ & $\eta_{IS}$ & 0.52 (0.49--0.54) & 0.48 (0.46--0.50) & $c \in U[0,1]$ \\ 
Lockdown effect, $S \rightarrow A$ & $\eta_{AS}$ & 0.52 (0.49--0.54) & 0.48 (0.46--0.50) & $c \in U[0,1]$ \\ 
Latent period, $E \rightarrow A$ [days] & $1/\sigma_A$ & 2.82 (2.76--2.89) & 2.89 (2.81--2.97) & $1/k, \, k  \in U[2,7]$ \\ 
Incubation period, $E \rightarrow I$ [days] & $1/\sigma_I$ & 4.38 (4.14--4.68) & 3.72 (3.55--4.00) & $1/k, \, k  \in U[2,7]$ \\ 
Infectivity period [days] & $1/M$ & 6.30 (6.24--6.37) & 6.13 (6.07--6.19) & $1/k, \, k  \in U[5,12]$ \\ 
Recovery period (asymptomatics), $A \rightarrow AR$ [days] & $1/M_{AR}$ & 6.95 (6.89--7.02) & 6.87 (6.80--6.96) & $1/k, \, k  \in U[5,12]$ \\ 
Recovery period (hospitalized), $H \rightarrow R$ [days] & $1/\chi$ & 6.36 (6.31--6.42) & 6.20 (6.16--6.26) & $1/k, \, k  \in U[5,20]$ \\ 
Period from hospitalized to deceased, $H \rightarrow D$ [days] & $1/\psi$ & 8.87 (8.75--9.00) & 8.76 (8.62--8.92) & $1/k, \, k  \in U[5,20]$ \\ 
Conversion fraction ($I \overset{\gamma}{\longrightarrow} H$, $I \overset{1-\gamma}{\longrightarrow} R$) & $\gamma$ & 0.44 (0.43--0.44) & 0.44 (0.43--0.45) & $c \in U[0.25,0.75]$ \\ 
Conversion fraction ($H \overset{\omega}{\longrightarrow} D$, $H \overset{1-\omega}{\longrightarrow} R$) & $\omega$ & 0.22 (0.21--0.22) & 0.21 (0.21--0.22) & $c \in U[0.1,0.5]$ \\ 
Initial population fraction, exposed & $E_0/I_0$ & 2.65 (2.39--3.01) & 2.71 (2.51--3.04) & $c \in U[1,5]$ \\ 
Initial population fraction, asymptomatic & $A_0/I_0$ & 2.84 (2.44--3.16) & 2.87 (2.59--3.18) & $c \in U[1,5]$ \\ 
Diffusivity, $S$ [km$^2$/day] & $\mathfrak{D}_S$ & 10 & 10 & \\
Diffusivity, $E$ or $A$ [km$^2$/day] & $\mathfrak{D}_E$ or $\mathfrak{D}_A$ & 100 & 100 & \\
\end{tabular} 

\end{ruledtabular}
\end{table}

The model-predicted evolution of the pandemic shown
in Fig.~\ref{fig:Greece0D} was calculated for the median parameters (solid blue line)
and for the cloud of the parameter variations represented as the shaded region in the figure.
Officially reported data are denoted by black dots. As in the case of Andalusia,
optimal parameters
are shown for two scenarios,
see Table~\ref{tab:TimeSequence} for the
events time sequence. The first scenario considers that quarantine
was strictly enforced at March 24, 2020 ($t_q=13$).
In reality, the lockdown in Greece
started at the end of March 22, 2020; it is reasonable to assume
that it was strictly enforced 1-2 days later.
In the second scenario the lockdown was considered to
have been imposed on April 3, 2020 ($t_q = 23$).
To account for the
change in parameters due to the lockdown, we imposed the  time dependence
of the transmission rates $\beta$ as shown in Eq.~(\ref{eq:TimeDependentBeta}) in the main text.
This time dependence forces the transmission rates $\beta_{IS}$ and $\beta_{AS}$ to decrease
by a factor $\eta_{IS}$ and $\eta_{AS}$, respectively,
relatively abruptly at the time the lockdown was imposed, $t_q$.

\textcolor{black}{The top  panels of Fig.~\ref{fig:Greece0D}  compare
model  predictions for the two scenarios for the total number of cases (left)
and fatalities (right).} Scenario-one
median parameters, i.e., imposing the
quarantine practically at the time when it was officially announced,
capture the data for fatalities in Greece fairly well.
However, the number of reported cases, top left panel, is
not that accurately reproduced.
We attribute this discrepancy to the previously mentioned
characteristic feature of the data (top left panel)
that model predictions fail to capture adequately: the  ``angle" in the semi-logarithmic
plot associated with the curbing of the cumulative number of infections $C(t)$
due to containment measures.
As the optimization algorithm
attempts to minimize the distance from the observed data,
initially it slightly over-predicts and then under-predicts the data
and eventually the long-term predictions seem to
over-predict the flattening of the cases curve.
Nevertheless the
overall differences are relatively small: the model
prediction flattening out (over 5 months) around 200 deceased and
slight over 3K infected individuals seem reasonable, were the
lockdown measures potentially extendable to such a long time
interval.
If, as in the case of Andalusia, were we to shift the time of the application of
the quarantine date by about 10 days later (scenario two), then we note in the
top panel of the figure a nontrivial difference. Most notably, without
significantly
missing on $D(t)$
we capture accurately the angle in the $C(t)$ data. The
relevant parameters (medians, interquartile range) are presented in the second column of
Table~\ref{tab:Gre_parameters}. As before, we justify our decision to consider
a second scenario in that lockdown measures have an almost
\textit{immediate} effect in model predictions, whereas in reality
there is a time lag before restrictive measures have a measurable effect.
Lastly, we note that the effect of the shift of $t_q$ is far less severe than
in the case of Andalusia (Fig.~\ref{fig:Andalusia0D}).}

\textcolor{black}{It is worthwhile
to compare the scenario-two median parameters (Table~\ref{tab:Gre_parameters},
second column) to those we found
for Andalusia.
There are no particularly noteworthy
differences, although some do exist.
For example, as in the case of Andalusia, the median incubation period is approximately 4 days (3.72),
the latent period approximately 3 days (2.89), the asymptomatic
infectious period about 7 days (6.87) and that of the infected
6 days (6.13). A slight difference is noted in
the fraction of infected that need to be hospitalized ($\gamma \approx 0.44$ instead of 0.55),
and the fraction of hospitalized that become fatalities ($\omega \approx 0.21$
instead of 0.25). As in the case of
Andalusia, we find $\beta_{IS} > \beta_{AS}$,
but note the proviso related to parameter identifiability reported later on, and
that the fraction of exposed who turn asymptomatic is approximately 0.56
while the ratio of asymptomatics to symptomatically infected is 1.29.
After lockdown measures are imposed, the two transmission rates decrease
by the same amount ($\eta_{IS} = \eta_{AS}$), again as we found for Andalusia.}

\begin{figure}[!htb]
\centering
\begin{tabular}{cc}
\multicolumn{2}{c}{\includegraphics[width=.45\textwidth]{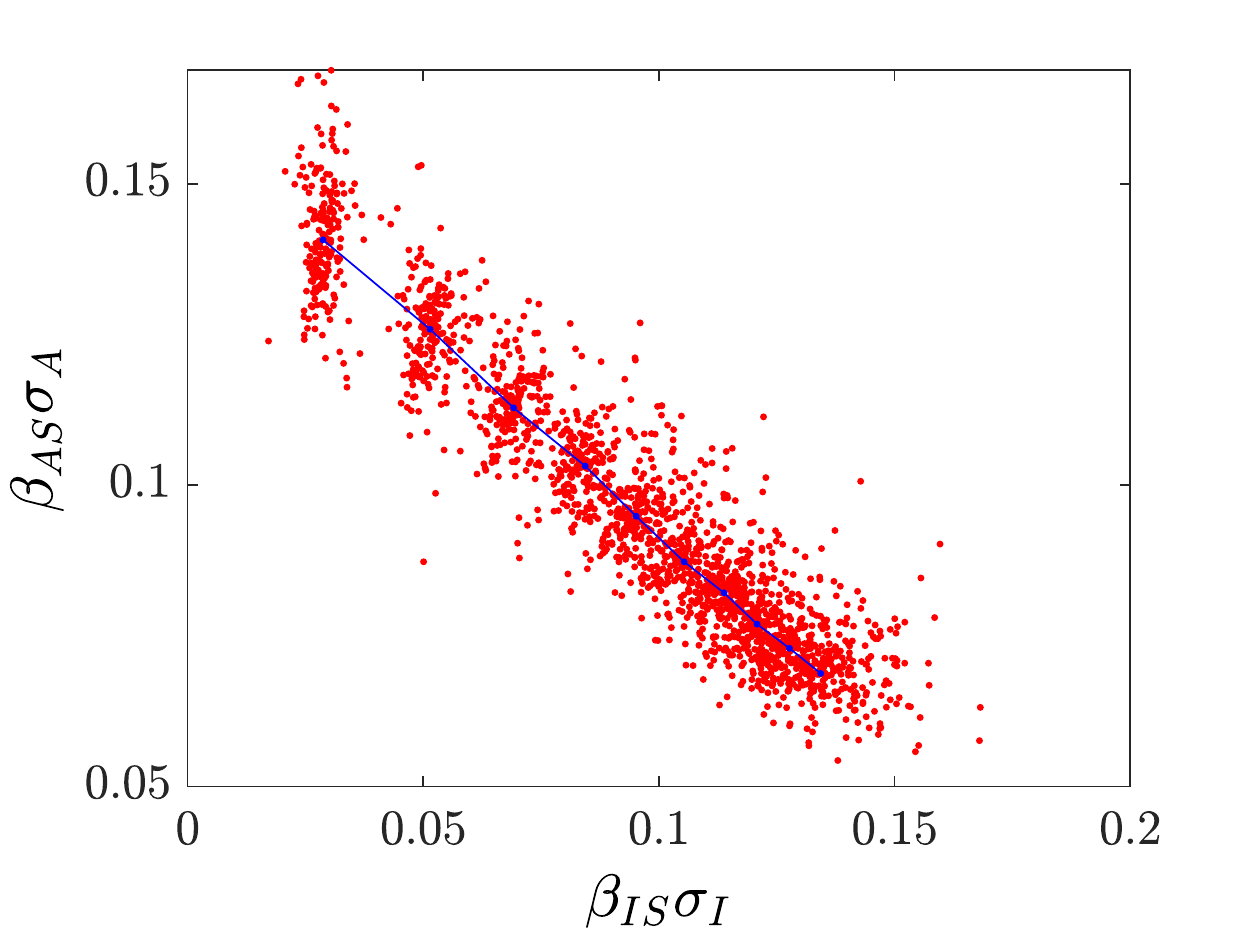}} \\ 
\includegraphics[width=.45\textwidth]{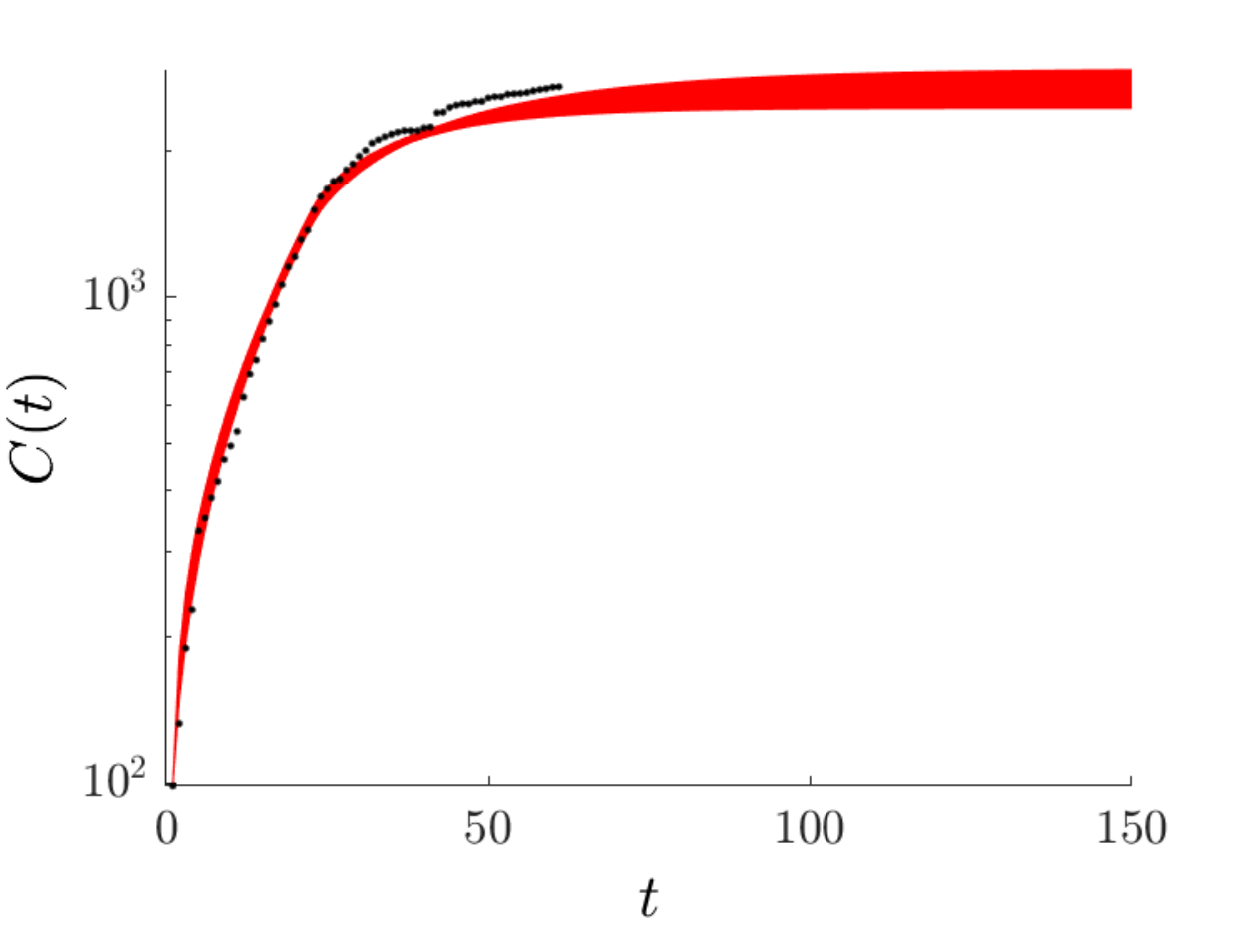} &
\includegraphics[width=.45\textwidth]{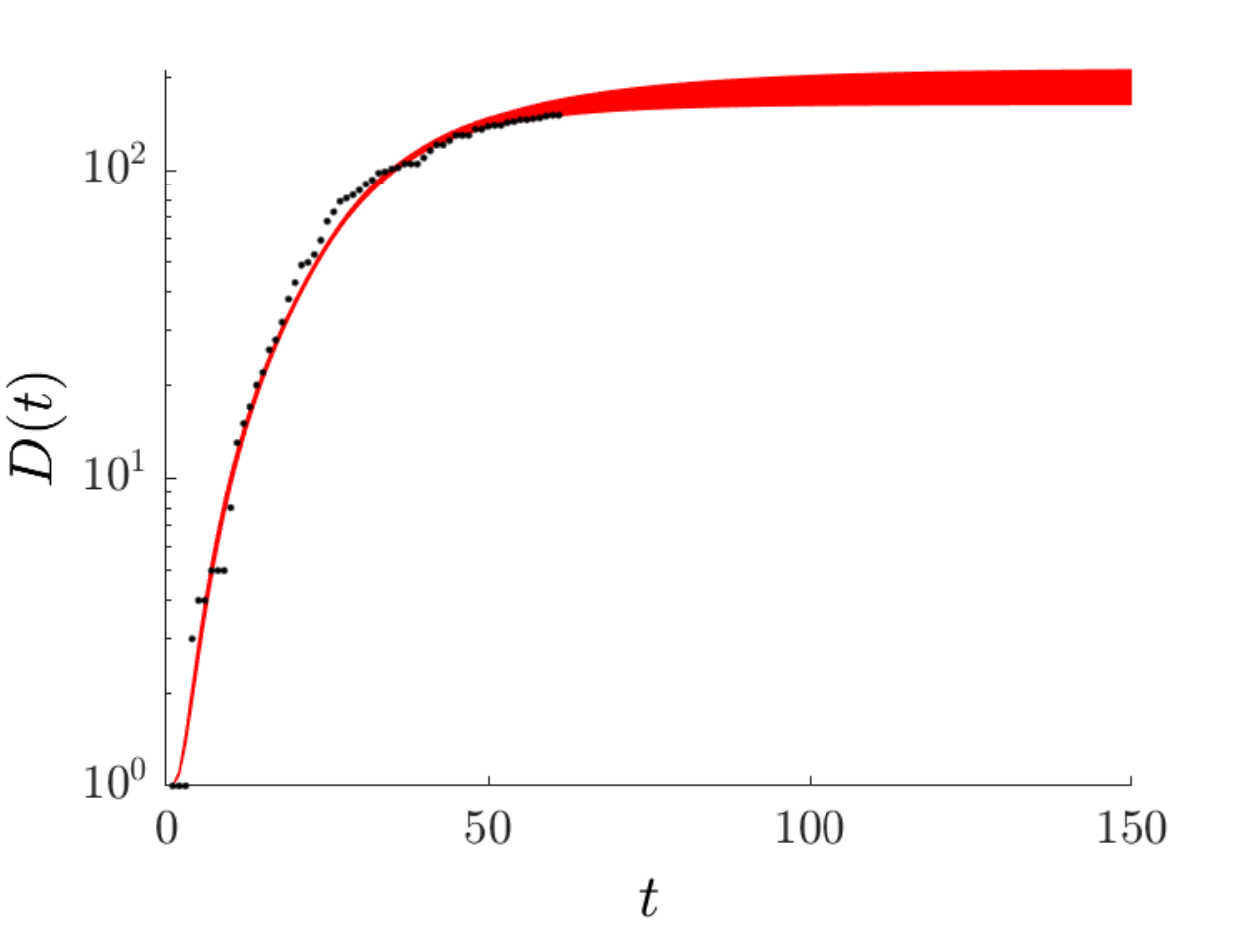}
\end{tabular}
\caption{\textcolor{black}{{(Color online.)} Sensitivity analysis for Greece (scenario two, $t_q=23$). The top panel shows the inverse
relationship between $\beta_{AS} \sigma_A$  and $\beta_{IS} \sigma_I$ for a series
of optimizations (200) performed by uniformly sampling model parameters  and
ten fixed equidistant ratios of $\beta_{AS}/\beta_{IS}$ within the range [0.2, 2].
The blue dots are the results of calculations with the median of the optimized
values (the line is a guide to the eye), while the red dots denote
the results of all the optimizations.
The two bottom panels compare the predicted future number of cases (left panel, red bundle)
and fatalities (right, red bundle) for all the optimizations shown in the top panel; black
dots are the reported numbers (and those used for model fitting). While the
transmission rates (times the associated time scales) vary significantly,
the forward, model-predicted evolution of the pandemic does not. This is
an indication that even though the model parameters are
not identifiable, they provide an adequate predictor of the evolution
of the pandemic if chosen within a
suitable range.}}
\label{fig:Greece_sens}
\end{figure}

The calculated  basic reproduction number $R_0$
(see Appendix~\ref{app:R0})
for the scenario-one pre-quarantine period
($t_q = 13$), i.e. calculated with data from the first column
of Table~\ref{tab:Gre_parameters} with the $\eta$ set
to unity, is \textcolor{black}{$R_0 = 1.64$ $(1.60 - 1.68)$, median and
interquartile range}. The effective reproduction number,
i.e., the reproduction number at the beginning of the quarantine with
the associated change of the transmission rates ($\eta$'s as
reported in column one) was calculated to be
\textcolor{black}{$R_{\textrm{eff}} = 0.849 (0.842 - 0.854)$, reflecting the
curbing of the epidemic curves, as shown in Fig.~\ref{fig:Greece0D}.
Similarly, the calculated pre-quarantine basic reproduction number
for scenario two ($t_q =23$) is $R_0=1.32$ $(1.31 - 1.34)$.
This post-quarantine
effective reproduction number decreases to $R_{\textrm{eff}} = 0.64$ $(0.63 - 0.65)$,
an indication that intervention measures lead to a curbing of the epidemic.}

\textcolor{black}{We conclude this section by a brief discussion of parameter identifiability and
the zero eigenvalues of the Hessian of the variation of the Euclidean norm with respect to
model parameters, in the spirit of the corresponding discussion for the case of Andalusia.
Figure~\ref{fig:Greece_sens} presents our calculations relevant to
parameter identifiability and sensitivity to parameter variations (for
the more reliable simulations of the second scenario). As for Andalusia,  we performed
200 optimization with parameters uniformly sampled within their variation
range with the ratio $\beta_{AS}/\beta_{IS}$ fixed at one of ten equidistant
values chosen within [0.2, 2]. The top panel in the figure shows the
inverse (apparently nearly linear)
relationship between properties of asymptomatics ($\beta_{AS} \sigma_A$)
and the corresponding properties of infected, a relationship that we argued may be
interpreted as the manifestation of the singular eigendirection of the Hessian. As the fraction
of asymptomatics to infected increases the asymptomatic transmission rate decreases.
Lastly, we note that the cloud of points for the total number of cases slightly overpredicts the
data initially en route to eventually (slightly) underpredicting their long-term evolution. The cloud of points follows rather
closely the data for total fatalities, with the data eventually lying
near the bottom edge of the prediction interval.}

\subsection{PDE model: Spatially distributed populations}

Results relevant to the scenario-two PDE simulations for the mainland of Greece
may be found in Fig.~\ref{fig:GreecePDE} for the same
parameters as for the 0D model, and in
Figs.~\ref{fig:GreecePDE_I}--\ref{fig:GreecePDE_C}
for the spatio-temporal simulations of the pandemic.

\begin{figure}[!ht]
\centering
\begin{tabular}{cc}
\includegraphics[width=.45\textwidth]{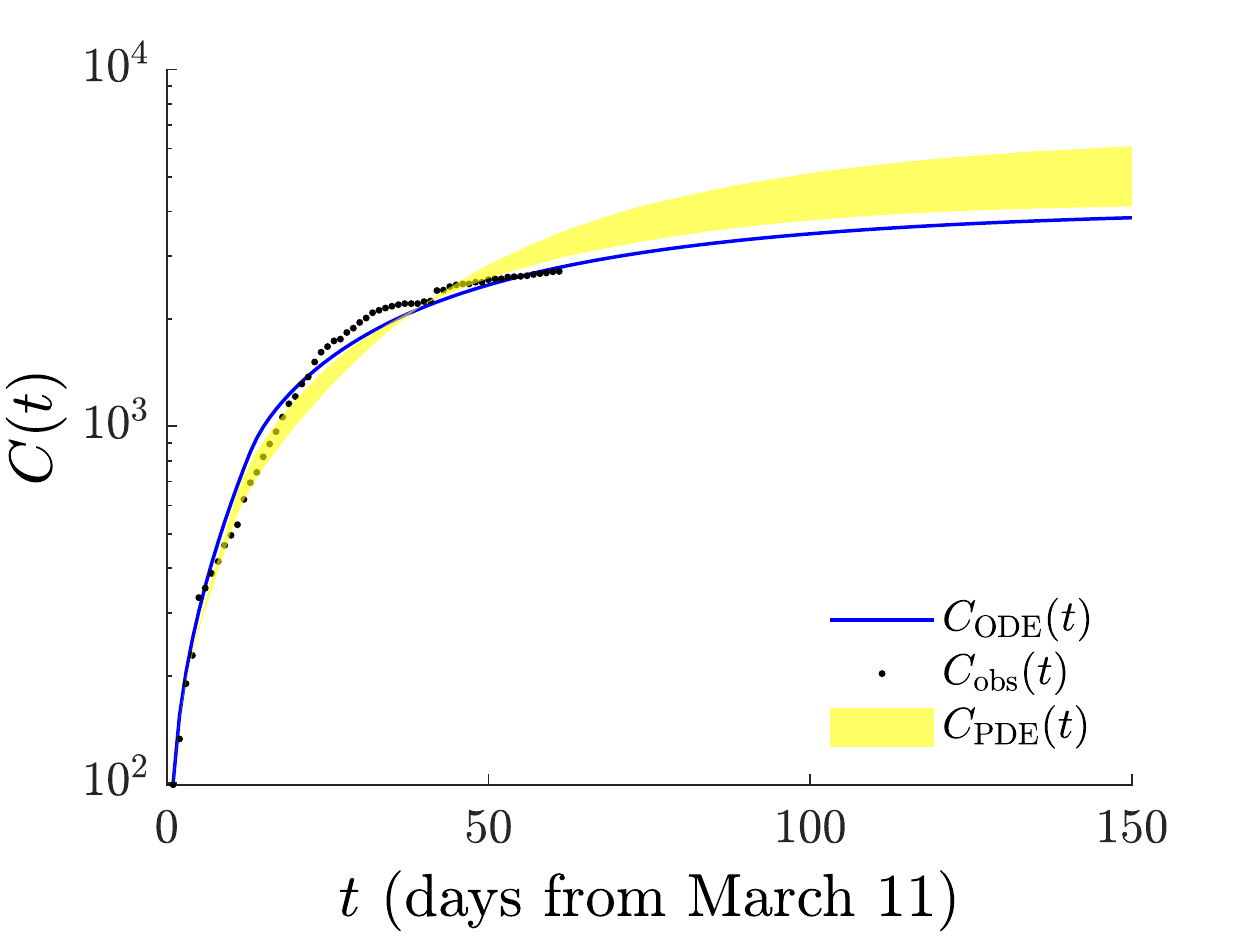} 
\includegraphics[width=.45\textwidth]{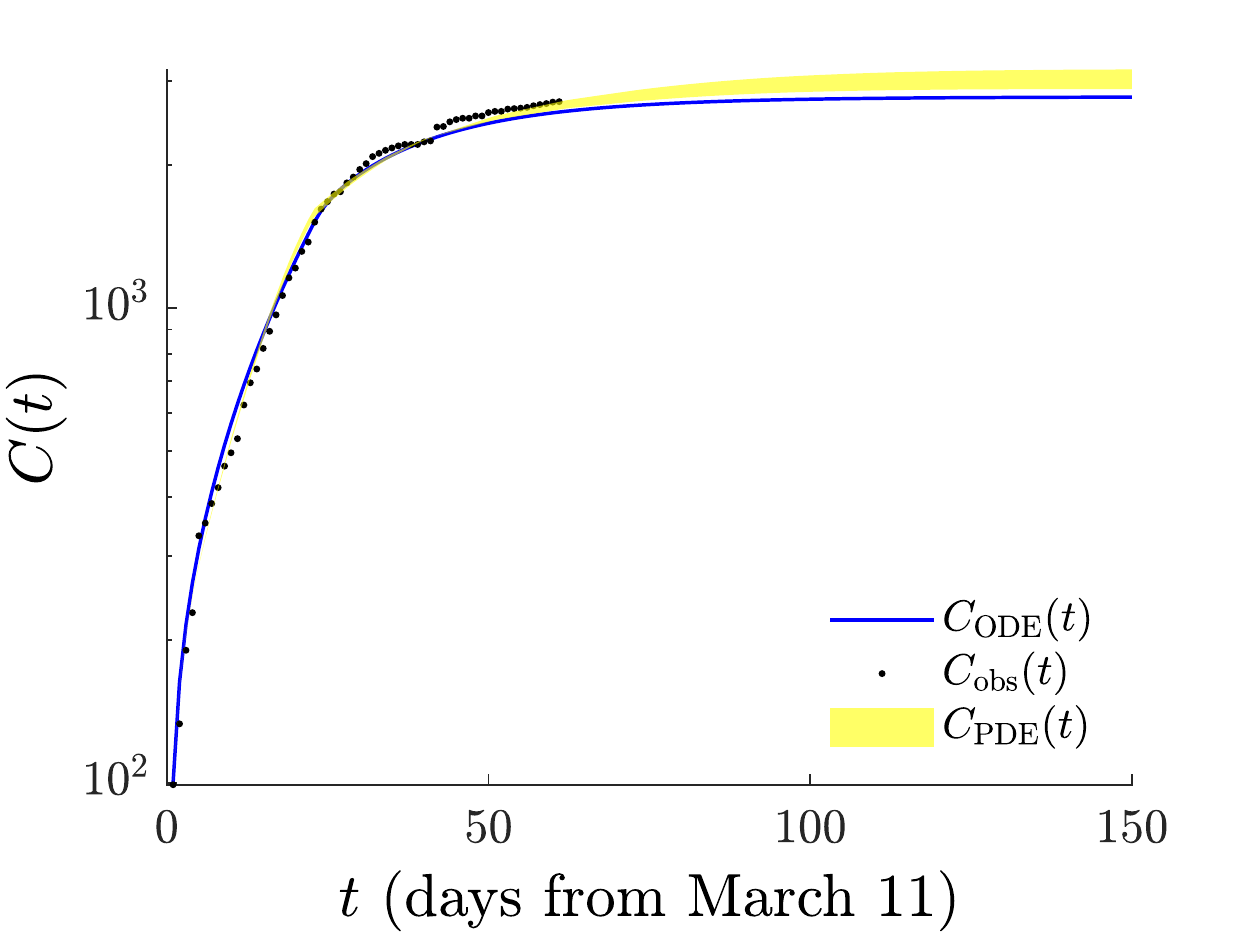}
\\
\includegraphics[width=.45\textwidth]{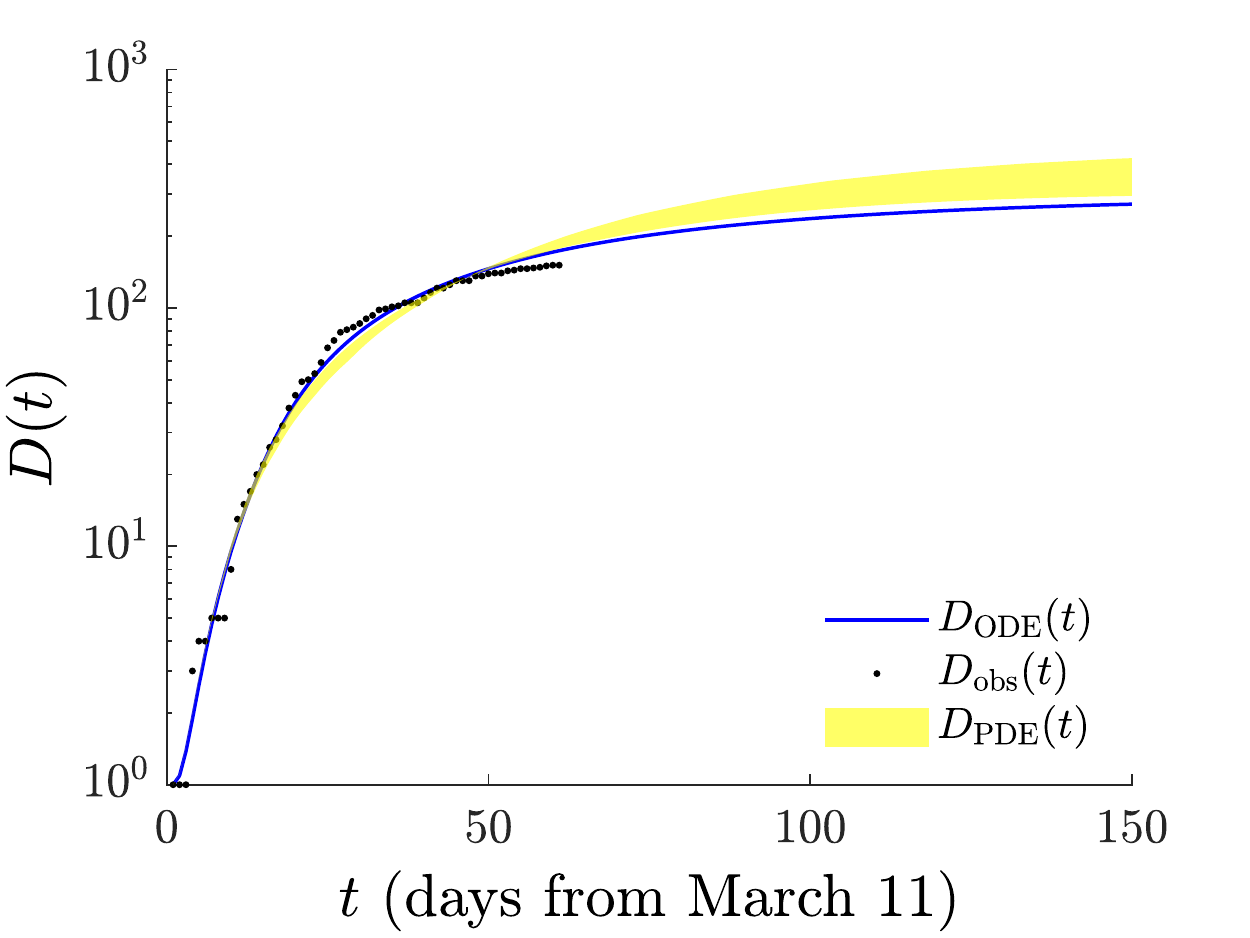} 
\includegraphics[width=.45\textwidth]{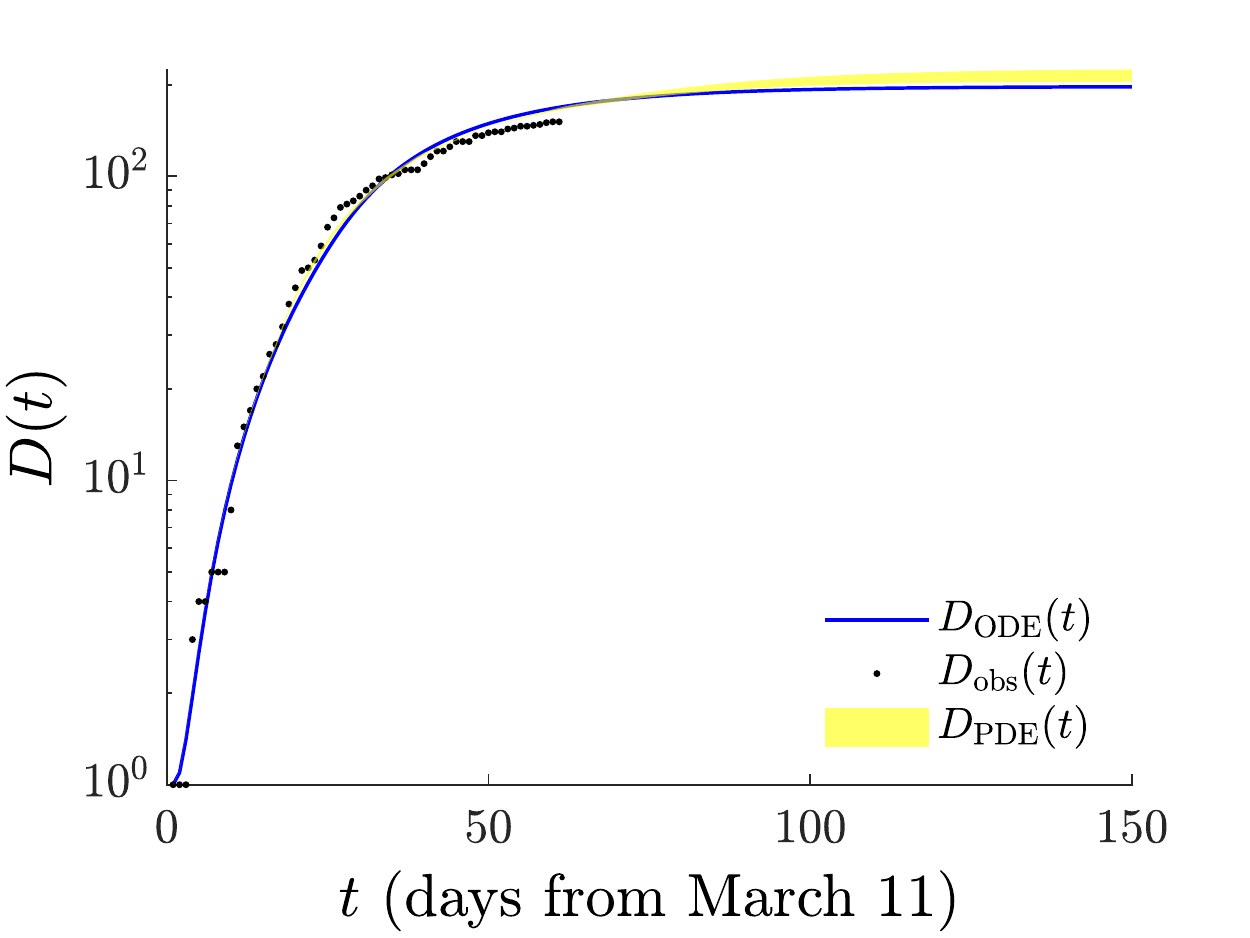}
\\
\includegraphics[width=.45\textwidth]{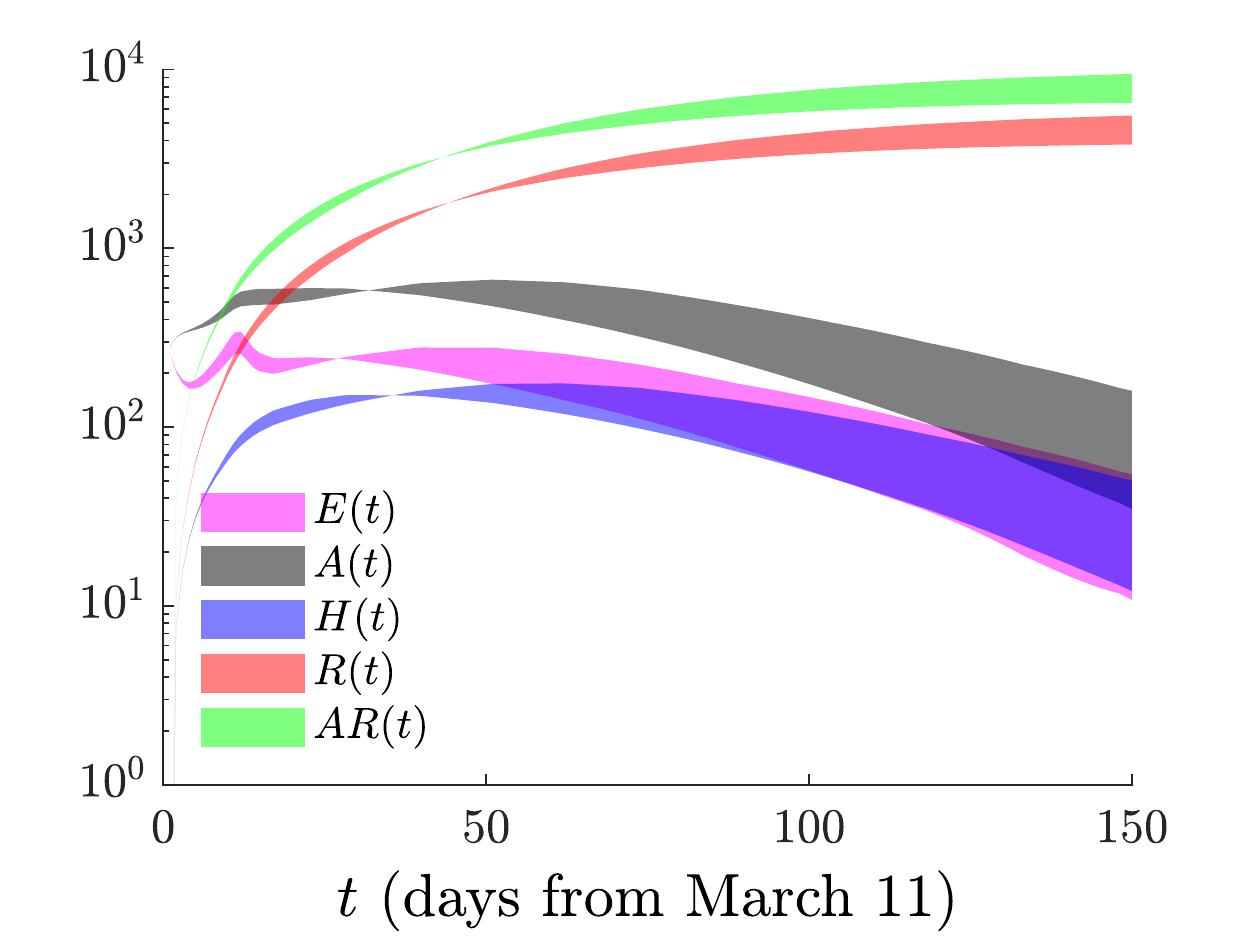} 
\includegraphics[width=.45\textwidth]{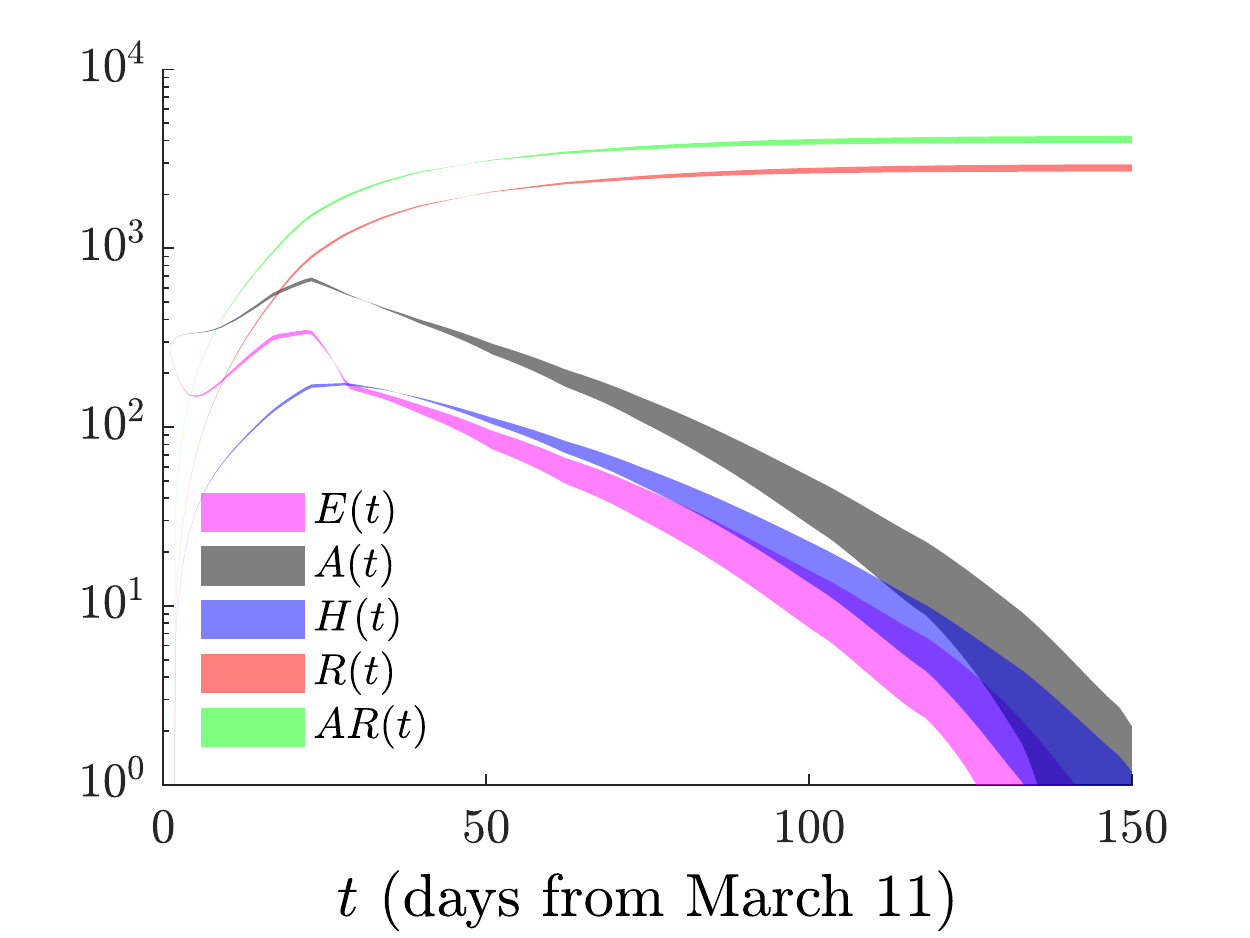}
\end{tabular}
\caption{{(Color online.)} PDE model for Greece with fitting to official data
from March 12, 2020 ($t=t_{\textrm{init}}=1$) to May 11, 2020 ($t_{\textrm{fit}}^{\textrm{end}} = 61$).
Confinement time starts at
March 22, 2020 ($t=11$). The plots displayed on the left panels correspond to $t_q=13$
(March 22, 2020)
and those on the right panels hold for $t_q=23$ (April 3, 2020).
The solid blue line (top two rows) reproduces the 0D simulation, cf. Fig.~\ref{fig:Greece0D}.
The median parameters
shown in Table~\ref{tab:Gre_parameters} are used, except for the
transmission rates that were scaled by $\xi \epsilon [0.00216, 0.00234]$
for $t_q =13$ and
$\xi \epsilon [0.00245, 0.00248]$ for $t_q =23$. The optimal scaling factor $\xi$
increases as the diffusion-coefficient reduction factor
$\eta_D$ increases. The latter was varied in the interval $[0.25, 0.50]$ in steps of $0.05$.
Shaded regions are delimited by the optimal plots for $\eta_D=0.5$ and $\eta_D=0.25$
{; their order at $t=150$ is the same as that of the bottom panels of Fig.~\ref{fig:Greece0D}.}}
\label{fig:GreecePDE}
\end{figure}

\begin{figure}[!ht]
\centering
\begin{tabular}{cc}
\includegraphics[width=.45\textwidth]{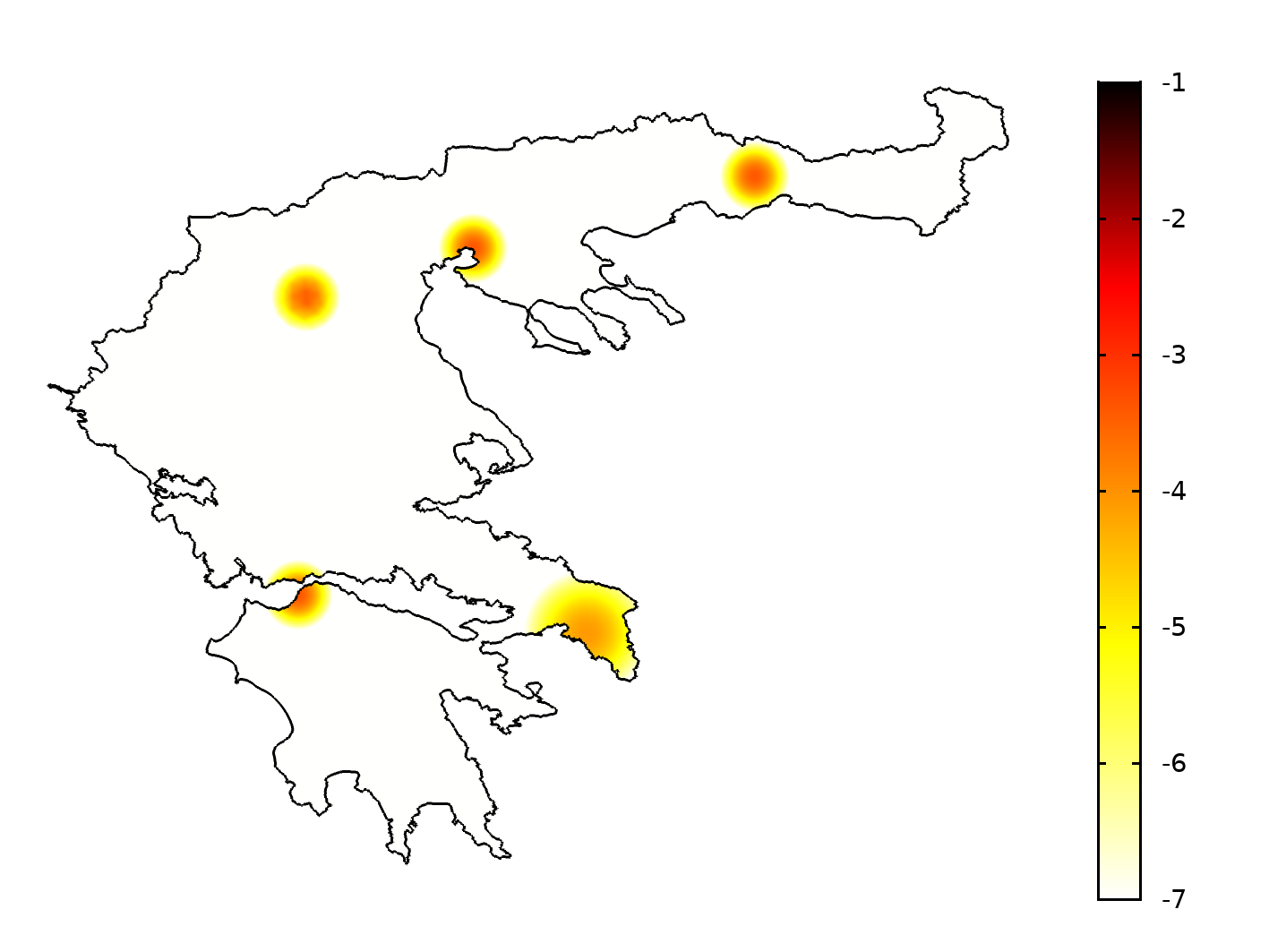} &
\includegraphics[width=.45\textwidth]{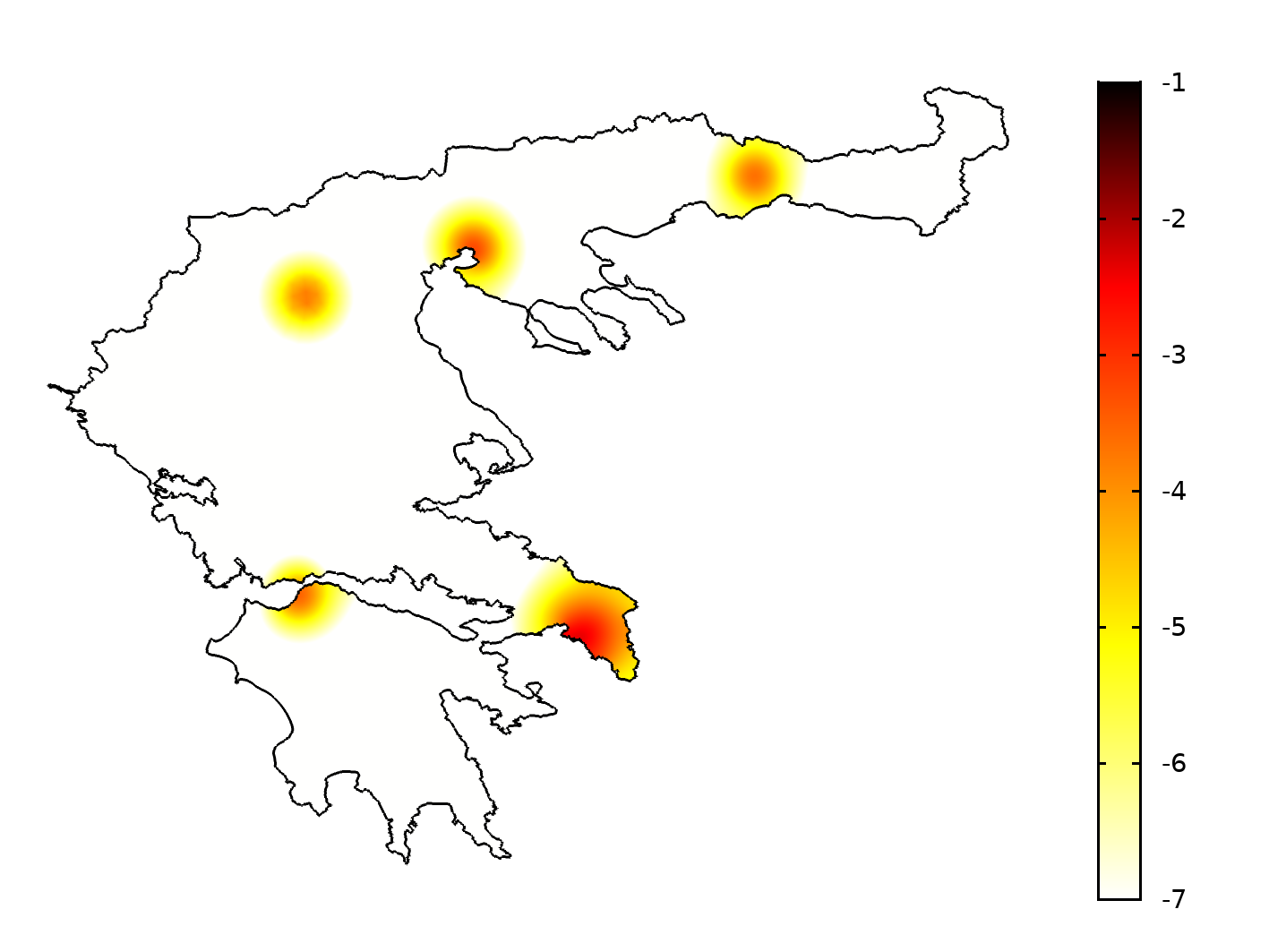} \\
\includegraphics[width=.45\textwidth]{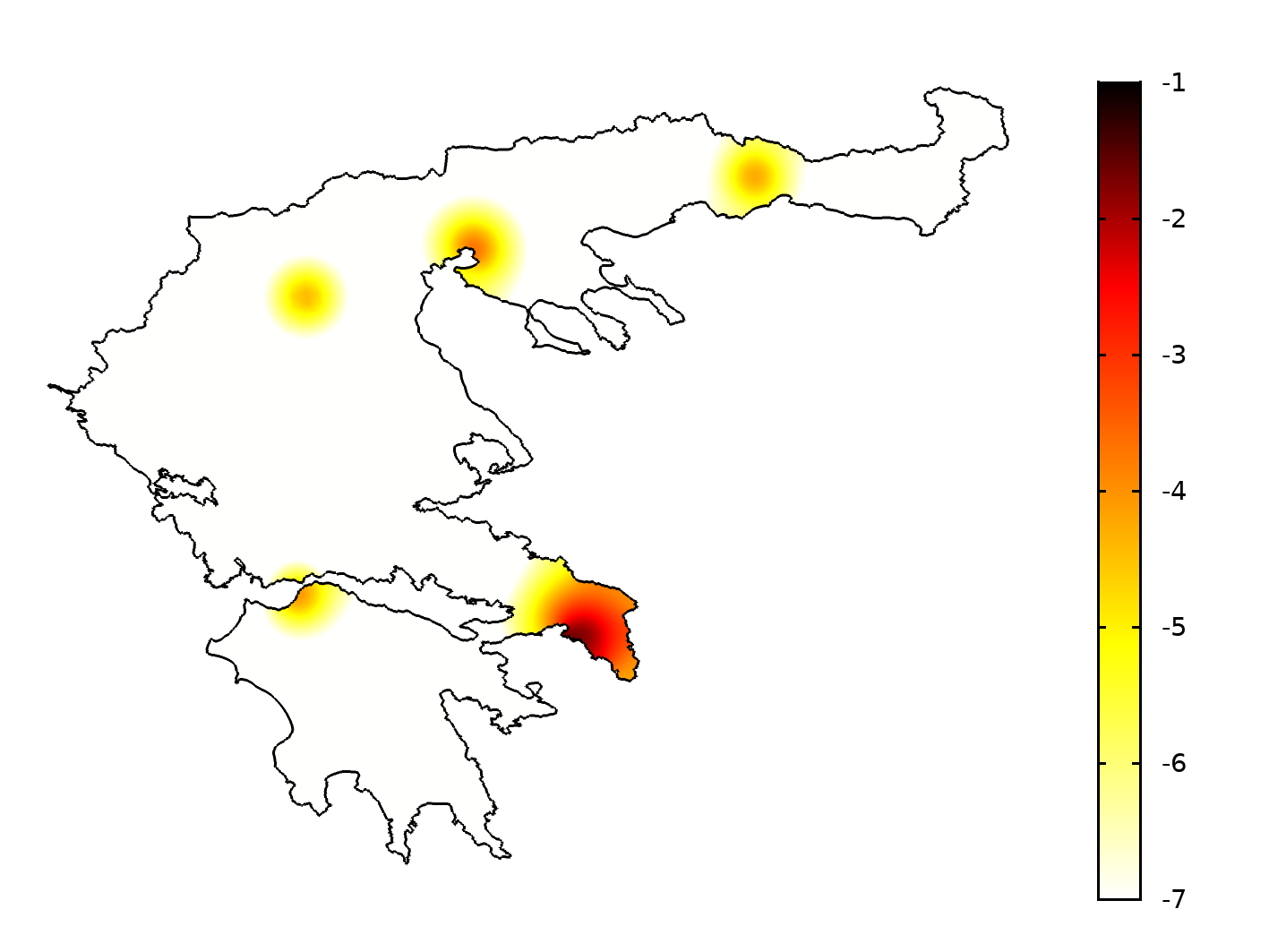} &
\includegraphics[width=.45\textwidth]{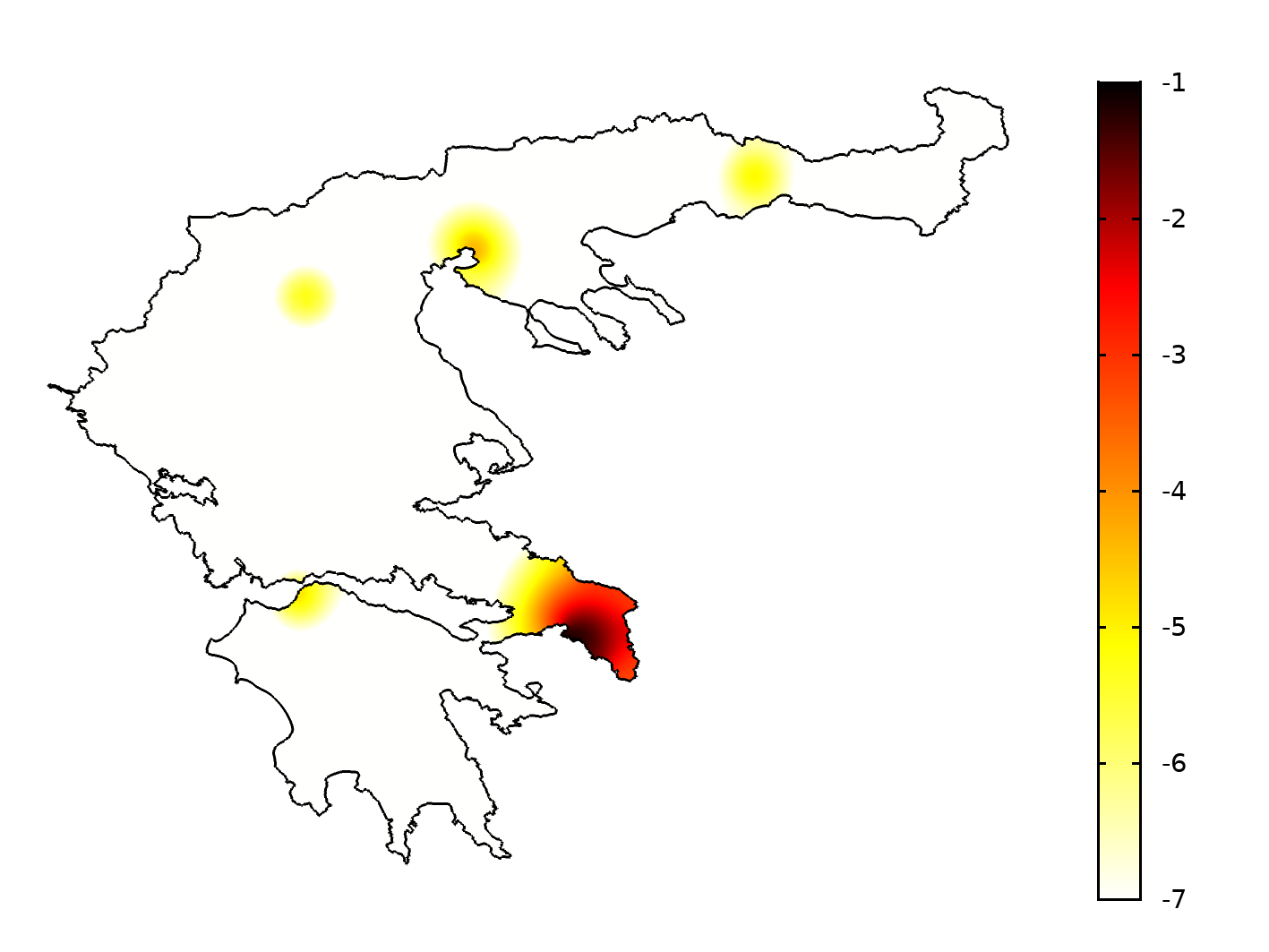} \\
\end{tabular}
\caption{{(Color online.)} Evolution of the mainland Greek infected
population density $\log_{10}I(x,y,t)$ for $t=1$ day (top left, March 12, 2020),
$t=6$ days (top right, March 17, 2020), $t=11$ days (bottom left, March 22, 2020), and
{$t=18$ days (bottom right,  March 29, 2020)}.
\textcolor{black}{$t_q=23$ (April 3, 2020), $\eta_D=0.35$ and scaling factor
$\xi = 0.00246$.}
{A logarithmic (base 10) colorbar scale is used.}}
\label{fig:GreecePDE_I}
\end{figure}

\begin{figure}[!ht]
\centering
\begin{tabular}{cc}
\includegraphics[width=.45\textwidth]{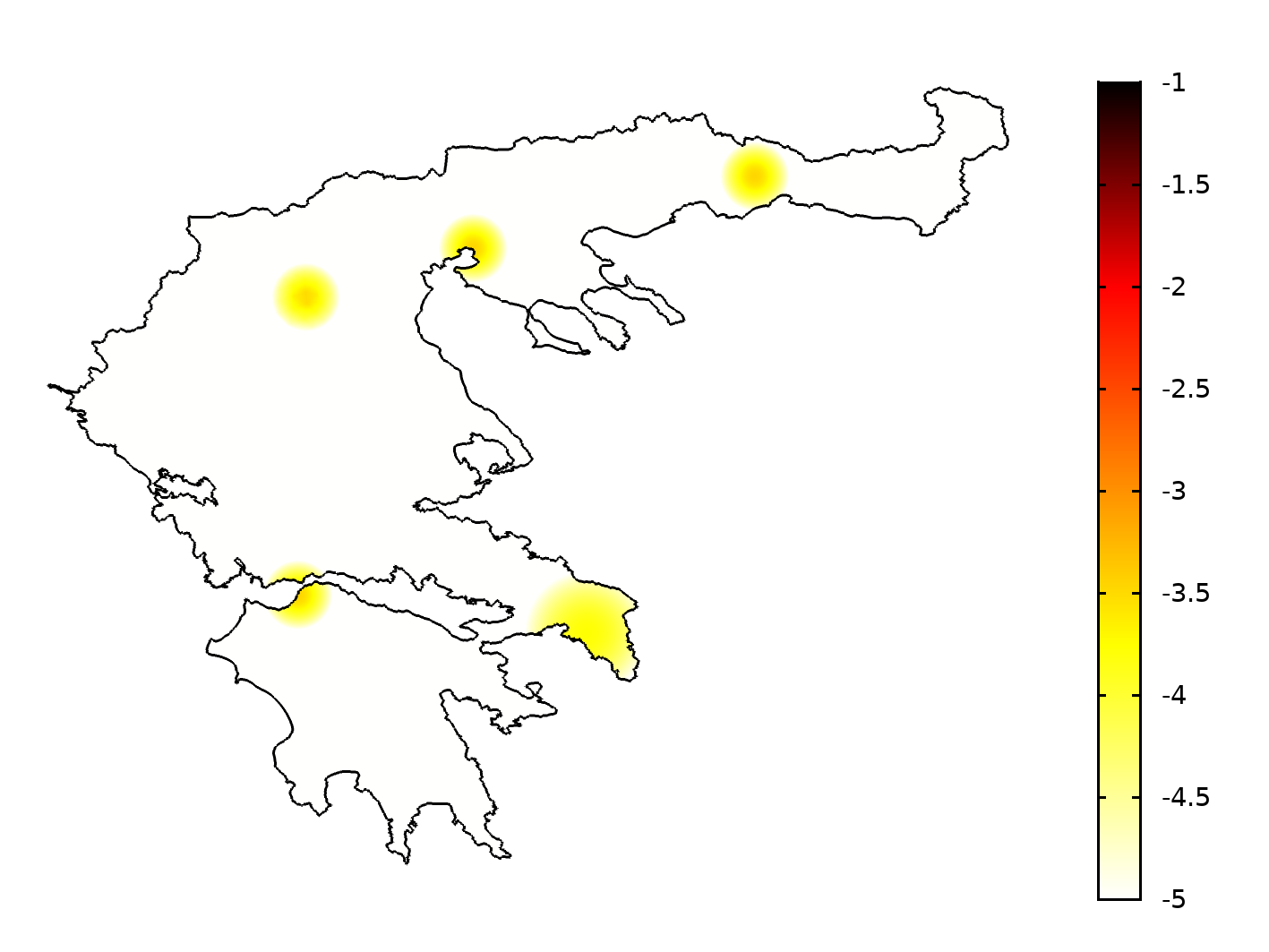} &
\includegraphics[width=.45\textwidth]{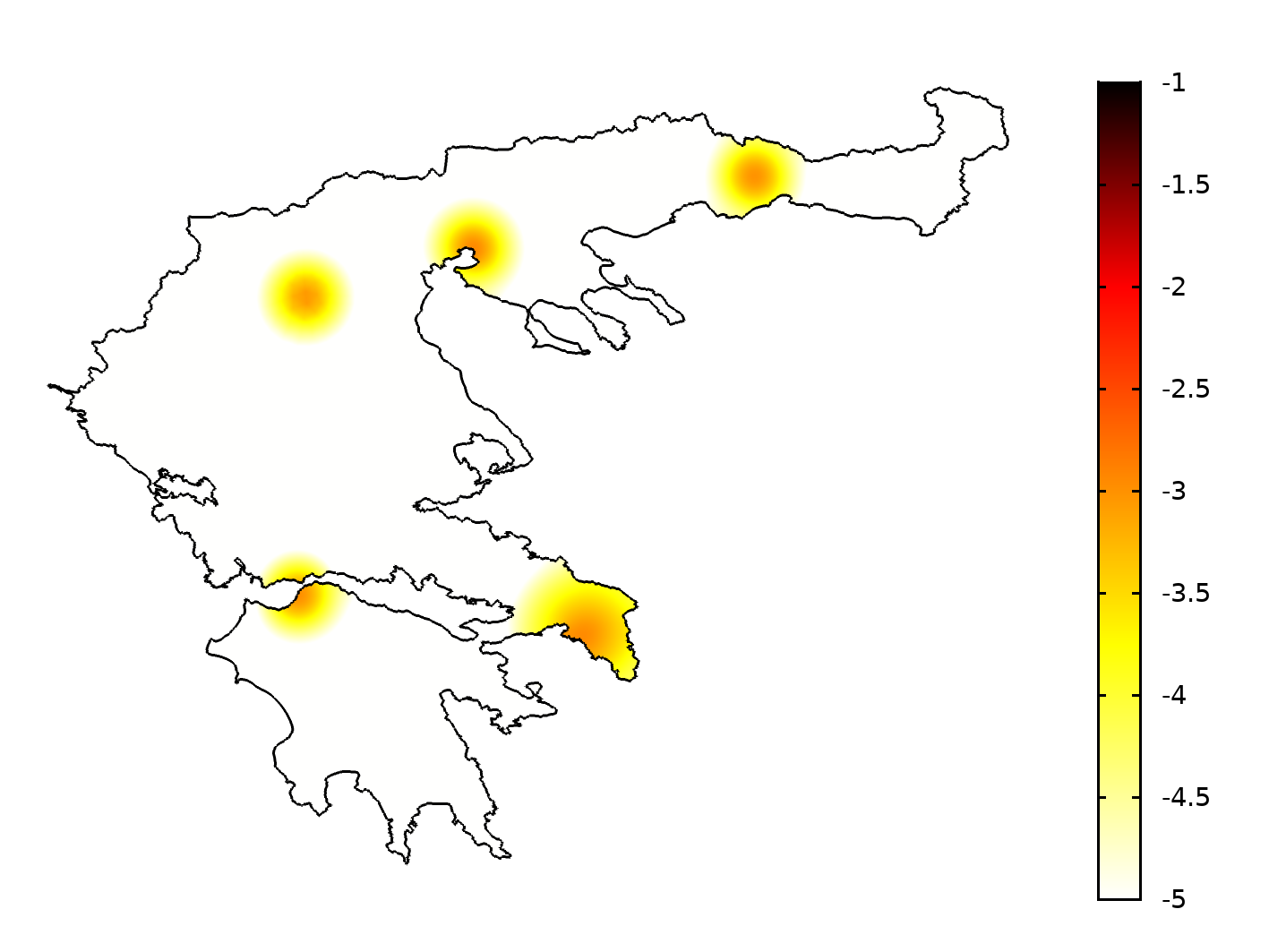} \\
\includegraphics[width=.45\textwidth]{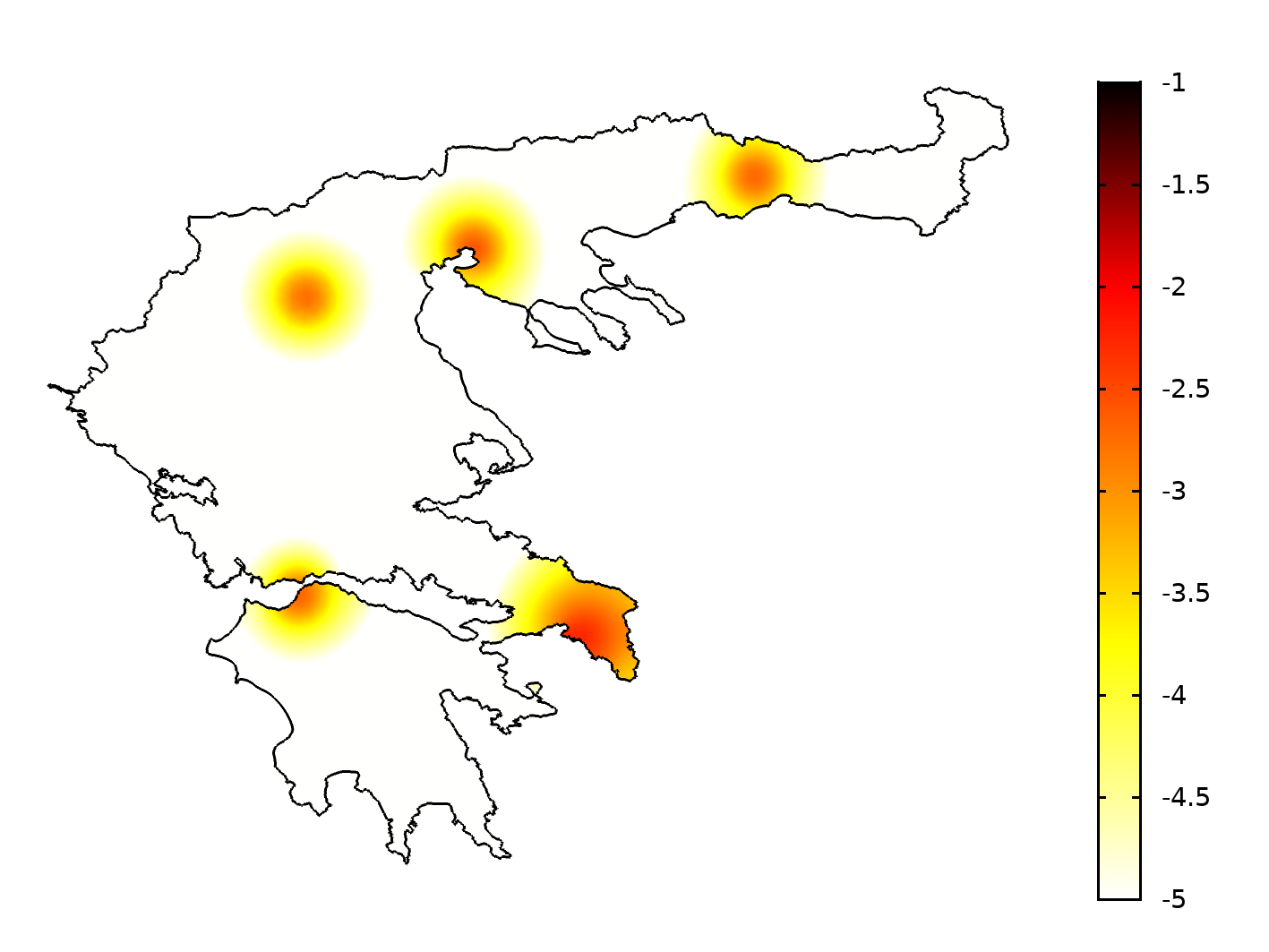} &
\includegraphics[width=.45\textwidth]{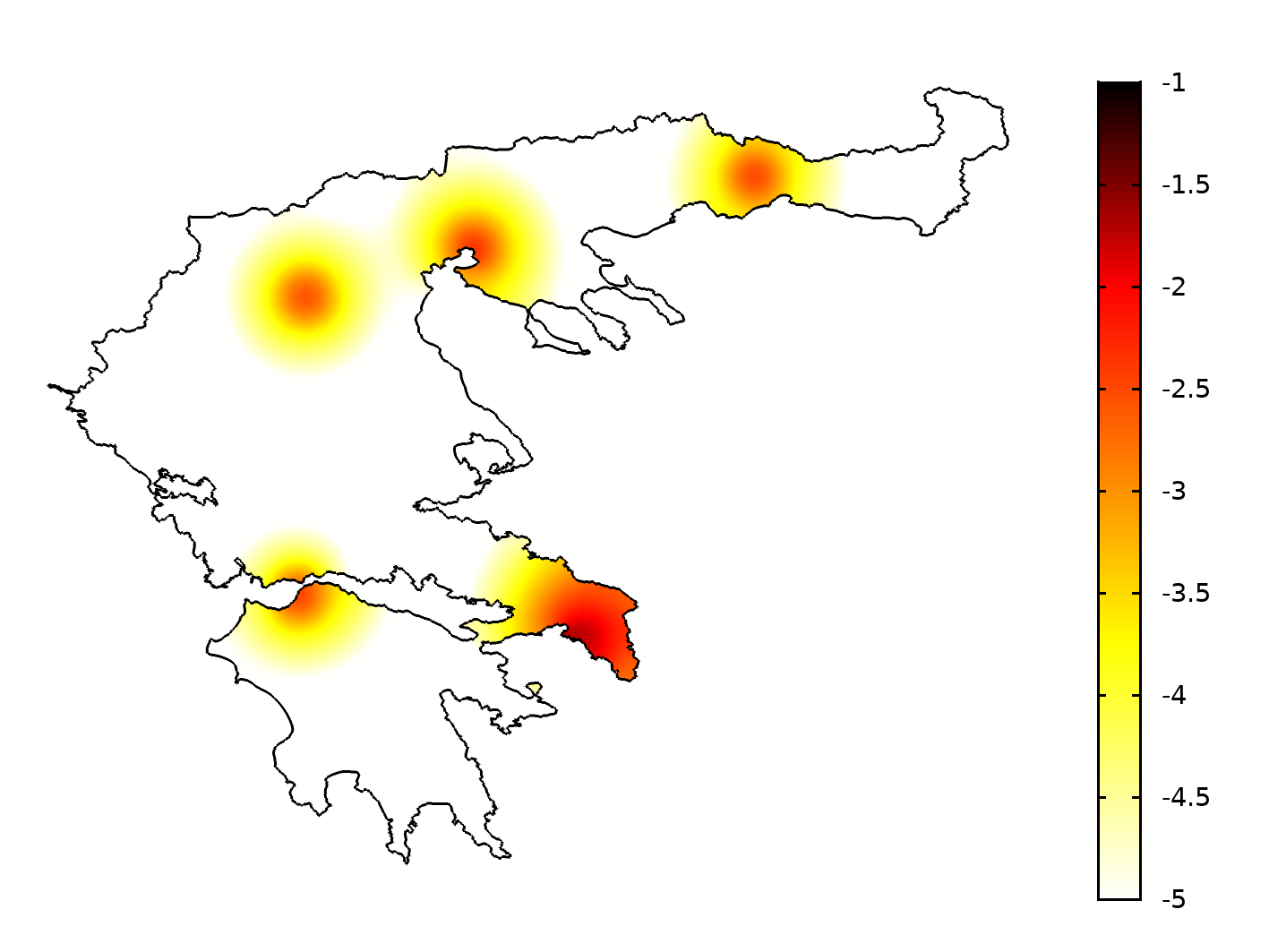} \\
\end{tabular}
\caption{{(Color online.)} Evolution of the Greek fatalities population density $\log_{10}D(x,y,t)$
for $t=1$ day (top left, March 12, 2020), $t=6$ days (top right, March 17, 2020),
$t=11$ days (bottom left, March 22, 2020), and {$t=18$ days (bottom right, March 29, 2020)}.
\textcolor{black}{$t_q=23$ (April 3, 2020), $\eta_D=0.35$ and scaling factor $\xi = 0.00246$.}
{A logarithmic (base 10) colorbar scale has been used.}}
\label{fig:GreecePDE_D}
\end{figure}

\begin{figure}[!ht]
\centering
\begin{tabular}{cc}
\includegraphics[width=.35\textwidth]{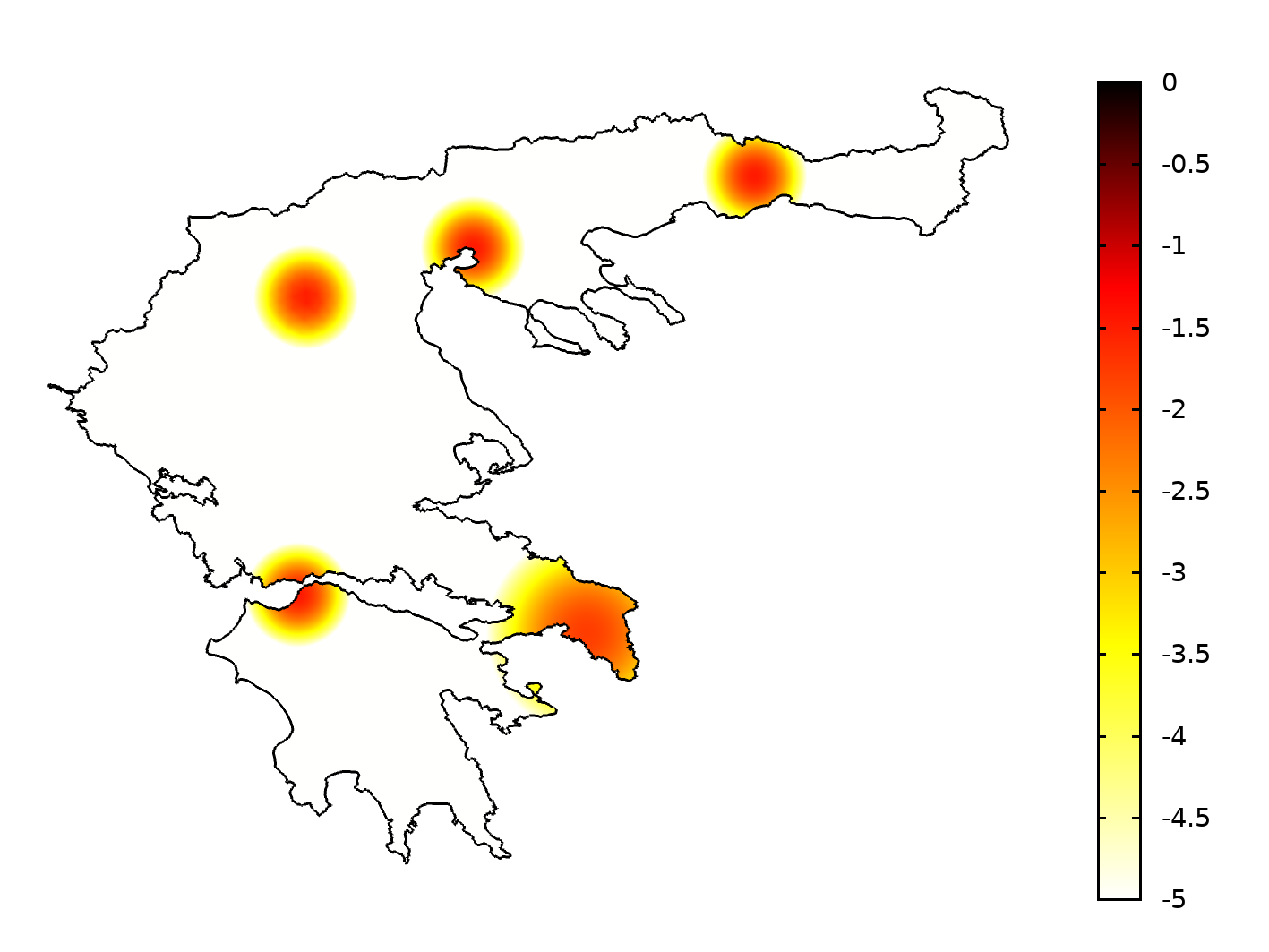} &
\includegraphics[width=.35\textwidth]{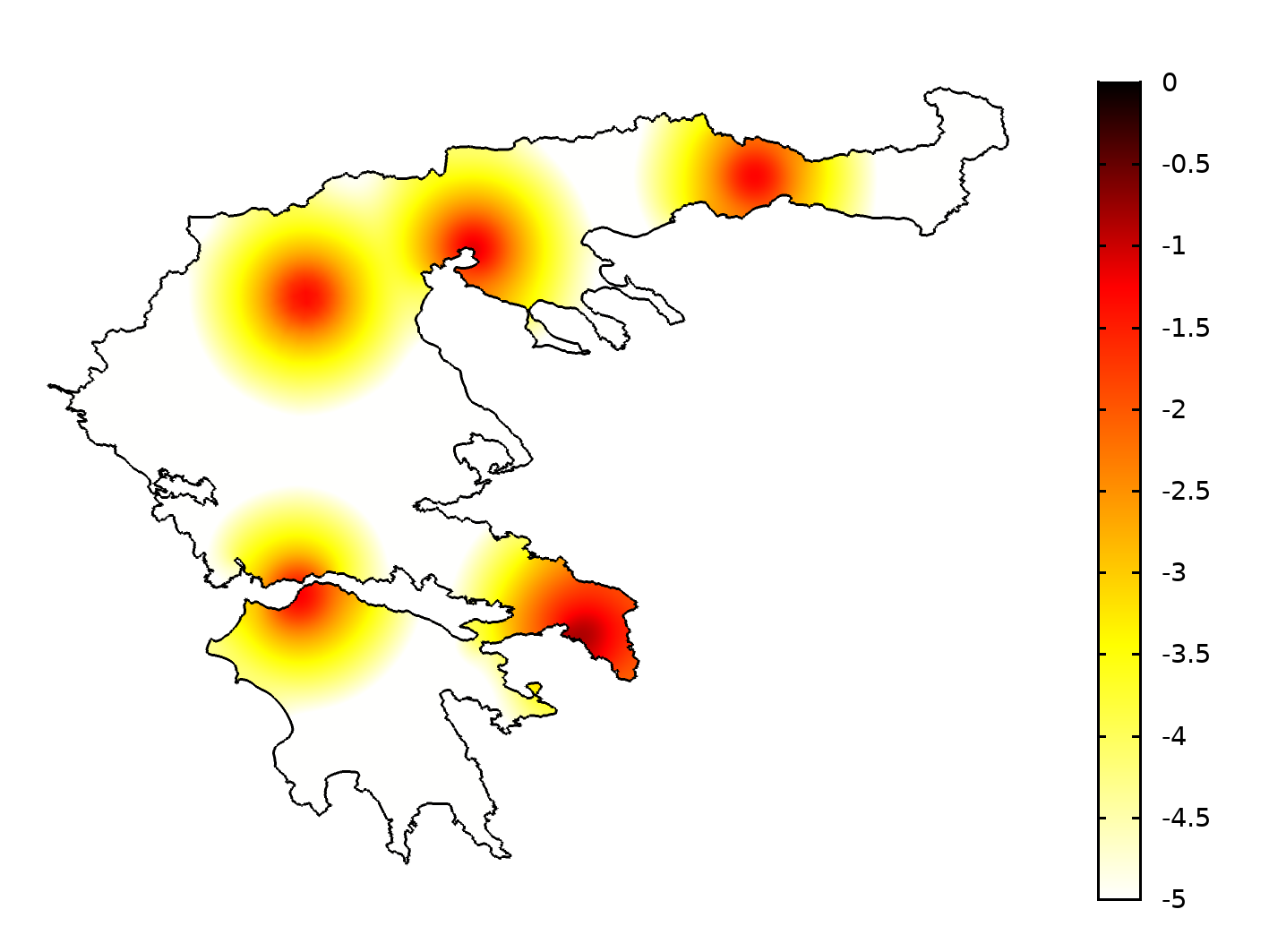} \\
\includegraphics[width=.35\textwidth]{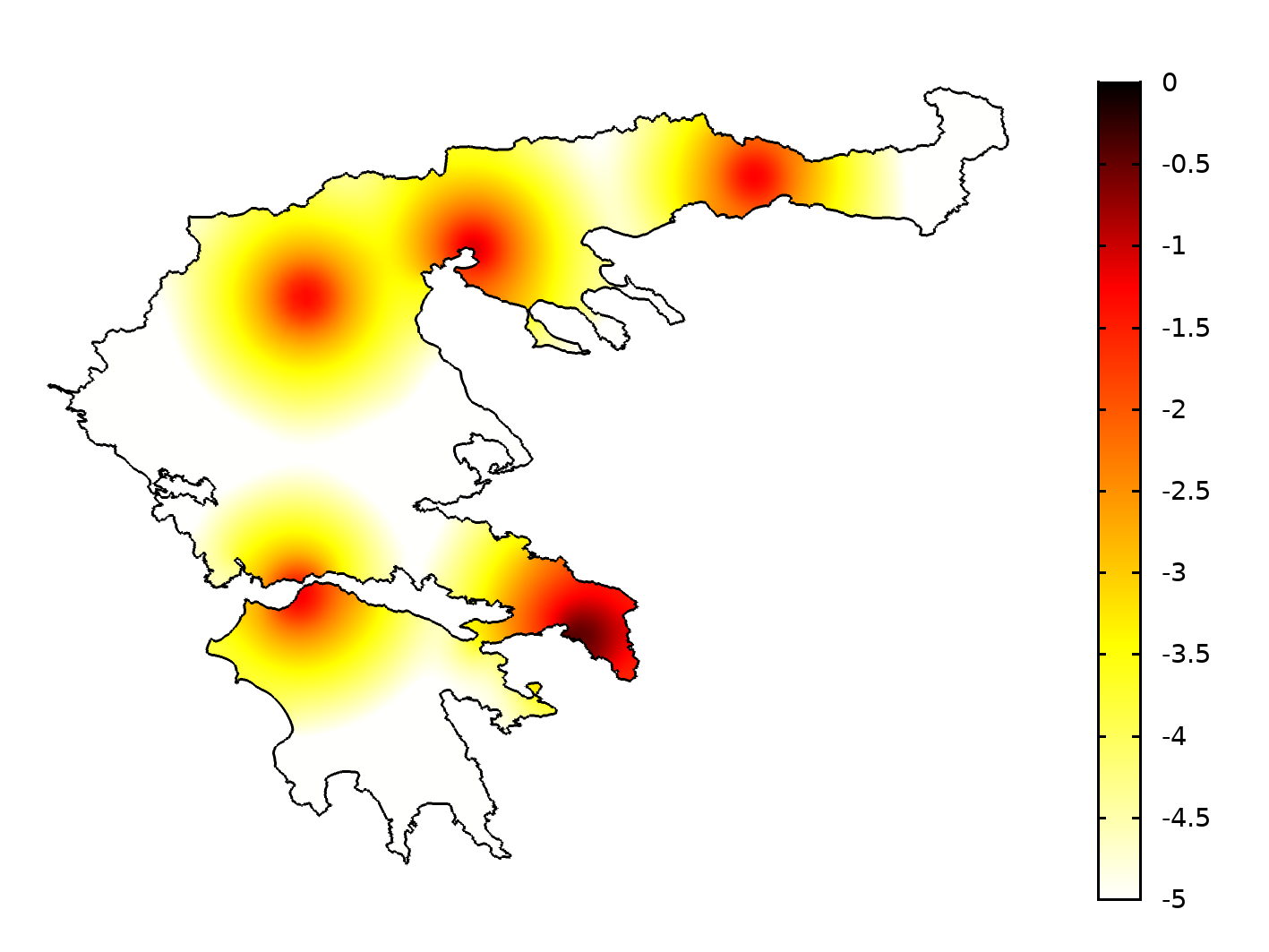} &
\includegraphics[width=.35\textwidth]{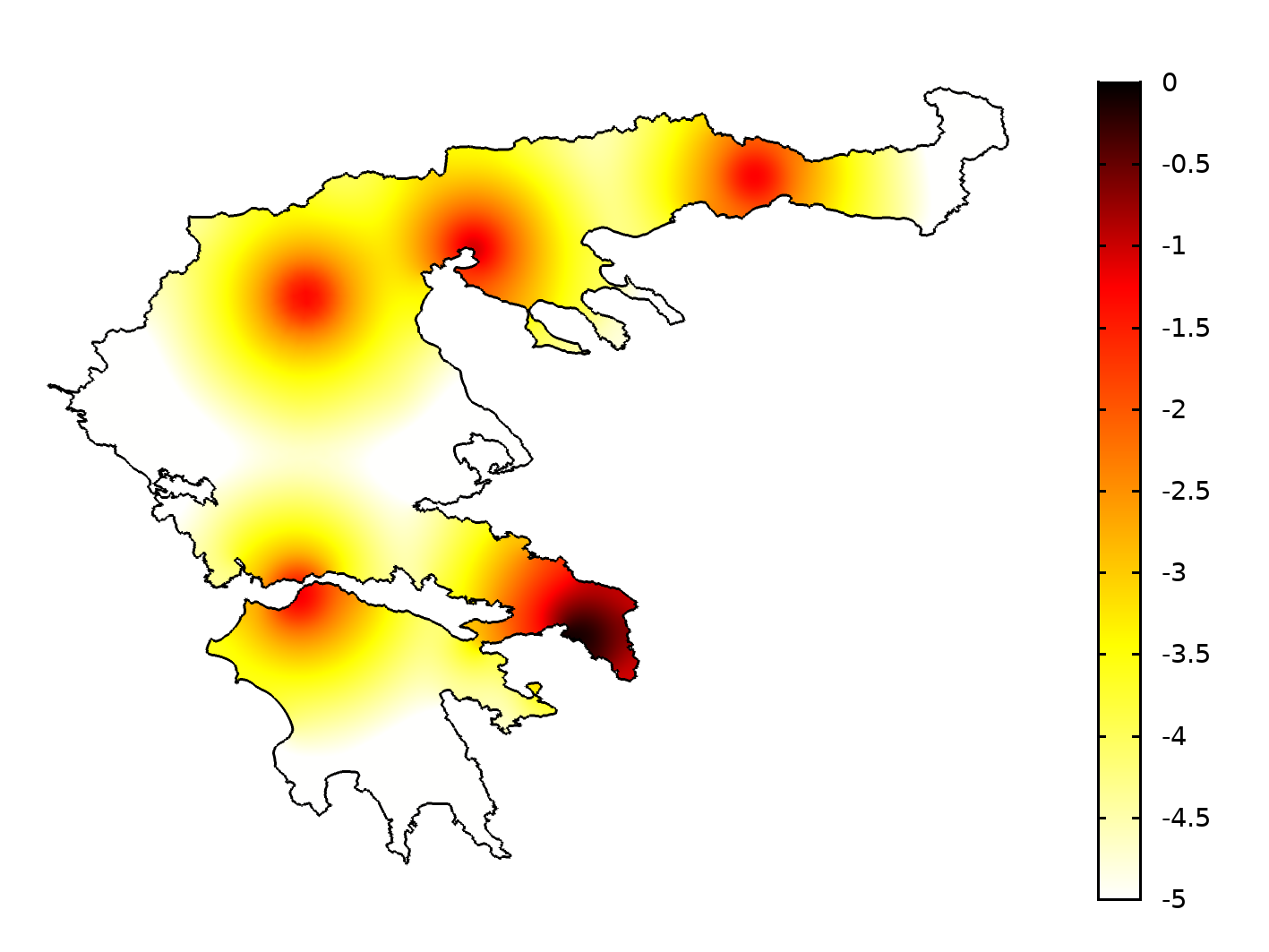} \\
\end{tabular}
\caption{{(Color online.)} Evolution of the Greek confirmed case density $\log_{10}C(x,y,t)$
for $t=1$ day (top left, March 12, 2020), $t=6$ days (top right, March 17, 2020),
$t=11$ days (bottom left, March 22, 2020), and {$t=18$ days (bottom right, March 29, 2020)}.
\textcolor{black}{$t_q=23$ (April 3, 2020), $\eta_D=0.35$ and scaling factor $\xi = 0.00246$.}
{A logarithmic (base 10) colorbar scale has been used.}}
\label{fig:GreecePDE_C}
\end{figure}

\begin{figure}[!htb]
\centering
\begin{tabular}{cc}
\includegraphics[width=.45\textwidth]{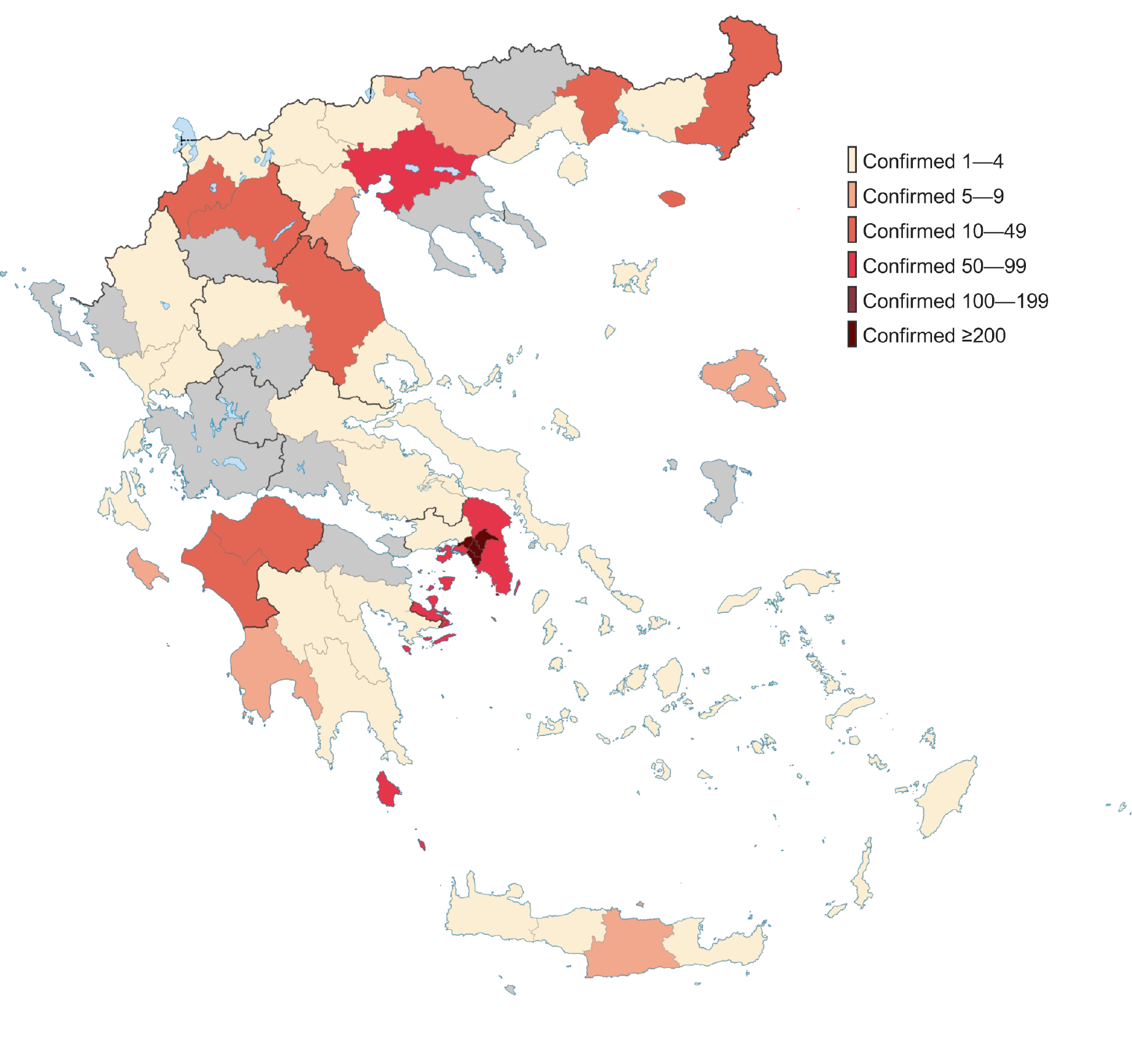}
\includegraphics[width=.45\textwidth]{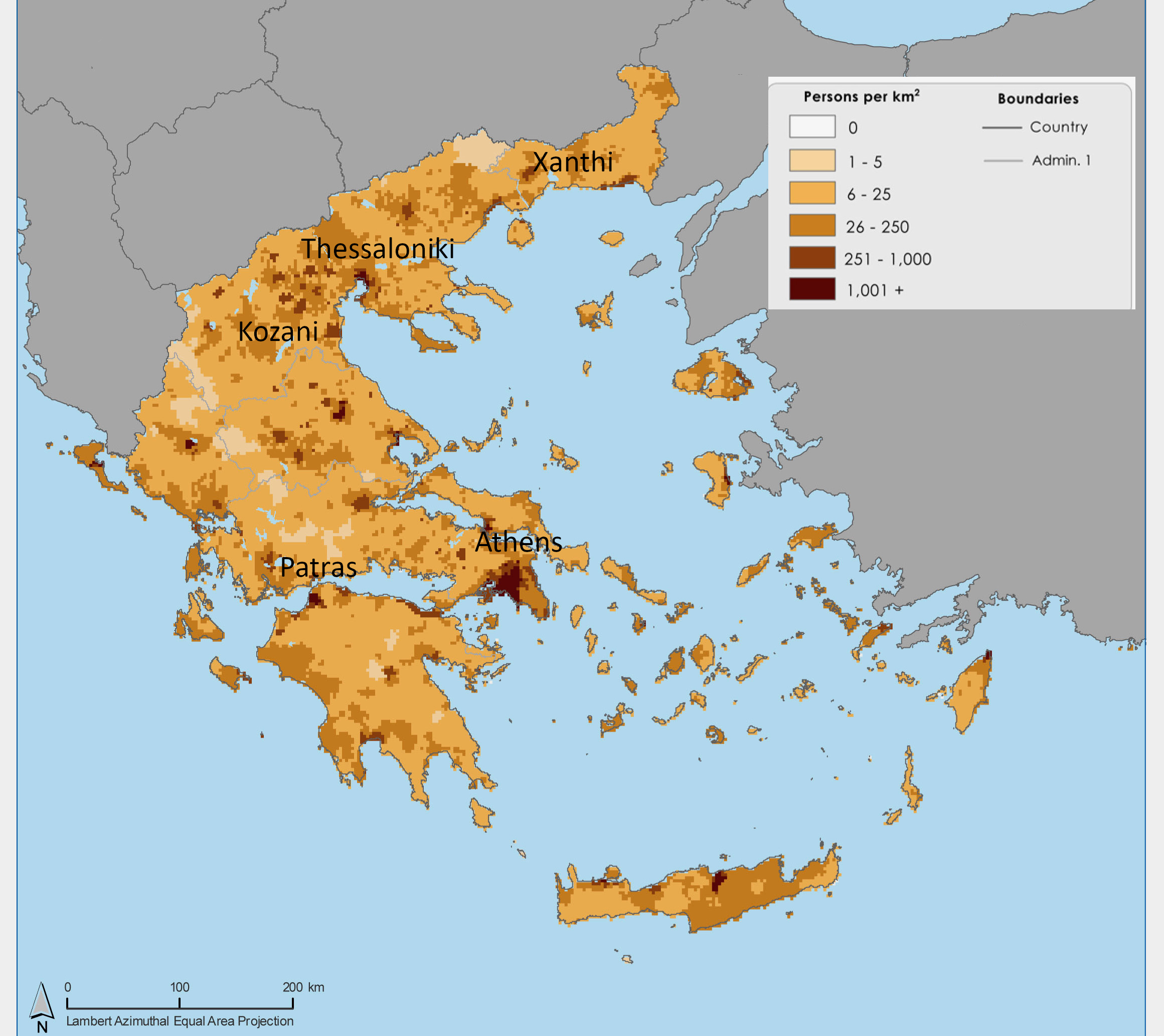}
\end{tabular}
\caption{{(Color online.)} Left panel: Map of COVID-19 outbreak in Greece as of March 29, 2020 ($t=18$) reproduced
from Ref.~\cite{greecewiki_Spatial}. Number of
confirmed cases per prefecture (Greece's regional units) denoted by color.
Compare and contrast with the right bottom
panel of Fig.~\ref{fig:GreecePDE_C}.
{Right panel: Population density, person per km$^2$, in Greece at
  2000,
  encompassing the identification of the cities where the
epidemic was initialized, i.e. Athens, Thessaloniki, Patras, Kozani, and Xanthi
(in decreasing order of the number of inhabitants). Modified image based on the
Wikimedia Commons image of~Ref.~\cite{GreecePopDensity}.}}
\label{fig:GreeceSpatialDist}
\end{figure}

The initialization of the model for Greece consisted of
selecting five of the key ``hotspots'' of
the infection, as they arose in Greece, to populate
initially.  We defined an infection radius of $10$ km
around the center
of Athens (largest city and capital), Thessaloniki (second largest
city and source of the first infection), Patras (third largest city and
the location where a key imported group of infected individuals was
transferred), as well as Kozani and Xanthi. {These cities where the
epidemic was initialized are identified in the 2000 population density
map shown in the right panel of Fig.~\ref{fig:GreeceSpatialDist}}. The latter two are two
significant peripheral centers where infections were seeded early on. In Athens,
we placed the largest (by a factor of two) source of infection, while
similar ``blobs'' of infection were initialized in the remaining four
cities. As in the case of Andalusia, these epicenters of infection were initiated via Gaussian
profiles whose spatial (variance) scale was selected as the infection
radius; their amplitude was chosen so that the total number of infections,
deaths, recoveries and hospitalizations, as calculated via the surface
integrals of the associated densities through the country, be the same
as the one reported in the original data. The population of
asymptomatics and exposed was, similarly to the ODE optimization, selected to be
proportional to the infected one with the proportionality ratios $A_0/I_0$ and
$E_0/I_0$ maintained as those of the ODE.

As for Andalusia, having initialized the PDE model, it was run {\it without} optimizing
\textcolor{black}{the
median parameters that are not expected to depend
on spatial scales at the PDE level. Instead, we varied the
diffusion-coefficient
reduction factor
$\eta_D$ in steps of $0.05$ in the interval $[0.25, 0.50]$ (six simulations in total).
For each simulation the scaling  factor was determined by minimizing the
$\ell^2$ norm, as previously discussed.
Similarly to the 0D simulations,
the comparison of the spatially integrated
PDE results to the data for Greece (see Fig.~\ref{fig:GreecePDE}) is
not particularly good for the first scenario ($t_q = 13$, left column) that
considers only a minor shift of the
quarantine time. In fact, the 0D results do not
fall within the range of the spatially averaged PDE simulations.}
Further modification of the quarantine time (scenario two)
can also help capture once again the ``angle'' in the relevant data (right column).
Clearly, the spatial model, via the surface integral of the population densities,
does a very adequate job at
capturing both the cumulative infections and the number of deaths.
Notice that in the bottom row of the figure, we illustrate the surface integrals
of each of the densities of $E$, $A$, $H$, $R$ and $AR$ as a function of time,
representing the evolution of the pandemic at the ``integrated'' level of
the entire country in an illustration similar to the one that we typically obtain from the ODE models.

That being said, of course, the PDE model provides considerable additional
information through its spatial resolution. In
Figs.~\ref{fig:GreecePDE_I}--\ref{fig:GreecePDE_C}, we can see the
spatio-temporal evolution of infections
(i.e., the spatial distribution at a few snapshots
over time), fatalities and cumulative infections $C(x,y,t)$,
respectively. These figures were generated using the median parameters
of  the second scenario, \textcolor{black}{and with the transmission
rates multiplied by the scaling factor $\xi = 0.00246$. The corresponding length scale
is $l = \sqrt{\xi N} \approx 0.163$ km, comparable to what we found for Andalusia
$l \approx 0.200$km. }
\textcolor{black}{In addition and as in the case of Andalusia,
if we associate the scaling factor with the inverse of the ``lived" population density
we find that in Greece~\cite{LivedDensityGreece} it is around $1/379 \approx 0.0026$,
remarkably close to the numerically-determined $\xi$. We believe that
the identification of the scaling factor as very closely matching
the inverse of the lived population density in two entirely
independent (and quite distinct in their number of infections) cases
suggests this scaling as a nontrivial insight stemming from these
studies
about the connection of ODE and PDE models.}

\textcolor{black}{We also produced movies of the corresponding
evolution that can be found in~\cite{movies}.}
It is important
to re-iterate here that we have not included the islands of Greece in this effort (i.e.,
we are looking at the mainland of Greece).
Obviously if one were to model the disease spread in each of these
islands it would be relevant to seed the infection in each island
individually and study the spreading there rather than together with
the spatially disconnected from the islands mainland of Greece.

We can clearly see how the infection spreads throughout the country,
affecting most significantly the regions of higher population density.
Indeed, it is clear that over time the infection is extinguished in
most regions and it finally persists chiefly in Attica, the region of
highest population density (and where the main metropolitan center of the
country, Athens, lies); see, in particular, the bottom panels of
Fig.~\ref{fig:GreecePDE_I}.
Nevertheless, it is evident from Fig.~\ref{fig:GreecePDE_D} that
a number of deaths develops in each of the 5
regions where the infection was initially seeded,
in line with the corresponding expectation from the country's data.
Indeed, also, each region features a discernible fraction of
cumulative infections in Fig.~\ref{fig:GreecePDE_C}, although clearly
once again the lion's share of infections pertains to Attica. The
second biggest fraction of infections pertains to Thessaloniki (the second
biggest metropolitan center) and so on.
It is clear from these figures that the model yields \textcolor{black}{a  reasonable prediction of the}
spatial evolution of the disease spread, in line with the cumulative
totals of deaths and infections throughout the country. \textcolor{black}{Nevertheless,
a comparison with the spatial distribution of the pandemic
throughout the country~\cite{greecewiki}, see Fig.~\ref{fig:GreeceSpatialDist}, suggests also some
limitations. The spatial snapshot of the number of confirmed cases per
prefecture shown in that figure was created on March 29, 2020, corresponding
to $t=18$. Hence the data are intermediate between the two bottom panels
of Fig.~\ref{fig:GreecePDE_C}, at $t=11$ and $21$ days. The persistence of
the epidemic in Attica (dark red) and to a lesser degree Thessaloniki (red) is evident in both
figures. The other areas seem to be slightly underpredicted. The comparison suggests
a partial time mismatch between the simulations and the data.}
One can also observe that since the initial spots for Andalusia were
spatially close, the infection spreads more easily in that region than
in the mainland of Greece.

\clearpage

{\it Acknowledgments.}
The authors are indebted to Dr. Maksym Bondarenko for his
substantial help with the WorldPop maps and setup, and also thank Dr. Jinlan Huang for
assistance in setting the relevant computation up in COMSOL.
\textcolor{black}{In addition, we thank  Francisco Rodr\'{i}guez S\'{a}nchez
for providing the data and R code used to generate the map with
the spatial distribution of the epidemic in Andalusia, Fig.~\ref{fig:AndalusiaSpatialDist}.}
PGK gratefully acknowledges discussions and input from Andy Ludu including
regarding the layout of Fig.~\ref{fig:SEAIHR}.
This material is based upon work supported by the US National Science Foundation
under Grants No. DMS-1815764 (ZR), PHY-1602994, and DMS-1809074 (PGK). J.C-M. also thanks the Regional Government of Andalusia under the project P18-RT-3480 and MICINN, AEI and EU (FEDER program) under the project PID2019-110430GB-C21.
\textcolor{black}{ZR and PGK also acknowledge support through the C3.ai Digital Transformation Institute.}
PGK also acknowledges support from the Leverhulme Trust via a Visiting Fellowship and
thanks the Mathematical Institute of the University of Oxford for its hospitality during  this work. 

\textit{Disclaimer}
The views expressed in this manuscript are purely
those of the authors and may not, under any circumstances, be regarded as an official position of the European Commission.


\begin{thebibliography}{99}

\bibitem{chowell2003}
\textcolor{black}{G. Chowell, P. W. Fenimore, M. A. Castillo-Garsow,
C. Castillo-Chavez, SARS outbreaks  in Ontario, Hong Kong and Singapore:
the role of diagnosis and isolation as  a control mechanism, Journal of
Theoretical Biology {\bf 224} 1--8 (2003).
}

\bibitem{breban2013}
\textcolor{black}{
R. Breban, J. Riou and A. Fontanet,
Interhuman transmissibility of Middle East respiratory syndrome coronavirus: estimation of pandemic risk,
Lancet {\bf 382} 694--699 (2013).
}

\bibitem{WHO_Count} \textcolor{black}{\url{https://covid19.who.int}}

\bibitem{lev4} B.S. Graham, J.R. Mascola, and A.S. Fauci,
Novel Vaccine Technologies: Essential Components of an Adequate Response to Emerging Viral Diseases,
JAMA {\bf 319}, 1431 (2018).

\bibitem{lev5} World Health Organization Writing Group, Emerg. Infect. Dis. {\bf 12}, 88 (2006).

\bibitem{lev3}  N.M. Ferguson {\it et al.}, Impact of non-pharmaceutical interventions (NPIs) to reduceCOVID-19
mortality  and  healthcare demand.  Imperial  College,  London,  (Mar.  2020),
\url{https://www.imperial.ac.uk/mrc-global-infectious-disease-analysis/covid-19/}

\bibitem{oxford} See, e.g., \url{https://www.covid19vaccinetrial.co.uk/about}

\bibitem{Editorialyd2020} Y. Drossinos and N.I. Stilianakis,
What aerosol physics tells us about airborne pathogen trannsmission, Aerosol Sci. Technol. {\bf 54}, 639 (2020).
\url{https://doi.org/10.1080/02786826.2020.1751055}.

\bibitem{ChineseRestaurant} J. Lu, J. Gu, K. Li, C. Xu, W. Su, Z. Lai, D. Zhou, C. Yu, B. Xu, and Z. Yang,
Estimating the asymptomatic proportion of coronavirus disease 2019 (COVID-19)
cases on board the Diamond Princess cruise ship, Yokohama, Japan, 2020, Emerg. Infect. Dis. \textbf{26}, 1268 (2020).
\url{https://doi.org/10.2807/1560-7917.ES.2020.25.10.2000180}

\bibitem{cruise} M. Kenji, K. Katsushi, Z. Alexander, C. Gerardo,
COVID-19 Outbreak Associated with Air Conditioning in Restaurant, Guangzhou, China, 2020,
Euro Surveill. \textbf{25}(10):pii=2000180 (2020).
\url{https://doi.org/10.3201/eid2607.200764}

\bibitem{beijing} P. Yang , J. Qi , S. Zhang, X. Wang, G. Bi, Y. Yang, B. Sheng, and G. Yang,
Feasibility study of mitigation and suppression strategies for controlling COVID\-19 outbreaks in London and Wuhan,
PLoS ONE \textbf{15}(8), e0236857 (2020). \url{https://doi.org/10.1371/journal.pone.0236857}

\bibitem{louisiana}  K.-M. Tam, N. Walker, and  J. Moreno, Projected Development of COVID-19 in Louisiana,
arXiv:2004.02859.

\bibitem{england} L. Danon, E. Brooks-Pollock, M. Bailey, and M. Keeling,
A spatial model of CoVID-19 transmission in England and Wales: early spread and peak timing,
\url{https://doi.org/10.1101/2020.02.12.20022566.}

\bibitem{arenas}
A. Arenas, W. Cota, J. G\'omez-Garde\~nes, S. G\'omez, C. Granell, J.T. Matamalas, D. Soriano-Pa\~nos, and B. Steinegger,
Modeling the Spatiotemporal Epidemic Spreading of COVID-19 and the Impact of Mobility and Social Distancing Interventions,
Phys. Rev. X \textbf{10}, 041055 (2020).
\url{http://doi.org/10.1103/PhysRevX.10.041055};
Derivation of the effective reproduction
number R for COVID-19 in relation to mobility restrictions and confinement,
\url{doi: https://doi.org/10.1101/2020.04.06.20054320}

\bibitem{tsironis} G.D. Barmparis and G.P. Tsironis, Estimating the infection horizon
of COVID-19 in eight countries with a data-driven approach, Chaos, Solitons and Fractals
\textbf{135}, 109842 (2020). \url{https://doi.org/10.1016/j.chaos.2020.109842}

\bibitem{SpainSpatial} \textcolor{black}{F. Ar\`{a}ndiga, A. Baeza, I. Cordero-Carri\'{o}n , R. Donat,
M. C. Mart\'{i}, P. Mulet, and D.F. Y\'{a}\~{n}ez, A spatial-temporal model for the evolution of the COVID-19
pandemic in Spain including mobility, Mathematics \textbf{8}, 1677 (2020).
}

\bibitem{country11} S. Flaxman, S. Mishra, A. Gandy, \textit{et al.},
Estimating the effects of non-pharmaceutical interventions on COVID-19 in Europe,
 Nature \textbf{584}, 257–261 (2020). \url{https://doi.org/10.1038/s41586-020-2405-7}

\bibitem{brazil} S.B. Bastos and D.O. Cajueiro, Modeling and forecasting the early
evolution of the Covid-19 pandemic in Brazil, Sci. Rep. \textbf{10}, 19457 (2020).
\url{https://doi.org/10.1038/s41598-020-76257-1}

\bibitem{sweden} C. Qi, D. Karlsson, K. Sallmen, and R. Wyss, Model studies on the COVID-19 pandemic in Sweden,
arXiv:2004.01575

\bibitem{albania} E. Gjini,  Modeling Covid-19 dynamics for real-time estimates and projections: an application to Albanian data,
\url{https://doi.org/10.1101/2020.03.20.20038141}

\bibitem{Italy} \textcolor{black}{M. Gatto, E. Bertuzzo, L. Mari, S. Miccoli, L. Carraro, R. Casagrandi, and A. Rinaldo,
Spread and dynamics of the COVID-19 epidemic in Italy: Effects of emergency containment measures, Proc. Natl.
Acad. Sci. U.S.A. \textbf{117}, 10484 (2020). \url{www.pnas.org/cgi/doi/10.1073/pnas.2004978117}}

\bibitem{NikosGreece} \textcolor{black}{I. Kioutsoukis and N.I. Stilianakis, On the transmission dynamics
of SARS-CoV-2 in a  temperate climate, Int. J. Environ, Res. Public Health \textbf{18}, 1660 (2021).
\url{https://doi.org/10.3390/ijerph18041660}}

 \bibitem{mbe} C. Yang and J. Wang, A mathematical model for the novel coronavirus epidemic in Wuhan, China,
 Math. Biosci. Eng. {\bf 17}, 2708 (2020). \url{https://doi.org/10.3934/mbe.2020148}

 \bibitem{TangRisk} \textcolor{black}{B. Tang, X. Wang, Q. Li, N.L. Bragazzi, S. Tang, Y. Xiao, and J. Wu,
 Estimation of the transmission risk of the 2019-nCoV and its implication for public health interventions,
 J. Clin. Med. ]\textbf{9}, 462 (2020). \url{http://doi.org/10.3390/jcm9020462}; B. Tang, N.L. Bragazzi,
 Q. Li, S. Tang, Y. Xiao, and J. Wu, An updated estimartion of the risk of transmission of the novel
 coronavirus (2019-nCov), Inf. Dis. Model. \textbf{5}, 248 (2020). \url{https://doi.org/10.1016/j.idm.2020.02.001}}

\bibitem{siettos} L. Russo, C. Anastassopoulou, A. Tsakris, G.N. Bifulco, E.F. Campana, G. Toraldo, and C. Siettos,
Tracing day-zero and forecasting the COVID-19 outbreak in Lombardy, Italy: A compartmental modelling
and numerical optimization approach, PLoS ONE \textbf{15}(10), e0240649 (2020).\url{https://doi.org/10.1371/journal.pone.0240649};
 C. Anastassopoulou, L. Russo, A. Tsakris, C. Siettos, Data-based analysis, modelling and forecasting of the COVID-19 outbreak
PLoS ONE \textbf{15}(3),  e0230405 (2020). \url{https://doi.org/10.1371/journal.pone.0230405}

\bibitem{vespignani2020}
\textcolor{black}{
A. Vespignani, H. Tian, C. Dye, J. O. Lloyd-Smith, R. M. Eggo, M. Shrestha,
S. V. Scarpino, B. Gutierrez, M. U. G. Kraemer, J. Wu, K. Leung and  G.  M. Leung,
Modeling COVID-19,  Nature Reviews Physics {\bf 2} 279--281 (2020).
}

\bibitem{anirudh2020}
\textcolor{black}{
A.  Anirudh, Mathematical modeling and the transmission dynamics in predicting
the Covid-19-What next in combating  the pandemic, Infectious Disease Modelling
{\bf 5} 366--374 (2020).
}

\bibitem{meehan2020}
\textcolor{black}{
M. T. Meehan,  D. P. Rojas, A.  I. Adekunle,O. A. Adegboye, J. M.  Caldwell,
E. Turek, B. M. Williams, B. J.  Marais, J. M. Trauer, E. S.  McBryde,
Modelling insights into the COVID-19 pandemic, Paediatric Respiratory Reviews
{\bf 35} 64--69 (2020).
}

\bibitem{holmdahl2020}
\textcolor{black}{
I. Holmdahl and C. Buckee, Wrong but useful--What COVID-19 epidemiological
models can and cannot tell us, N. England J. Med. {\bf 383} 303--305 (2020).
}

\bibitem{burki2021}
\textcolor{black}{
T. Burki, Understanding variants of SARS-CoV-2, The Lancet
{\bf 397}, P462 (2021), \url{https://doi.org/10.1016/S0140-6736(21)00298-1}}

\bibitem{MathematicsInfectiousDiseases} \textcolor{black}{H.W. Hethcote, The mathematics of infectious diseases,
SIAM Review \textbf{42}:4, 599-653 (2000);  N.T. J. Bailey,
\textit{The mathematical theory of epidemics}, Charles Griffin \& Co., Ltd., London (1957);
V. Capasso, \textit{Mathematical structures of epidemic systems}, Springer-Verlag Berlin Heidelberg (1993).}

\bibitem{theo} C. Gai, D. Iron, and T. Kolokolnikov, Localized outbreaks in an S-I-R model with diffusion,
 Math. Biol. {\bf 80}, 1389 (2020). \url{https://doi.org/10.1007/s00285-020-01466-1}

\bibitem{reluga2004} T. Reluga, A two-phase epidemic driven by diffusion, J. Theor. Biol. {\bf 229}, 249 (2004).
\url{https://doi.org/10.1016/j.jtbi.2004.03.018}

\bibitem{BooksSIRDiffusion} \textcolor{black}{J. Murray, \textit{Mathematical biology II: Spatial models and biomedical
applications}, Vol. 3, Springer-Verlag (2001); M.J. Keeling and P. Rohani, \textit{Modeling infectious
diseases in humans and animals}, Princeton University Press (2011).}

\bibitem{Malaria} \textcolor{black}{J. Gaudart, M. Ghassani, J. Mintsa, M. Rachdi, J. Waku, and J. Demongeot,
Demography and Diffusion in Epidemics: Malaria and Black Death Spread, Acta Biotheor. \textbf{58} 277–305 (2010)
\url{https://doi.org/10.1007/s10441-010-9103-z}}

\bibitem{DropletDynamics} N.I. Stilianakis and  Y. Drossinos,  Dynamics of infectious
disease transmission by inhalable respiratory droplets, J. R. Soc. Interface \textbf{7}, 1355
(2010), \url{https://royalsocietypublishing.org/doi/10.1098/rsif.2010.0026}.

\bibitem{Noble} \textcolor{black}{J.V. Noble, Geographic and temporal development of plagues. Nature \textbf{250}, 726–729 (1974).
\url{https://doi.org/10.1038/250726a0}}

\bibitem{YanScalingLaws} \textcolor{black}{X.-Y. Yan, W.-X. Wang, Z.-Y. Gao, and Y.-C. Lai, Universal model
of individual and populatrion mobility on diverse spatial scales, Nat. Commun. \textbf{8}, 1639 (2017).
\url{https://doi.org/10.1038/s41467-017-01892-8}}

\bibitem{france}
\textcolor{black}{
Y. Mammeri, A reaction-diffusion system to better comprehend the unlockdown:
Application of SEIR-type model with diffusion to the spatial spread of
COVID-19 in France, \url{https://arxiv.org/pdf/2005.03499.pdf}}

\bibitem{italyViguerie}
\textcolor{black}{
A. Viguerie, G. Lorenzo, F. Auricchio, D. Baroli, T.R.J. Hughes, A. Patton,
A. Reali, T.E. Yankeelov, A. Veneziani, Simulating the
spread of COVID-19 via spatially-resolved susceptible - exposed - infected -
recovered - deceased (SEIRD) model with heterogeneous
diffusion, Appl. Math. Lett. {\bf 111}, 106617 (2021). }

\bibitem{vespignani} V. Colizza and A. Vespignani, Epidemic modeling in metapopulation
systems with heterogeneous coupling pattern: Theory and simulations, J. Theor. Biol. {\bf 251}, 450 (2008).
\url{https://doi.org/10.1016/j.jtbi.2007.11.028}

\bibitem{Keeling2002} \textcolor{black}{M.J. Keeling and P. Rohani, Estimating spatial coupling in epidemiological
systems: A mechanistic approach, Ecol. Lett. \textbf{5}, 20 (2002). \url{ https://doi.org/10.1046/j.1461-0248.2002.00268.x}}

\bibitem{Vespignani2011} \textcolor{black}{S. Meloni, N. Perras, A. Arenas, S. G\'{o}mez, Y. Moreno, and A.
Vespignani, Modeling human mobility responses to the large-scale spreading of infectious diseases, Sci. Rep.
\textbf{1}, 62 (2011). \url{https://doi.org/10.1038/srep00062}}

\bibitem{eqnfree} C.W. Gear, J.M. Hyman, P.G. Kevrekidis, I.G. Kevrekidis, O. Runborg, and
C. Theodoropoulos, Equation-Free, Coarse-Grained Multiscale Computation:
Enabling Mocroscopic Simulators to Perform System-Level Analysis Comm. Math. Sci. {\bf 1}, 715 (2003).
\url{https://dx.doi.org/10.4310/CMS.2003.v1.n4.a5}

\bibitem{epigraph} G. Mart{\'i}n, D.E. Singh, M.-C. Marinescu, and J. Carretero,
Parallel algorithm for simulating the spatial transmission of influenza in EpiGraph, Parallel Computing {\bf 42}, 88 (2015).
\url{https://doi.org/10.1145/2488551.2488585}

\bibitem{arons2020} M. M. Arons {\it et al.},  Presymptomatic SARS-CoV-2 infections and transmission in a skilled nursing facility,
N. Engl. J. Med. \textbf{382}:2081-2090 (2020). \url{http:/doi.org//10.1056/NEJMoa2008457}

\bibitem{elife} See, e.g., Y.M. Bar-On, A. Flamholz, R. Phillips, and R. Milo, SARS-CoV-2 (COVID-19) by the numbers,
eLife \textbf{9}, e57309 (2020). \url{https://doi.org/10.7554/eLife.57309}

\bibitem{SEAIHRversionMERS} \textcolor{black}{Z.-Q. Xia, J. Zhang, Y.-K. Xue, G.-Q. Sun, and Z. Jin,
Modeling the transmission of Middle East Respiratory Syndrome corona virus in the Republic of Korea,
PLoS ONE \textbf{10}(12),  e0144778 (2015). \url{https://doi.org/10.1371/journal.pone.0144778}}

\bibitem{FastDiffusion} \textcolor{black}{H. Berestycki, J.M. Roquejoffre, and L. Rossi,
Propagation of epidemics along lines of fast diffusion, Bull. Math. Bio. \textbf{83}, 2 (2021).
\url{https://doi.org/10.1007/s11538-020-00826-8}}

\bibitem{greecewiki} \textcolor{black}{\url{https://en.wikipedia.org/wiki/COVID-19_pandemic_in_Greece}}

\bibitem{time} See, e.g., \url{https://time.com/5824836/greece-coronavirus/}, and also \\
\url{https://www.bloomberg.com/opinion/articles/2020-04-10/greece-handled-coronavirus-crisis-better-than-italy-and-spain}

\bibitem{Comsol} COMSOL Multiphysics$^\circledR$ v. 5.5. \url{https://www.comsol.com}. COMSOL AB, Stockholm, Sweden.

\bibitem{worldpop} \url{https://www.worldpop.org/}

\bibitem{randomf} F.R. Stevens, A.E. Gaughan, C. Linard, and A.J. Tatem, Disaggregating census data for
population mapping using random forests with remotely-sensed and ancillary data,
PLoS ONE {\bf 10} (2), e0107042 (2014). \url{http://doi.org/10.1371/journal.pone.0107042}

\bibitem{mit} R. Dandekar and G. Barbastathis, Neural network aided quarantine control model estimation of global
Covid-19 spread, arXiv:2004.02752.

\bibitem{another} J.P. Arcede, R.L. Caga-anan, C.Q. Mentuda, and Y. Mammeri,
Accounting for Symptomatic and Asymptomatic in a SEIR-type model of COVID-19, Math. Model. Nat.
Phenom. \textbf{15}, 34 (2020). \url{https://doi.org/10.1051/mmnp/2020021}

\bibitem{MargueriteSpatialDynamics} M. Robinson, N.I. Stilianakis, and Y. Drossinos, J. Theor. Biol. \textbf{297}, 116 (2012),
\url{https://doi.org/10.1016/j.jtbi.2011.12.015}

\bibitem{datos_Junta} \url{http://www.juntadeandalucia.es/institutodeestadisticaycartografia/salud/index.htm}

\bibitem{MathBio2021} J. Cuevas-Maraver, P. Kevrekidis, Q.-Y. Chen, G.Kevrekidis, V. Villalobos-Daniel, Z. Rapti, Y. Drossinos,
Lockdown Measures and their Impact on Single- and Two-age-structured Epidemic Model for the COVID-19 Outbreak in Mexico,
Math. Biosci. (accepted). medRxiv 2020.08.11.20172833; \url{https://doi.org/10.1101/2020.08.11.20172833.}

\bibitem{Farmer2008} \textcolor{black}{T.G. Farmer, T.F. Edgar, and N.A. Peppas, Parameter set uniqueness and
confidence limits in model identification of insulin transport models from simulation data.
Diabetes Technol Ther. \textbf{10}(2):128-41 (2008). \url{https//doi.org/10.1089/dia.2007.0254}}

\bibitem{AndalusiaAsymptomatics}
\url{https://www.juntadeandalucia.es/organismos/saludyfamilias/actualidad/noticias/detalle/234496.html}

\bibitem{LivedDensityCovid19}  \textcolor{black}{P. Garland, D. Babbitt, M. Bondarenko, A. Sorichetta,
A.J. Tatem, and O. Johnson, The COVID-19 pandemic as experienced by the individual,
arXiv:2005.01167}

\bibitem{LivedDensityAndalusia} \url{http://www.juntadeandalucia.es/institutodeestadisticaycartografia/distribucionpob/}

\bibitem{movies} {See Supplemental Material at [URL will be inserted by publisher] for animations on the time evolution of the predictions of  infections, deaths and total cases for Andalusia and Greece}

\bibitem{AndalusiaSpatial} \textcolor{black}{\url{https://pakillo.github.io/COVID19-Andalucia/evolucion-coronavirus-andalucia.html}}

\bibitem{newer} S. Flaxman, S. Mishra, A. Gandy, A. et al.,  Estimating the number of infections and the impact of non-pharmaceutical interventions on COVID-19 in European countries: technical description update, arXiv:2004.11342

\bibitem{newer0} D. Tsiotas and L. Magrafas, The effect of anti-COVID-19 policies on the
evolution of the disease: A complex network analysis of the successful case of Greece,
Physics \textbf{2}(2), 325-339 (2020). \url{https://doi.org/10.3390/physics2020017}

\bibitem{andal} T. Chac{\'o}n Rebollo and D. Franco Coronil, Predictive data assimilation through
reduced order modeling for epidemics with data uncertainty, arXiv:2004.12341.

\bibitem{he2020}  X. He, E.H.Y. Lau, P. Wu, et al., Temporal dynamics in viral shedding and transmissibility of
COVID-19, Nat Med \textbf{26}, 672–675 (2020). \url{https://doi.org/10.1038/s41591-020-0869-5}

\bibitem{li2020} R. Li, S. Pei, B. Chen, Y. Song, T. Zhang, W. Yang, and J. Shaman,
Substantial undocumented infection facilitates the rapid dissemination of novel
coronavirus (SARS-CoV-2) Science \textbf{368}, 489--493 (2020). \url{https//doi.org/10.1126/science.abb3221}

\bibitem{zhao2011}
\textcolor{black}{Y. Lou and X.-Q. Zhao, A reaction-diffusion malaria model with
incubation  period in the vector population, J.  Math. Biol. {\bf 62}, 543--568 (2011).}

\bibitem{red} See, e.g., section 6.4 in
  \url{http://physics.bu.edu/~redner/896/spin.pdf}

\bibitem{LivedDensityGreece}
\url{https://theconversation.com/think-your-country-is-crowded-these-maps-reveal-the-truth-about-population-density-across-europe-90345}

\bibitem{greecewiki_Spatial}
\textcolor{black}{\url{https://es.wikipedia.org/wiki/Archivo:COVID-19_Outbreak_Cases_in_Greece_per_regional_unit_(prefecture).svg}}

\bibitem{GreecePopDensity} {\url{https://commons.wikimedia.org/wiki/File:Greece_Population_Density,_2000_(6172438874).jpg}}

\end{thebibliography}
\end{document}